\documentclass[twocolumn]{aastex631}
\usepackage{soul}
\usepackage{booktabs}
\usepackage{hyperref}

\def\hii{\mbox{H\,{\sc ii }}}

\begin{document}

\title{Star Formation at the Periphery of a Molecular Superbubble: The Case of G12.79+0.43}

\author[0009-0009-6941-7131]{Arun Seshadri}
\affiliation{Indian Institute of Space Science and Technology (IIST), Thiruvananthapuram, Kerala, India -  695547}

\author[0000-0002-0801-8550]{Veena V. S.}
\affiliation{Department of Astronomy \& Astrophysics, Tata Institute of Fundamental Research,
        Mumbai 400005, India}
\affiliation{Max Planck Institute for Radio Astronomy, P.O. Box 20 24
D-53010 Bonn, Germany}

\author[0000-0002-3477-6021]{Sarita Vig}
\affiliation{Indian Institute of Space Science and Technology (IIST), Thiruvananthapuram, Kerala, India -  695547}

\author[0009-0003-2273-1090]{Ashish P John}
\altaffiliation{Intern at IIST under the National Initiative on Undergraduate Science (NIUS) scheme of HBCSE, TIFR, Mumbai, India. }
\affiliation{Cochin University of Science and Technology, Kochi, Kerala, India - 682022}



\begin{abstract}
We present a multiwavelength investigation of the molecular cloud complex G12.79+0.43, which extends over $\sim18'$ on the sky. Several infrared- and radio-bright regions are arranged along an irregular rim, surrounding a central region characterised by diffuse 24~$\mu$m emission. CO molecular line observations reveal three prominent velocity components along the line of sight. Low-frequency radio continuum observations at 666 and 1300~MHz show diffuse emission spanning $\sim10.5'$ ($\sim$7.3~pc), predominantly filling the central region enclosed by the infrared-bright structures. We identify 70 compact radio sources and six \hii~regions across the cloud complex, which are likely powered by early B-type ZAMS stars. Using infrared data, we identify a total of 82 YSO candidates, including 28 Class~I sources, distributed across the cloud complex. On larger scales, the kinematics of the molecular gas over a $2^\circ\times2^\circ$ field indicate that G12.79+0.43 is located along the rim of a larger molecular superbubble (diameter $\sim50$~pc) that also encompasses the well-known W33 region. The inferred expansion age of this superbubble is $\sim0.3$~Myr. While the spatial association between G12.79+0.43 and the superbubble is evident, the current data do not allow us to establish a clear causal connection between the superbubble evolution and the ongoing star formation within G12.79+0.43.
\end{abstract}

\keywords{ISM: clouds--(ISM:) HII regions--ISM: kinematics and dynamics--stars: protostars
}


\section{Introduction} \label{sec:intro}
 Massive stars (M $\gtrsim 8~M_{\odot}$) are formed deeply embedded in their natal molecular clouds. Young massive stars and their lower mass counterparts are under active investigation to understand the conditions under which star formation is initiated and subsequently proceeds \citep{2018ARA&A..56...41M, 2023ASPC..534..153H, 2025arXiv250116866B}. Massive stars are often formed in the densest regions of giant molecular clouds (GMCs), such as infrared dark clouds (IRDCs) and central hubs of hub-filament systems (HFS), making direct observations of their formative stages difficult \citep{2019A&A...629A..81T, 2021A&A...652A..71S}. As a result, the exact mechanism of massive star-formation is not yet completely understood. Other factors that compound our difficulties in discerning a clear observational evolutionary sequence in massive star formation are their short formation timescales, their rarity, and relatively large distances. To add to this, massive stars form in clusters, along with other low-mass companions and resolving individual protostars starts becoming difficult as distance to these regions increases.
 
At the earliest stages of evolution, the radiation from a massive young stellar object (MYSO) heats the surrounding dust, ionises atoms, and photo-dissociates molecules, while stellar winds clear circumstellar material and erode the natal cloud \citep[e.g.,][]{2021A&A...654A.109K}. During the main sequence phase, intense UV radiation photoionises the parent molecular cloud, and fast stellar winds carve out cavities or bubbles, injecting turbulence into the surrounding medium \citep{{2008ApJ...681.1341W},{2006ApJ...649..759C}}. In the final stages of their evolution, massive stars explode as supernovae, injecting $\sim10^{51}$~ergs of kinetic energy into the ISM. Stellar feedback from massive stars dramatically alters the structure and dynamics of the ISM, creating interstellar shells up to $\sim10^3$ pc in size \citep[e.g.,][]{{2008MNRAS.387...31D},{2012MNRAS.421.3159M},{2022A&A...668A..44S},{2023ApJ...944L..22B}} and producing a multiphase medium composed of regions with distinct temperatures, densities, and ionization states \citep{2011BSRSL..80..297C}. To better understand the interplay between stellar feedback and the ISM, it is essential to study regions actively forming massive stars, where the impact of feedback processes can be directly observed. Cloud complexes encompassing a range of evolutionary stages offer crucial insights into how young stellar populations interact with and reshape their environments.

\begin{figure}
    \centering
    \hspace*{-0.9cm}
    \includegraphics[width = 1.2\columnwidth]{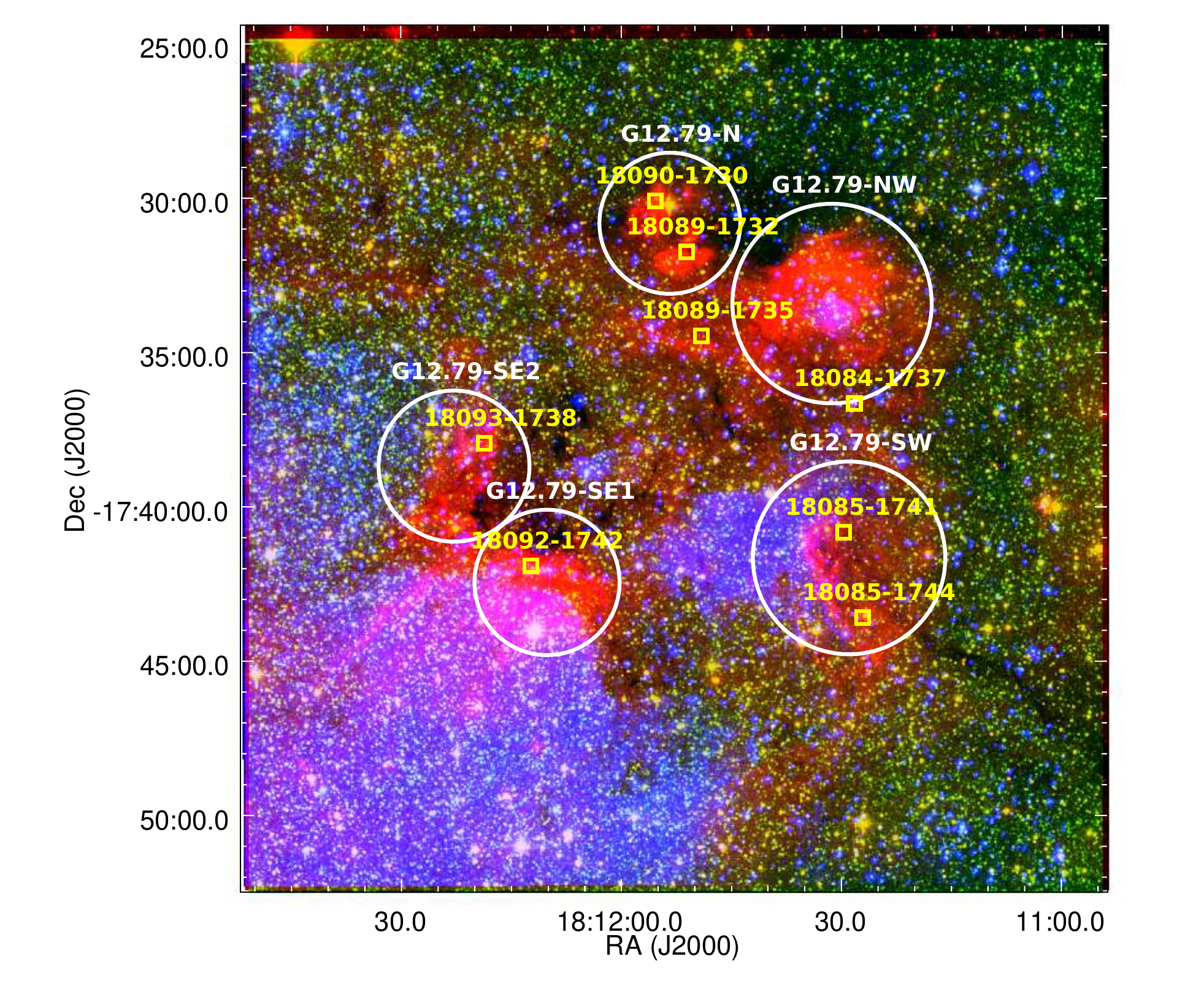}
    \caption{The three colour composite image for the cloud complex G12.79 + 0.43, showing the GLIMPSE $5.8~\mu$m in red, 2MASS K-band image in green, and DSS2 image in blue. The subregions are marked with white dotted circles. The IRAS objects associated with the complex are indicated as yellow squares.}
    \label{fig:1}
\end{figure}

 The cloud complex G12.79 + 0.43 (hereafter G12.79) with an angular extent of $15' \times 18'$, hosts the optical nebula RCW~155 ($\alpha_{\textrm{J}2000} = 18^{h}12^{m}12^{s}$ and $\delta_{\textrm{J}2000} = -17^{\circ}40'0.0"$) \citep{1960MNRAS.121..103R}. There are multiple IRAS objects (shown in figure~\ref{fig:1}), the most prominent ones being IRAS~18089-1732 and IRAS~18092-1742. An arc-shaped structure is located to the west, housing two IRAS objects, IRAS~18085--1741 and IRAS~18085--1744. The source IRAS~18092$-$1742 exhibits diffuse emission extending both to the south and northeast, coinciding with IRAS~18093$-$1738. In addition, a bright-rimmed structure with central diffuse emission is observed to the north in \textit{Spitzer} images and has been identified as an infrared bubble by \citet{2019MNRAS.488.1141J}.

Using the ATLASGAL $870~\mu m$ survey data,   \citet{2018MNRAS.473.1059U} identified 19 clumps within the complex, and  \citet{2017MNRAS.471..100E} have identified a total of 69 Hi-GAL sources using the \textit{Herschel} space telescope.. \citet{2014ApJS..212....1A} have identified 16 \hii~regions using the infrared WISE data.  Besides, \citet{2021MNRAS.500.3027D} have utilised the SEDIGISM CO survey and identified 13 clouds in the region using the SCIMES algorithm. Collectively, these studies indicate the presence of multiple active star-forming sites within the G12.79+0.43 complex. The Local Standard of Rest (LSR) velocities of different clouds within this complex, as reported in these studies, vary between $\sim$16 and 35~km/s, corresponding to kinematic distances ranging from $\sim$2 to 3.5~kpc \citep{2013ApJS..209....2S, 2016PASA...33...30R, 2018MNRAS.473.1059U}. Trigonometric parallax of masers associated with the northern region ($l=12.88^{\circ}, b=0.49^{\circ}$) yields a distance estimate of $\sim2.4$~kpc \citep{2011ApJ...733...25X, 2013A&A...553A.117I}. We therefore adopt a distance of 2.4~kpc for the G12.79+0.43 complex, which is further justified in  Sect.~\ref{sec:discussions}.

 We identify five prominent mid-infrared (MIR) bright regions within the cloud complex and label them based on their relative locations as G12.79-N, G12.79-NW, G12.79-SW, G12.79-SE1, and G12.79-SE2. The cloud complex contain an MIR bright-rimmed structure, multiple ATLASGAL and Hi-Gal clumps, YSOs, and possibly multiple interacting components. Motivated by these characteristics, we explore whether recent star formation in the region may have been influenced by feedback from massive stars. We analyse the region across a broad range of wavelengths, from radio to optical, and attempt to reconstruct a plausible evolutionary scenario for the cloud complex.

The article is organised as follows. Section~\ref{sec:data} describes the observations, data analysis, and archival datasets used in this study. Section~\ref{sec:results} presents the results. Section~\ref{sec:discussions} discusses these results in the context of the broader Galactic-plane environment, and Section~\ref{sec:conclusions} summarises the main conclusions.

\section{Observations and data} \label{sec:data}
In order to probe the star formation activity towards G12.79+0.43, we used observations and archival data spanning radio, submillimetre, infrared, and optical wavelengths.

\subsection{Radio continuum observations} \label{subsec:radio_obs}

The radio continuum observations were carried out using the upgraded Giant Metrewave Radio Telescope \citep[uGMRT; ][]{2017CSci..113..707G}, located in Khodad, India, at 550–750~MHz (Band~4). The sources 3C48 and 3C286 were observed as primary flux calibrators, and 1822$-$096 was used as the phase calibrator. Details of the observations are listed in Table~\ref{tab:1}. The total bandwidth was 200~MHz, comprising 2048 channels. The edge channels were noisy and were excluded from further analysis. The remaining inner channels were divided into six spectral sub-bands of $\sim$32~MHz each, which were flagged and calibrated separately using the Astronomical Image Processing Software (\texttt{AIPS}) tasks \texttt{UVFLG}, \texttt{TVFLG}, \texttt{CALIB}, and \texttt{CLCAL}. Multiple rounds of self-calibration and imaging were performed for each dataset using the \texttt{IMAGR} and \texttt{CLCAL} tasks, resulting in six images. Two sets of images were generated: one using \texttt{UVTAPER} = $-30~30$ and another using \texttt{UVTAPER} = $-50~50$. The former enhances sensitivity to diffuse emission at the expense of angular resolution, while the latter samples the full uv-range, providing higher angular resolution and more uniform sensitivity across spatial scales. All images were corrected for the primary beam response using the \texttt{PBCOR} task. The resulting synthesized beam sizes range from $\sim4\arcsec$–$7\arcsec$ for the low-resolution images and $\sim3\arcsec$–$6\arcsec$ for the high-resolution images. All the  images were convolved to a beam of full-width half-maximum (FWHM) $4.8\arcsec \times 6.5\arcsec$ (high resolution), and $8\arcsec \times 8\arcsec$ (low resolution)  using the task \texttt{CONVL}. The first spectral band image had significantly higher rms compared to the other images and was excluded from further analysis. The 5 images were combined using the task \texttt{SUMIM}, to obtain the final image. The rms of the final high resolution and low resolution  images are $\sim 60$ and $\sim 95$~$\mu$Jy/beam, respectively.

\subsection{Archival Data} \label{subsec:archival_data}
We have utilised archival data in the optical, infrared as well as radio wavebands in obtain a comprehensive picture of the cloud complex. The details of the archival data used in this work are presented in Table~\ref{tab:2}. The Digitized Sky Survey (DSS) image is used for locating the regions associated with optical emission.  The \textit{Spitzer} Galactic Legacy Infrared Midplane Survey Extraordinaire (GLIMPSE) has been used for investigating the warm dust emission in mid-infrared regime, while the cold dust emission has been mapped using the \textit{Herschel} infrared Galactic Plane Survey (Hi-GAL). For probing the gas kinematics, we have used a CO survey of the Galactic plane: the FOREST Unbiased Galactic plane Imaging survey (FUGIN) that covers $^{12}$CO, $^{13}$CO and C$^{18}$O emission corresponding to the \textit{J}=1--0 transition. The radio emission at $1.3$~GHz towards the region has been investigated using the SARAO MeerKAT Galactic Plane Survey (SMGPS).

\begin{table}[]
    \centering
    \begin{tabular}{lc}
        \hline \hline
        Source & RCW 155 \\
        Observation date & 5 December, 2021 \\
        Frequency & 550$-$750 MHz (Band-4) \\
        On source time & 270~min \\
        Flux calibrator & 3C286, 3C48 \\
        Phase calibrator & 1822$-$096 \\
        Central frequency & 666.07~MHz \\
        $uvrange=0-30~k\lambda$ (low resolution) &  \\
        Synthesised beam & $8\arcsec \times 8\arcsec$ \\
        Pixel size & $2\arcsec$ \\
        Noise & $95$~$\mu$Jy/beam \\
        $uvrange=0-50~k\lambda$ (high resolution) &  \\
        Synthesised beam & $6.5\arcsec \times 4.8\arcsec$ \\
        Pixel size & $1.2\arcsec$ \\
        Noise & $60$~$\mu$Jy/beam \\
        \hline \hline
    \end{tabular}
    \caption{Details of radio continuum observations carried out using uGMRT, Pune, and image parameters}
    \label{tab:1}
\end{table}

\begin{table*}[]
    \centering
    \begin{tabular}{cccc}
        \hline \hline
        Survey & Wavelength/frequency & Resolution & References \\
        \hline \hline
        DSS & 630 - 690~nm & $1\arcsec$ & \citet{10.26131/irsa441} \\
        GLIMPSE & $3.6, 4.5, 5.8, 8.0~\mu$m & $1.7\arcsec$, $1.7\arcsec$, $1.9\arcsec$, $2.0\arcsec$ & \citet{2003PASP..115..953B} \\
        Hi-GAL & $70, 160, 250, 350, 500~\mu$m & $6\arcsec$, $12\arcsec$, $18\arcsec$, $24\arcsec$, $35\arcsec$ & \cite{2016AnA...591A.149M} \\
        MAGPIS & 20~cm & $6.2\arcsec \times 5.1\arcsec$ & \citet{2006AJ....131.2525H} \\
        FUGIN & $^{12}$CO, $^{13}$CO and C$^{18}$O (J=1-0) & $\theta = 30\arcsec$, $\Delta v = 0.63$~km/s & \citet{2017PASJ...69...78U} \\
        SMGPS & 1.3~GHz & $8\arcsec$ & \citet{2024MNRAS.531..649G} \\
        \hline \hline
    \end{tabular}
    \caption{Details of archival data used in the current work}
    \label{tab:2}
\end{table*}

\section{Results} \label{sec:results}

\subsection{Molecular Line Emission} \label{subsec:molecular}
We utilise CO emission data from the FUGIN  survey to probe molecular gas towards the entire complex. The FUGIN survey covers $^{12}$CO, $^{13}$CO and C$^{18}$O emission corresponding to the \textit{J}=1--0 transition. We find multiple velocity components in $^{12}$CO, at about 9.5, 19, 24, 31, 47 and 56 km/s with varying intensity across the regions. In this work, we consider velocity components between 15-38 km/s that encompasses the strongest emission associated with the cloud complex.  

\begin{figure}
    \centering
    \includegraphics[width=\columnwidth]{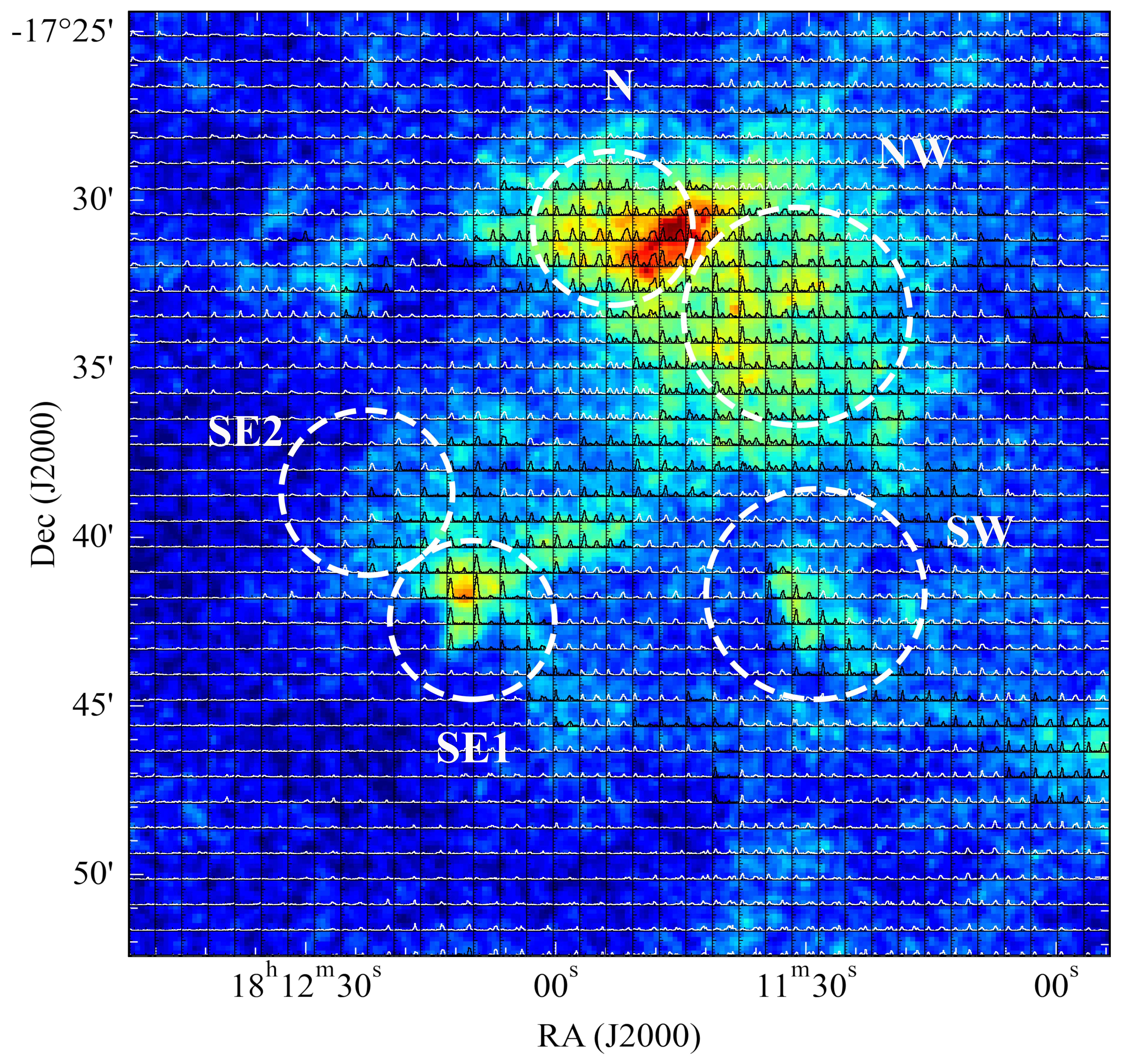}
    \caption{The $^{13}$CO spectra for $6 \times 6$ pixels, overplotted on the $^{12}$CO moment-0 map. The spectral X-axis ranges between 15-38~km/s, while the Y-axis ranges between -0.5-16~K.}
    \label{fig:2}
\end{figure}

Figure~\ref{fig:2} shows the intensity map for $^{13}$CO integrated within this velocity range. To gain an insight into the line profiles across the region, we have overlaid spectra corresponding to $6 \times 6$ pixels on the moment-0 image. We notice that 2-3 velocity components of varying intensity are seen across the region.
     
\begin{figure*}
    \centering
    \includegraphics[width=\textwidth]{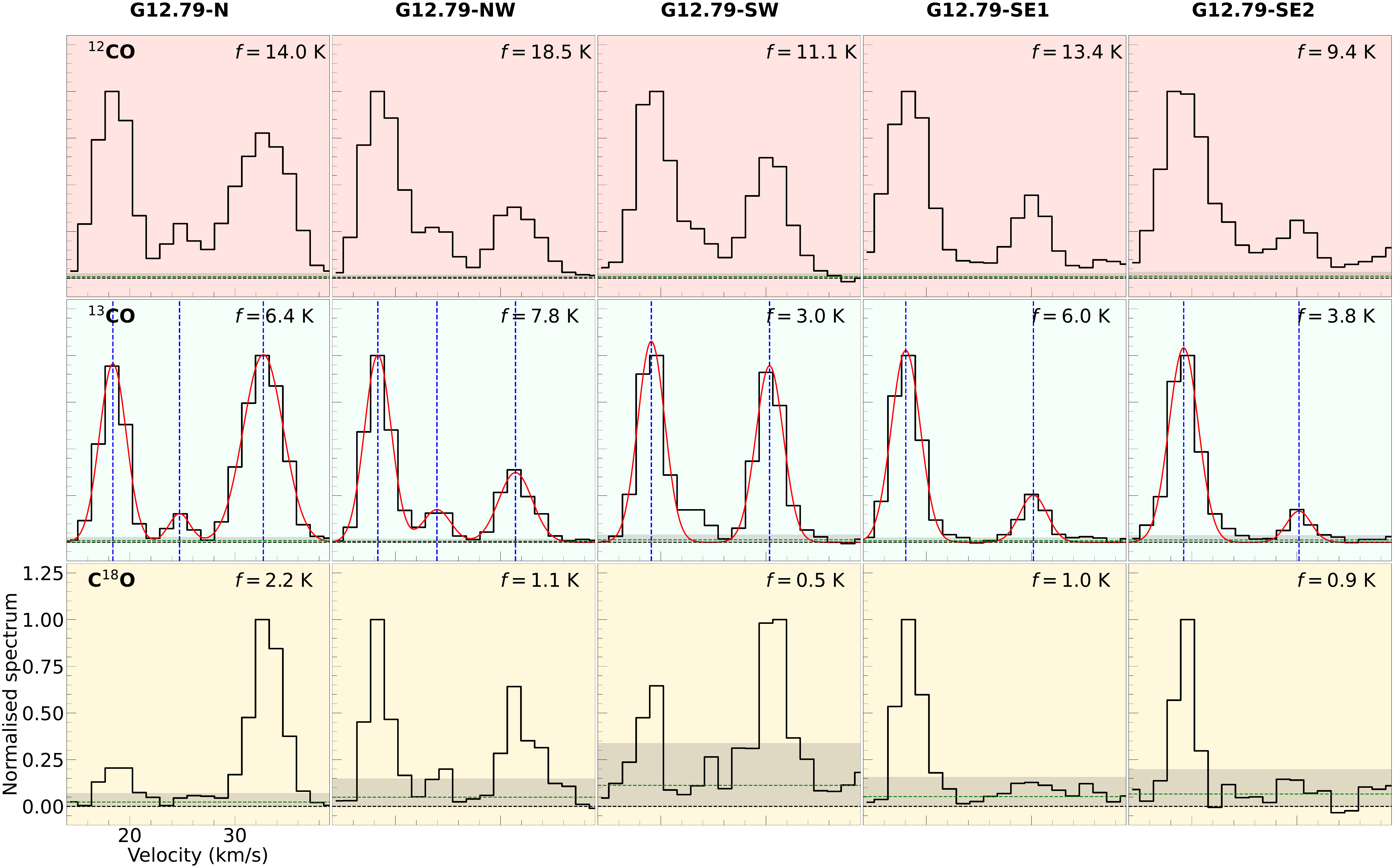}
    \caption{Normalised spectra of $^{12}$CO (top), $^{13}$CO (middle), and C$^{18}$O (bottom) extracted towards the five subregions marked in Figure~\ref{fig:1}.  The spectrum has been smoothed by averaging over two consecutive channels. The factor $f$, displayed at the top right corner, represents the normalisation with respect to the peak value. For $^{13}$CO, the red line corresponds to the total fit, consisting of multiple gaussian functions. The vertical dashed lines represent the LSR velocities corresponding to the identified central velocities, from the gaussian fitting. The black dashed line corresponds to the zero line. The green dashed line is the rms value for the spectra, indicating the noise level. The gray shaded region denotes 3 times the rms levels.}
    \label{fig:3}
\end{figure*}

\begin{figure*}
    \centering
    \includegraphics[width=0.85\textwidth]{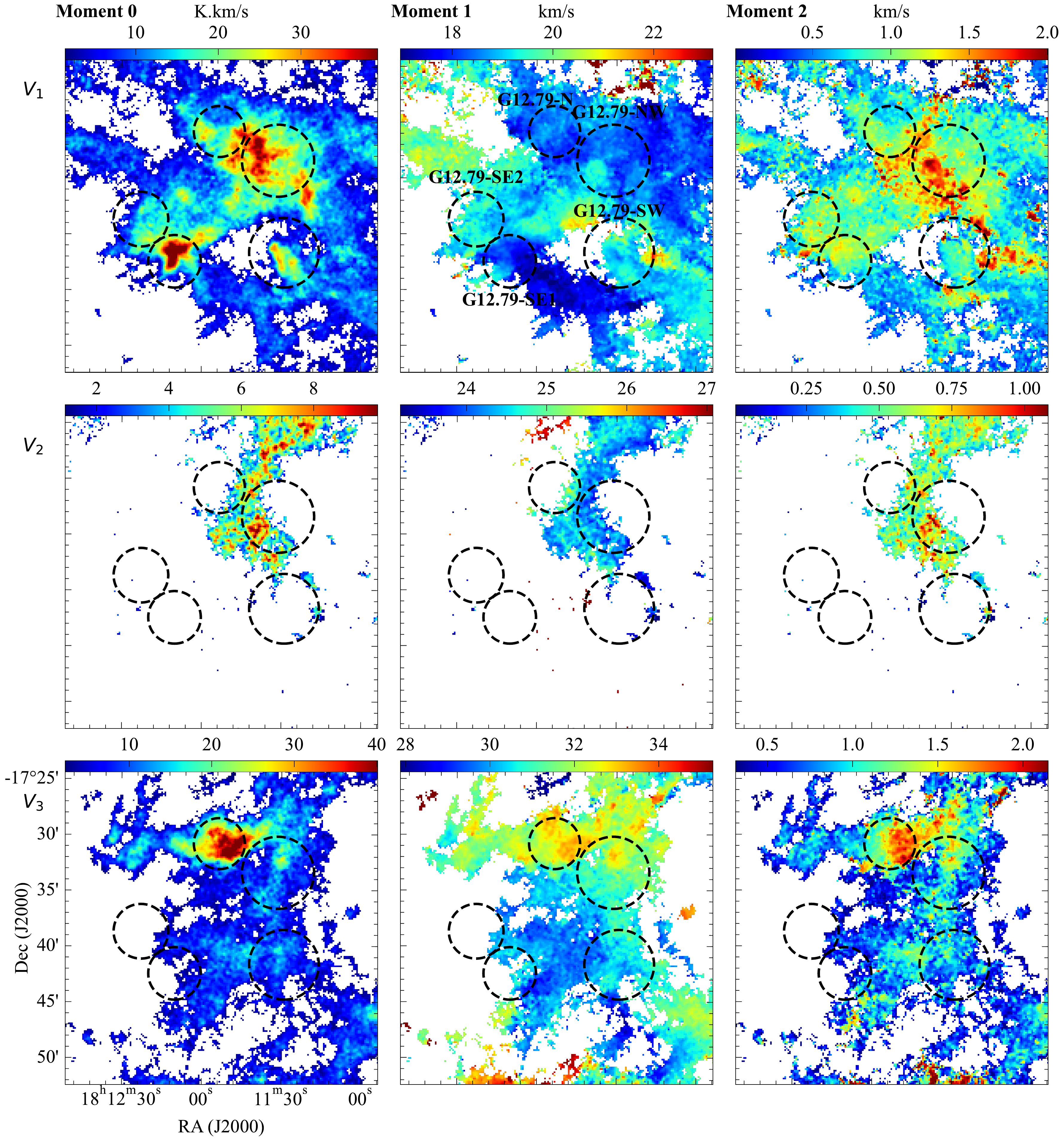}
    \caption{The moment-0, moment-1, and moment-2 maps for $^{13}$CO corresponding to the three components. The five sub regions are overlaid as black dashed circles.}
    \label{fig:4}
\end{figure*}

\begin{figure*}
    \centering
    \includegraphics[width=\textwidth]{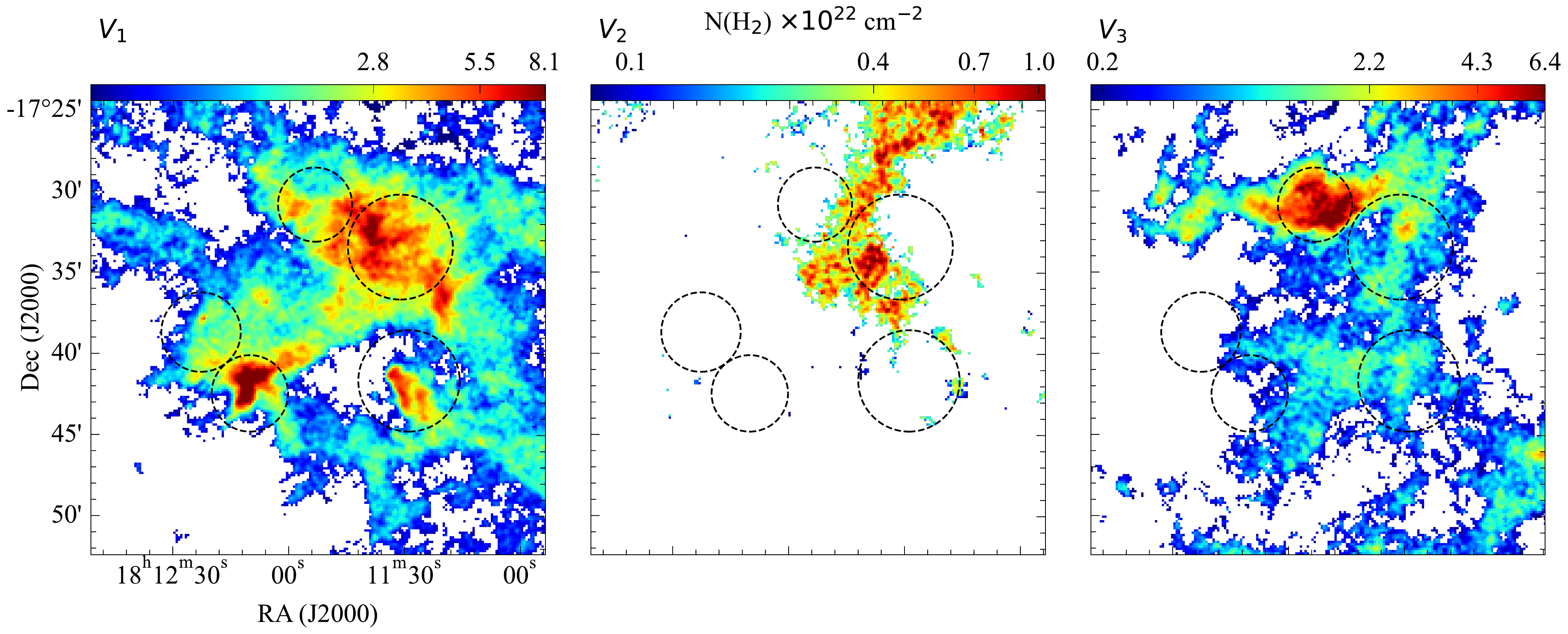}
    \caption{The column density images corresponding to the different components, generated using the $^{12}$CO and $^{13}$CO data.}
    \label{fig:5}
\end{figure*}

In order to identify the clouds corresponding to these components, we have extracted the $^{12}$CO, $^{13}$CO and C$^{18}$O spectra within circular apertures corresponding to the five subregions shown in Figure~\ref{fig:2}. These are displayed in Figure~\ref{fig:3}. Two prominent velocity components at $\sim 19$ and $\sim 31$~km/s are clearly visible. Towards G12.79-SE1, G12.79-SE2, and G12.79-SW, the component at $\sim 19$~km/s has larger intensity, whereas both the components are of nearly equal intensity towards G12.79-N. An additional intermediate velocity component at $\sim 24$~km/s is visible towards the western region, i.e towards G12.79-NW and G12.79-SW, albeit with lower intensity compared to the other two components. We designate the components as $V_1\sim19$~km/s, $V_2\sim24$~km/s and $V_3\sim31$~km/s.
Amongst the three CO lines, we analyse the $^{13}$CO lines and fit Gaussian function to the velocity components associated with $V_1$, $V_2$ and $V_3$. We choose $^{13}$CO as it is more optically thin compared to $^{12}$CO, while sampling more diffuse gas as compared to C$^{18}$O. The best fit parameters are listed in the Appendix.

We have generated the moment maps corresponding to each component for $^{13}$CO in the velocity range 15-21 ($V_1$), 21-25 ($V_2$), 26-38 ($V_3$) km/s by masking out noisy pixels using the method outlined by \citet{2006PASP..118..590R}. These are displayed in Figure~\ref{fig:4}. The emission from component $V_1$ emission is found across the complex, with strong emission observed predominantly towards SE1 and NW. The component $V_2$ is largely observed towards NW, while component $V_3$ is strongest towards N. In component $V_1$, filamentary structures are seen connecting NW and SE1. 

For $V_1$ and $V_2$ components, there does not appear to be a clear trend in velocity gradient across the region. However, for the $V_3$ component we note that there is a distinctive velocity gradient with North regions (N and NW) red-shifted compared to the South regions (SE1, SE2 and SW). The velocity dispersion or the moment-2 maps suggest that the velocity dispersion in the region is about $0.5-2$~km/s. The velocity dispersion is larger towards the peak emission in moment-0 maps, suggesting that the onset of star formation activity is plausibly responsible for the broadening.

Column density maps corresponding to the 3  components have been generated using the masked $^{12}$CO and $^{13}$CO FUGIN maps using the methodology \citep[][and references therein]{2016MNRAS.460...82S} summarised below. It is assumed that $^{12}$CO is optically thick as compared to $^{13}$CO. The $^{12}$CO excitation temperature, corresponding to each pixel in the $^{12}$CO map is calculated using
 \begin{equation}
     T_{\mathrm{ex}}=5.5 \ln \left(1+\frac{5.5}{T_{\mathrm{b}, \text { peak }}^{12}+ c_1}\right)^{-1}
 \end{equation}

Here, $T_{b,peak}^{12}$ is the  brightness temperature of $^{12}$CO at peak emission for the pixel and 5.5~K = $h \nu(^{12}\mathrm{CO})/k_B$. Assuming that $^{12}$CO and $^{13}$CO have equal excitation temperatures along the line of sight, the optical depth of $^{13}$CO line is estimated as:

\begin{equation}
    \tau_{13}(\nu)=-\ln \left[1-\frac{T_{\mathrm{b}}^{13}(\nu)}{5.3}\left\{\exp \left(\frac{5.3}{T_{\mathrm{ex}}}-1\right)^{-1}-c_2\right\}^{-1}\right]
\end{equation}

where, $T_{\mathrm{b}}^{13}(\nu)$ is the peak brightness temperature of $^{13}$CO in the pixel, and 5.3~K = $h \nu(^{13}\mathrm{CO})/k_B$. $\nu(^{12}\mathrm{CO})$ and $\nu(^{13}\mathrm{CO})$ are 115.271 and 110.201~GHz, respectively for $J=1-0$ transitions. The background emission is corrected using the constants $c_1$ and $c_2$, each having values $0.82$ and $0.16$, respectively. Further, it is assumed that the kinetic and excitation temperatures are equal, and that the energy levels are populated according to the Boltzmann distribution. Then, $^{13}$CO column density is estimated using the equation:

\begin{equation}
    N\left(^{13} \mathrm{CO}\right)=3.0 \times 10^{14} \times \frac{\tau_{13}}{1-e^{-\tau_{13}}} \times \frac{\int T_{\mathrm{b}}^{13}(v) \mathrm{d} v}{1-e^{-5.3 / T_{\mathrm{ex}}}} ,
\end{equation}
where $\int T_{\mathrm{b}}^{13}(v) \mathrm{d} v$ is the integrated intensity in units of K~km/s. Assuming the [$^{12}$CO/$^{13}$CO] isotopic ratio to be 77 \citep{1994ARA&A..32..191W}, and the [H$_2$/$^{12}$CO] abundance ratio as $1.1 \times 10^{-4}$ \citep{1982ApJ...262..590F}, the final $N(H_2)$ images are generated, and are displayed in Figure~\ref{fig:5}. For 3 pixels near the peak emission of the region SE1, we find that the excitation temperature is not high enough relative to brightness of $^{13}$CO due to optical depth effects. These pixel values are replaced using linear two-dimensional interpolation from the neighboring pixels. 

We take a look at the column density values for the different components. For component $V_1$, the maximum column density of $2.4 \times 10^{23}$~cm$^{-2}$ is observed towards SE1, while towards the other regions, the column densities are observed to be of the order of $10^{22}$~cm$^{-2}$. Component $V_2$ shows a maximum column density of $1.2 \times 10^{23}$~cm$^{-2}$ towards N. We also find extended gas distribution towards N, with higher column density values of the order of $10^{22}$~cm$^{-2}$. For component $V_3$, we observe a maximum column density of $9.5 \times 10^{22}$~cm$^{-2}$ towards NW, which shows clumpy emission. These values are in agreement with the values seen towards other star-forming regions \citep{2025A&A...699A.137B, 2025MNRAS.tmp.1315P, 2024MNRAS.528.2199R}.

\subsection{The distribution of dust} \label{subsec:dust_distribution}
\begin{figure}
    \centering
    \includegraphics[width=0.91\columnwidth]{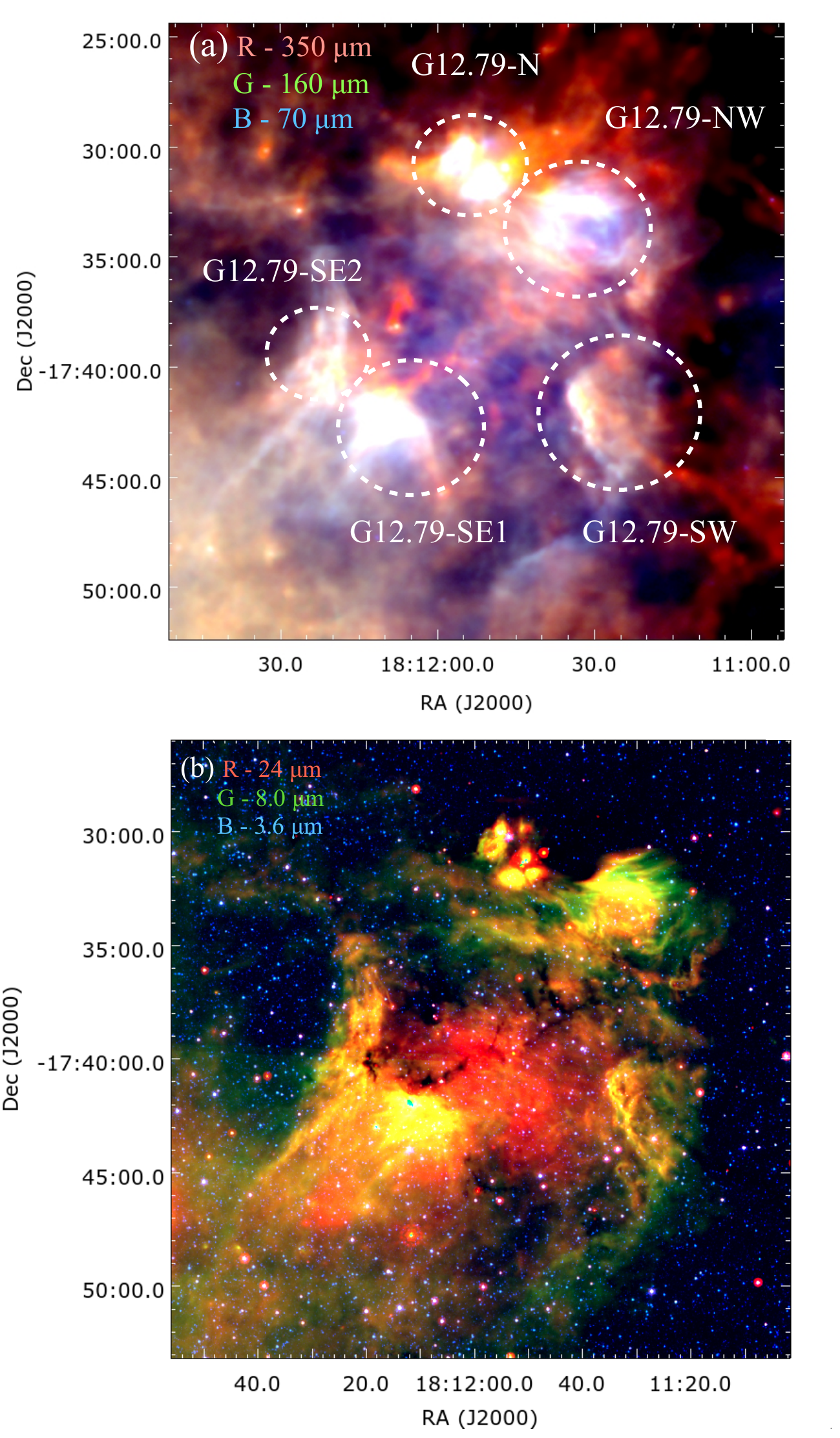}
    \caption{(a)The colour-composite image showing the \textit{Herschel} $350~\mu$m (red), $160~\mu$m (green), and $70~\mu$m (blue). (b) The \textit{Spitzer} RGB colour composite image showing MIPS $24~\mu$m (red), $8.0 \mu$m (green), and $4.5~\mu$m (blue).}
    \label{fig:6}
\end{figure}

The \textit{Herschel} Hi-Gal images map cold dust emission at five wavelength bands in the range $70 - 500~\mu$m, where the cloud spectral energy distribution (SED) is known to peak. The \textit{Herschel} colour-composite image of our region of interest using 70, 160 and 350~$\mu$m is shown in Figure~\ref{fig:6} (a). Two bright nebulous regions ($\sim 22''$ and $\sim 37''$) are observed towards N region. Towards NW, we observe faint circular features, resembling a cavity, with $70$~$\mu$m emission towards the centre. Towards SW region, we find an arc-like emission extending further south of size $\sim 6''$, with $70$~$\mu$m emission at the outer rim of the arc. SE1 features bright emission, with a ridge of size $\sim 8''$ towards the south that is dominated by $70$~$\mu$m emission. We observe faint filamentary arc-like structures towards SE2. A few faint extended structures are also  visible towards the central region at $350$~$\mu$m (visible in red), suggesting the presence of very cold dust structures. 

\begin{figure}
    \centering
    \includegraphics[width=\columnwidth]{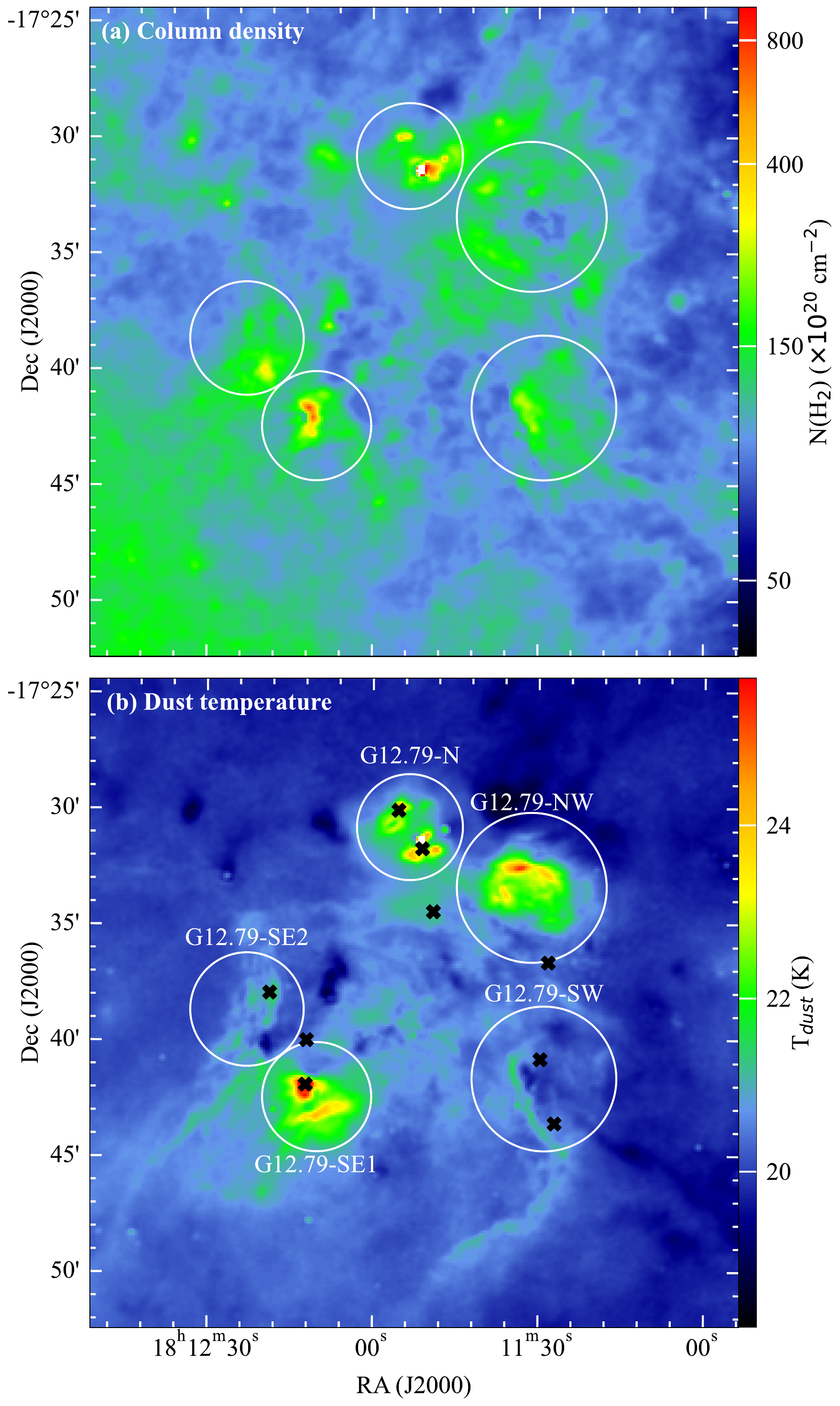}
    \caption{The \textit{Herschel} column density (\textit{top}) and dust temperature (\textit{bottom}) maps obtained from the PPMAP archive. The IRAS objects towards the regions are shown on the dust temperature map as black crosses. The five subregions are shown as white circles.}
    \label{fig:7}
\end{figure}

The Herschel maps can be used for estimating the gas column density and the dust temperature in the region. We use the high-resolution ($12''$) ``Point Process MAPping" (PPMAP) column density and dust temperature images, provided by \citet{2017MNRAS.471.2730M}. The technique uses the Bayesian procedure and assumes that the dust temperature is non-uniform along the line of sight. Further, the total column density is inferred by integrating the differential column densities across the temperature `bins'. We present the PPMAP images of the column density and dust temperature towards this region in Figure~\ref{fig:7}. A comparison with the column density image derived from CO in the previous section shows broadly similar large-scale morphological structures, although variations are observed toward a few high-density regions. A one-to-one correspondence between the two maps is not necessarily expected, because the dust-based column density traces the total line-of-sight material and depends on the adopted dust opacity and temperature, whereas the CO-based column density relies on assumptions such as LTE excitation, optical depth, and the adopted CO abundance. These differences, together with CO depletion in the densest regions, can account for the observed deviations \citep{2009ApJ...692...91G, 2022ApJ...931....9L}.

The highest column density peaks is observed towards the N region with a peak value of $\sim10^{23}$~cm$^{-2}$ towards IRAS~18089-1732. High values of column density are also found towards 
SE1 $\sim7 \times 10^{22}$~cm$^{-2}$.
Towards NW, we observe a structure resembling a broken bubble with enhanced values of column density as compared to the central region. Overall, we find that all the regions show clumpy distribution of high column density. In addition, filamentary structures are observed within these regions as well as towards the central region of the field of interest. Across our selected circular regions, the column density ranges from $\sim6.5 \times 10^{21}$ to $\sim1.2 \times 10^{23}$~cm$^{-2}$. 
The column density values are in general agreement with those found towards other massive star-forming regions \citep[for Aquila, Corona Australis, and RCW~117, respectively]{2011A&A...529L...6A, 2018A&A...615A.125B,  2024MNRAS.527.4244S}. 

The dust temperature maps display peaks towards specific locations in the circular regions of interest. The temperature peaks towards SE1 with a value of $\sim27$~K, while NW and N regions have peak temperatures of $\sim25$~K. SE2 and SW exhibit relatively lower peak temperatures $\sim21.5$~K. Towards N and SE1, localized regions are discerned with higher temperature indicative of warmer dust associated with star formation. Towards NW, we find a circular feature, akin to a broken bubble with higher temperature observed towards the northern portion of the feature. A faint arc-like structure is observed towards SW, that is elongated and wave-like. Across the circular regions, we find that the dust temperature is between $18 - 27~K$. The values agree well with typical dust temperature values found in star-forming regions \citep[towards W3, NGC~7538, and Aquila, respectively]{2013ApJ...766...85R, 2013ApJ...773..102F, 2015A&A...584A..91K}.

Warm dust emission, in the mid infrared (MIR) wavelength range $3.6 - 8.0~\mu$m, as well as at $24~\mu$m, is examined using \textit{Spitzer} GLIMPSE and MIPSGAL images, respectively. We note a striking resemblance of the features with those in the dust temperature map - bright compact regions towards N and SE1, the broken-bubble feature in NW, and the arc-like feature towards SW. The extended wave-like structure towards the south is not observed. In particular, we notice diffuse warm dust emission at $24~\mu$m (seen in red) towards the central region and southwest of the field. Further, this $24~\mu$m emission is also fills the broken bubble in the NW region whose northern edge is bright in 8~$\mu$m emission (green). A few dark features - enclosed within ellipses - are observed near the central region that display low temperature and high column densities. 

\begin{figure*}
    \centering
    \includegraphics[width=\textwidth]{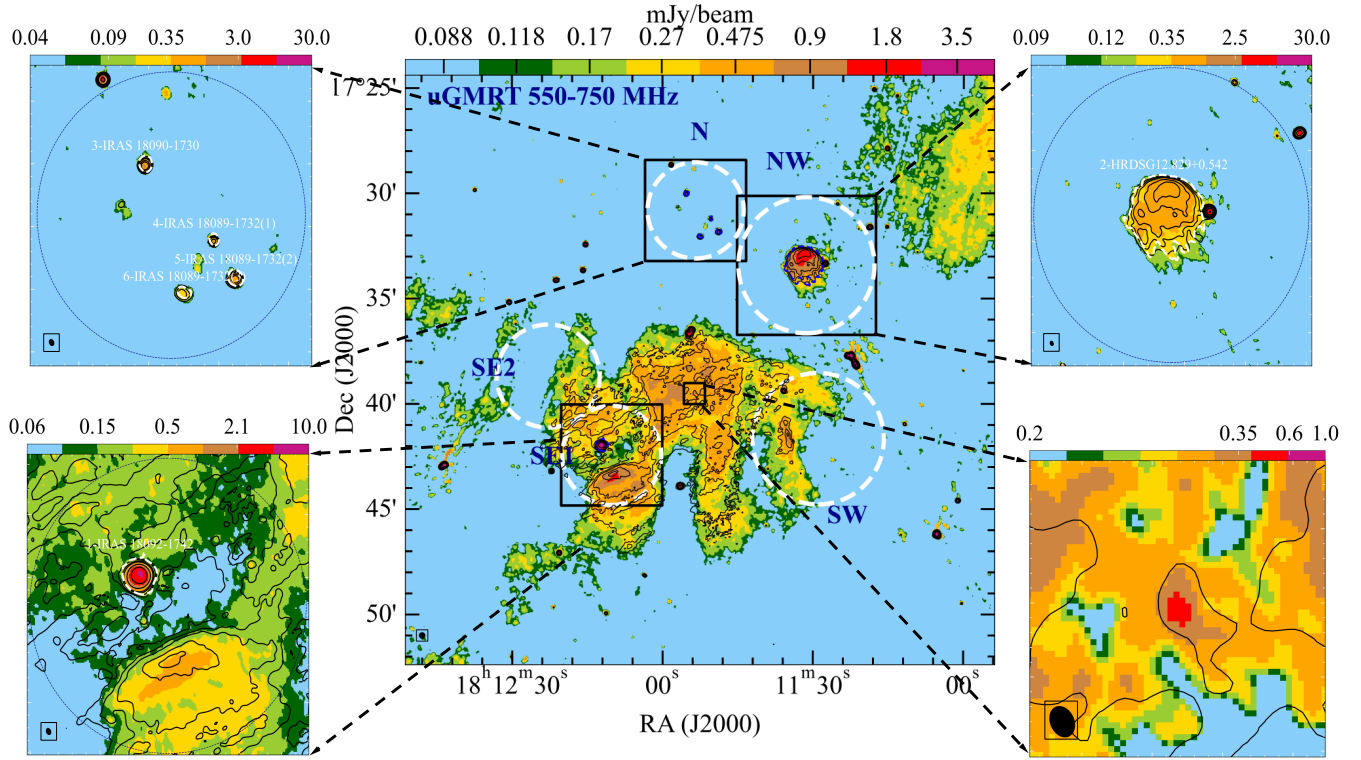}
    \caption{The radio continuum image at band-4 of uGMRT (666~MHz), with the beam measuring $8\arcsec$, shown in the bottom left corner. The SMGPS 1.3 GHz image is shown as black contours with levels 0.4, 0.6, 0.8, 1.6, 3.2, 6.4, 1.28 mJy/beam. The beam, measuring $8\arcsec \times 8\arcsec$, is shown as a black solid circle, enclosed within a box, in the bottom left corner. Identified \hii region candidates with thermal spectral indices are shown as insets.}
    \label{fig:8}
\end{figure*}

\subsection{Radio continuum emission} \label{subsec:radio_continuum}
The radio continuum emission from uGMRT at band 4 (effective frequency of 666~MHz) is shown in Figure~\ref{fig:8}. The emission is dominated by an extended emission feature toward the central region of the cloud complex. Toward the northern region (G12.79-N), three compact radio sources are detected, consistent with the distribution of warm dust emission seen in the MIR maps and with elevated dust temperatures. In the north-western region (G12.79-NW), the radio emission exhibits a cometary morphology, with the bright head oriented toward the north. The south-eastern region SE1 shows bright, approximately spherical radio emission, suggestive of a compact ionised structure. Immediately to its south, diffuse emission with a sharp intensity gradient is observed, spatially coincident with the ridge-like structure seen in infrared images. In the south-western region (G12.79-SW), radio emission is detected primarily to the east of the arc-like structure traced in far-infrared data. The south-eastern region SE2 is characterized by diffuse radio emission located predominantly on its western side. Additionally, low-level diffuse emission is present in the area between the SE1 and SW regions. 

 A number of compact radio sources are detected across the field, several of which are embedded within diffuse emission. To identify and quantify these sources, we generated a higher-resolution image using a uv-taper of 50 k$\lambda$. Compact sources were selected using a detection threshold of peak intensity $\geq 3\sigma$, where $\sigma \approx 60~\mu$Jy~beam$^{-1}$. Using this criterion, a total of 70 compact sources were identified in the uGMRT image. Applying the same selection method to the SMGPS 1.3 GHz image resulted in 47 detected sources, all of which have counterparts in the uGMRT data. Spectral indices were estimated for sources detected at both frequencies using the relation $S_\nu \propto \nu^\alpha$, where $S_\nu$ is the flux density at frequency $\nu$ and $\alpha$ is the spectral index. The derived spectral indices span a range from $-1.6$ to $+1.7$, indicating the presence of both thermal and non-thermal emission mechanisms. Adopting $\alpha < -0.5$ as a criterion for non-thermal emission, we identify 22 non-thermal and 25 thermal radio sources. The measured flux densities and spectral indices are listed in Table~\ref{tab:C_3}.

\begin{splitdeluxetable*}{ccccccccBccccccccccc}
    \tablecaption{Parameters estimated for identified \hii regions}
    \tablehead{\colhead{No.} & \colhead{Object} & \colhead{RA} & \colhead{Dec} & \colhead{Area} & \colhead{$S_\nu$} & \colhead{$N(H_2)$} & \colhead{$\theta_{src}$} & \colhead{No.}  & \colhead{$EM$} & \colhead{$n_{e}$} &  \colhead{$log_{10}(N_{Lyc})$} & \colhead{ZAMS\tablenotemark{*}} & \colhead{$R$} & \colhead{$n_{0}$} & \colhead{$R_{S}$} & \colhead{$t_{dyn}$} \\ \colhead{} & \colhead{} & \colhead{h:m:s} & \colhead{d:m:s} & \colhead{$(arcsec^{2})$} & \colhead{$(mJy)$} & \colhead{$(\times 10^{23}cm^{-2})$} & \colhead{$\arcsec$} & \colhead{} & \colhead{$(\times 10^{3}pc$ $cm^{-6})$} & \colhead{$(cm^{-3})$} & \colhead{} & \colhead{(Sp. type)} & \colhead{$(pc)$} & \colhead{$(\times 10^{5}cm^{-3})$} & \colhead{$(\times 10^{-4}pc)$} & \colhead{$(Myr)$}}
    \startdata
    1 & HRDSG12.829+0.542 & 18:11:31.81 & -17:33:35.11 & $7977.6$ & $103.0$ & $23.39$ & $100.8$ & 1 & $6.9$ & $76.9$ &  $46.7$ & B0.5 & $0.58$ & $19.4$ & $7.5$ & $4.85$ \\
    2 & IRAS 18089-1732(2) & 18:11:48.88 & -17:31:50.21 & $187.2$ & $2.8$ & $1.04$ & $15.4$ & 2 & $8.0$ & $211.15$ &  $45.14$ & B1 & $0.09$ & $5.6$ & $5.1$ & $0.24$ \\
    3 & IRAS 18089-1732(1) & 18:11:50.42 & -17:31:13.42 & $83.5$ & $1.1$ & $1.59$ & $10.3$ & 3 & $7.4$ & $247.94$ &  $44.75$ & B2 & $0.06$ & $12.9$ & $2.2$ & $0.23$ \\
    4 & IRAS 18089-1732(3) & 18:11:52.54 & -17:32:04.31 & $110.9$ & $0.8$ & $0.65$ & $11.9$ & 4 & $3.8$ & $166.32$ &  $44.59$ & B2 & $0.07$ & $4.6$ & $3.9$ & $0.19$ \\
    5 & IRAS 18090-1730 & 18:11:55.32 & -17:30:00.82 & $159.8$ & $2.5$ & $1.65$ & $14.3$ & 5 & $8.3$ & $223.92$ &  $45.08$ & B1 & $0.08$ & $9.7$ & $3.4$ & $0.28$ \\
    6 & IRAS~18092-1742 & 18:12:12.23 & -17:41:59.07 & $941.8$ & $41.9$ & $6.45$ & $34.6$ & 6 & $23.9$ & $243.8$ &  $46.3$ & B0.5 & $0.20$ & $15.6$ & $6.4$ & $0.84$ \\
    \enddata
    \tablenotetext{*}{From \citet{1973AJ.....78..929P}}
    \label{tab:3}
\end{splitdeluxetable*}

 While some of these non-thermal compact sources have angular sizes consistent with compact or ultra-compact \hii~regions at a distance of 2.4~kpc, their steep radio spectra are not expected for pure thermal free--free emission ($\alpha \approx -0.1$ to $+2$). We therefore interpret most of the sources with $\alpha < -0.5$ as likely synchrotron-dominated emitters, plausibly dominated by background extragalactic sources seen through the Galactic plane, with a subset potentially associated with Galactic non-thermal processes (e.g. shocks). Previous GMRT studies have shown that some \hii~regions can exhibit a mixture of thermal and non-thermal emission in spectral index maps \citep{{2016MNRAS.456.2425V},{2016AJ....152..146N}}. However, disentangling such contributions requires spatially resolved spectral index information and broader frequency coverage than available in the present study. We refrain from estimating spectral indices for the diffuse radio emission, as both the uGMRT and SMGPS may be affected by missing short spacings. This can lead to spatial filtering of extended emission, potentially biasing spectral index estimates. Consequently, a reliable spectral index determination for the extended emission is not possible with the present data. 
 
 Compact radio sources associated with massive star formation are expected to show infrared counterparts due to heated dust. To identify such associations, we cross-matched the positions of compact radio sources with the GLIMPSE '07 8~$\mu$m, MIPSGAL $24~\mu$m, and the Hi-Gal $70~\mu$m point source catalogues using \texttt{TOPCAT} with a search radius of $2\arcsec$. This resulted in 16 GLIMPSE, three MIPSGAL $24~\mu$m, and one Hi-GAL $70~\mu$m counterparts, as listed in Table~\ref{tab:C_3}. We note that this catalogue-based comparison is not exhaustive, as MIPSGAL and Hi-GAL sources are often difficult to identify in regions of bright nebulosity. We therefore performed a visual inspection of the 24~$\mu$m and 70~$\mu$m images and identified four compact radio sources with clear infrared counterparts. All four have been previously identified as \hii~regions: G12.879+0.496, G12.891+0.495, G12.883+0.480, and G12.918+0.487 \citep{2006AJ....131..939Z}. In addition to these compact sources, diffuse radio emission with a nearly spherical morphology is detected toward two other known \hii~regions, HRDS~G012.825+00.542 and IRAS~18092$-$1742 \citep{1996A&AS..115...81B}. All six \hii~regions are associated with molecular gas and far-infrared dust emission, supporting their identification as sites of massive star formation.

 We estimate the emission measure, electron density, Lyman continuum photon rate, Str"omgren radius, and dynamical age for these \hii~regions using the 1.3~GHz flux densities following the formulations of \citet{2016A&A...588A.143S} and \citet{1997pism.book.....D}. An average electron temperature of $T_e \sim 7000$~K is adopted based on the Galactic electron temperature gradient presented by \citet{2006ApJ...653.1226Q}. The derived physical parameters are listed in Table~\ref{tab:3}. Based on their sizes, and following the classification scheme of \citet[their Table~3]{2005IAUS..227..111K}, we classify IRAS~18092$-$1742 and HRDS~G012.825+00.542 as compact \hii~regions, while the remaining four are identified as ultracompact \hii~regions. The inferred Lyman continuum photon rates indicate that the ionising sources are likely early B-type ZAMS stars \citep{1973AJ.....78..929P}.

\begin{figure}
    \centering
    \includegraphics[width=\columnwidth]{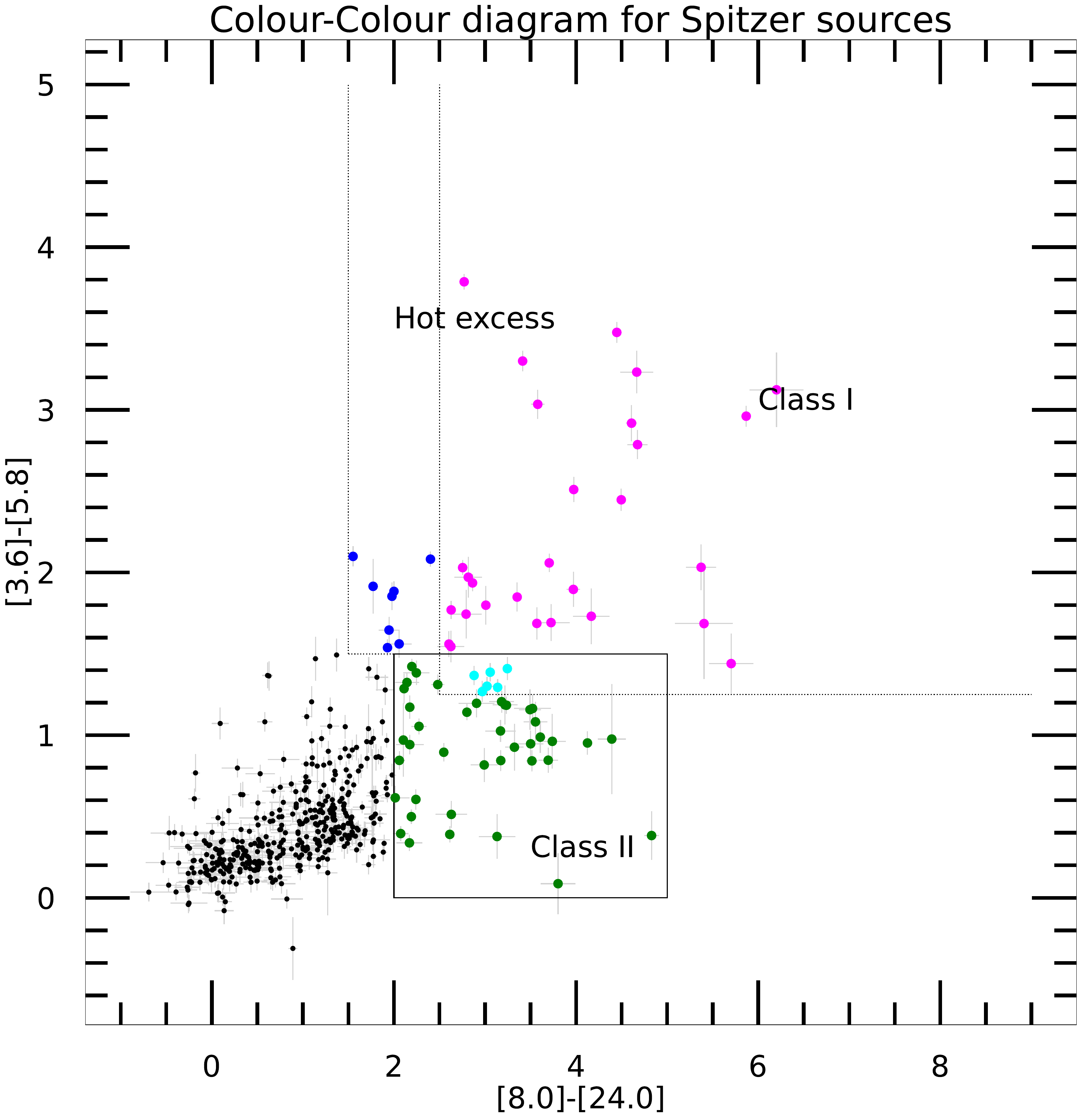}
    \caption{The [3.6]-[5.8] vs [8.0]-[24.0] colour-colour diagram showing the various YSO classes. The classification is carried out for the YSO candidates identified using the [3.6]-[4.5] vs [5.8]-[8.0] colour-colour diagram.}
    \label{fig:9}
\end{figure}

\subsection{Young stellar objects}
Young stellar objects (YSOs) are surrounded by circumstellar disks, and depending on their evolutionary stage, may also be embedded within infalling envelopes. Consequently, YSOs are most effectively identified through infrared excess emission arising from their disks and envelopes. We identify YSOs toward G12.79+0.43 region using the [3.6]-[5.8] vs [8.0]-[24.0] colour-colour diagram \citep[e.g.,][]{2004ApJS..154..379M, 2018ApJ...852...93V}. Inclusion of the 24~$\mu$m band allows us to preferentially identify younger, more deeply embedded YSOs, which typically exhibit rising flux densities toward longer infrared wavelengths in their spectral energy distributions (SEDs).

\begin{figure}
    \centering
    \includegraphics[width=\columnwidth]{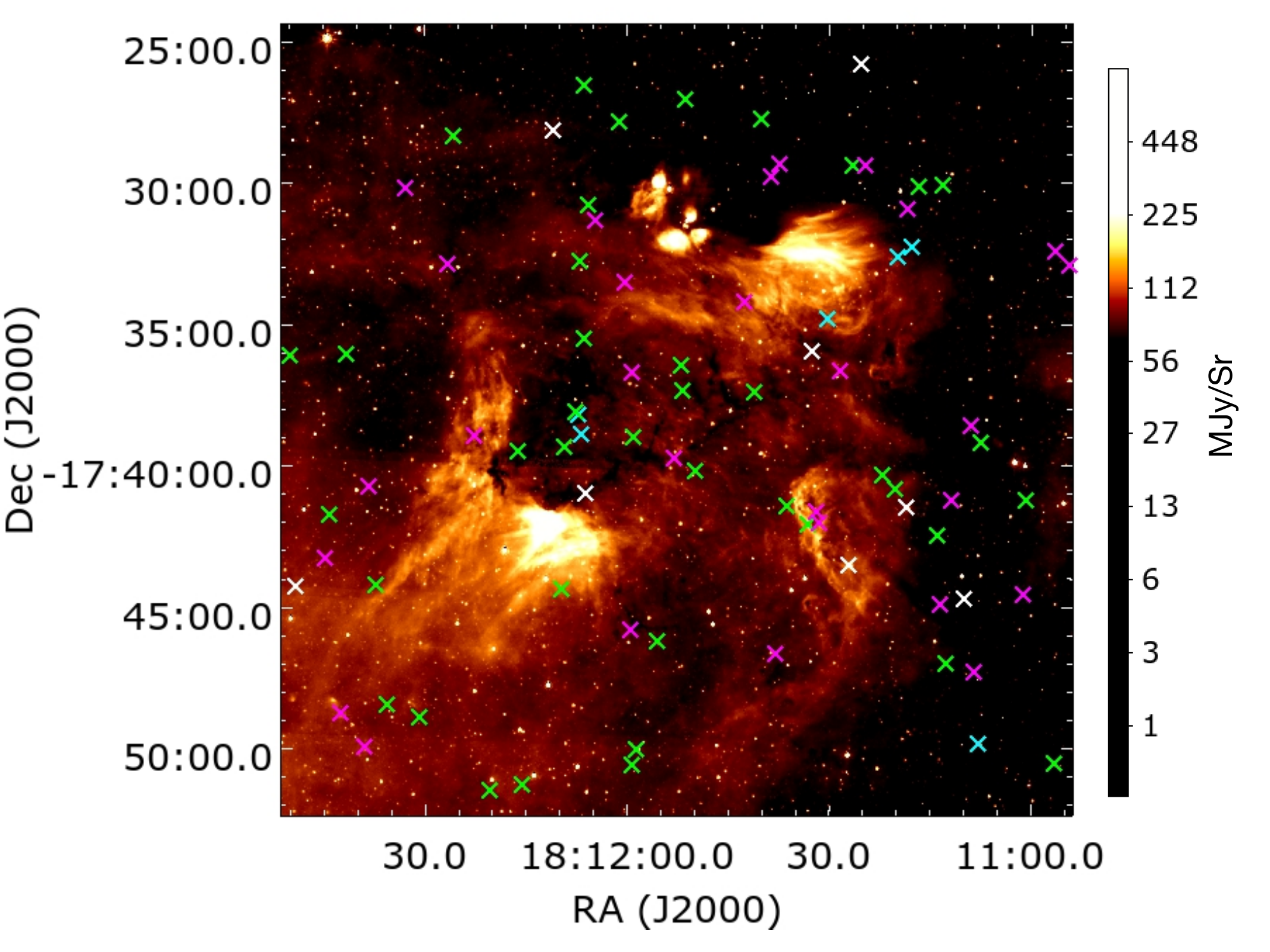}
    \caption{ The \textit{Spitzer} IRAC 4 image, showing the Class I (magenta),  Class I/II (cyan) objects, Class II (green), and hot excess (blue) objects.}
    \label{fig:10}
\end{figure}

The sources detected at 3.6, 5.8, 8.0 and 24~$\mu$m bands were extracted from the GLIMPSE I Spring '07 catalog \citep[]{10.26131/irsa405} and the MIPSGAL catalog \citep[]{2015AJ....149...64G, 10.26131/irsa435}. We identify a total of 82 YSOs, with 28 Class I sources, 40 Class II sources, 6 Class I/II sources, and 8 hot excess candidates (see Figure~\ref{fig:9}).The spatial distribution of the identified YSOs is shown in Figure~\ref{fig:10}. No clear clustering or spatial segregation between different YSO classes is observed across the region. We also searched for 70~$\mu$m counterparts using the Hi-GAL catalog \citep{2024A&A...688A.203M} and find that 14 sources have associated 70~$\mu$m emission, consistent with the presence of relatively young and embedded sources. The YSO population in this region was previously investigated by \citet{2013MNRAS.435..663B} using IRAC colour–colour criteria. They identified a total of 56 YSOs, including 22 Class~I and 36 Class~II sources. A comparison between the two samples reveals 25 common YSOs, with the classifications agreeing for approximately 86$\%$ of these sources.

\section{Discussion} \label{sec:discussions}
\subsection{A global view of star-formation towards the region}

\begin{figure}
    \centering
    \includegraphics[width=\columnwidth]{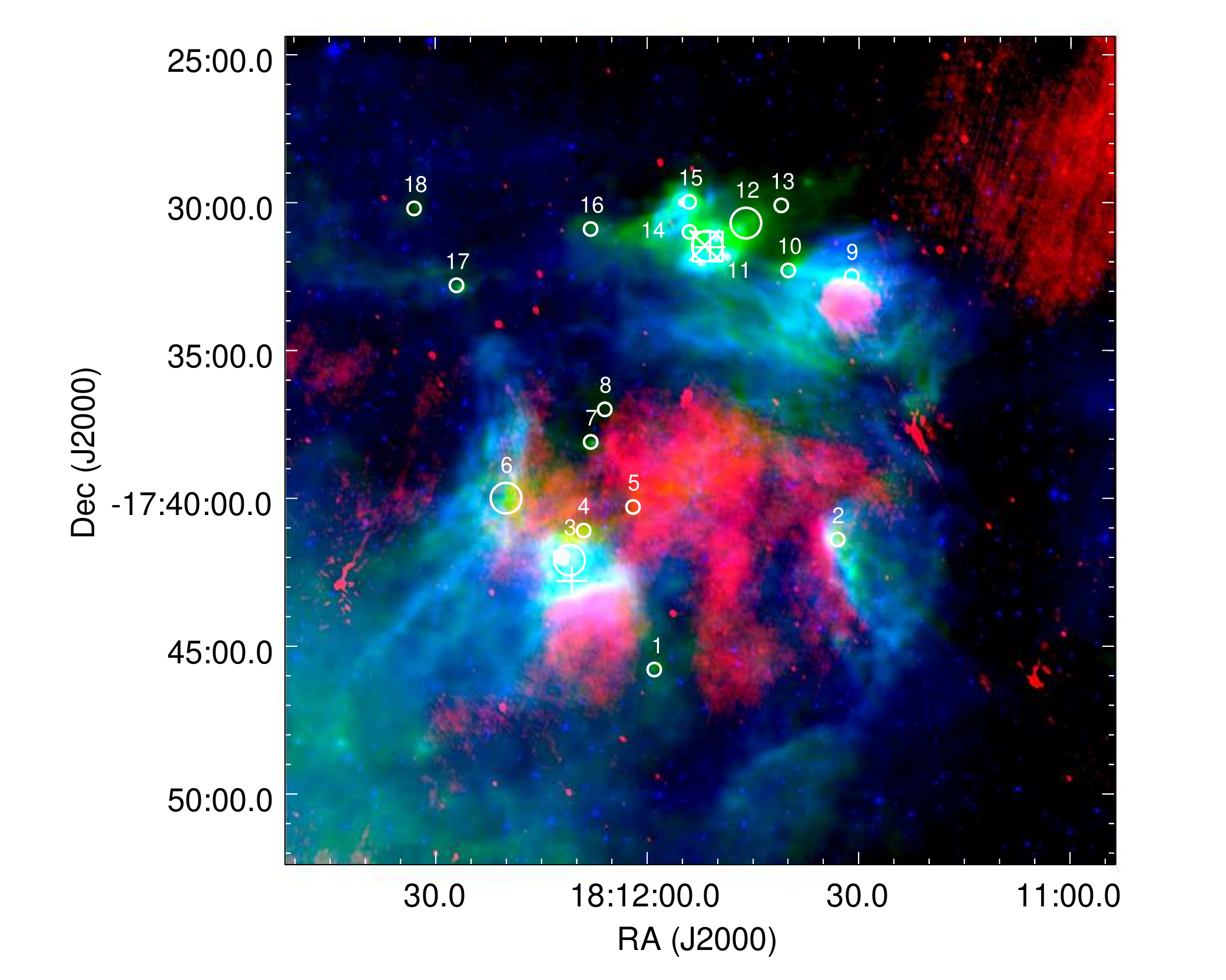}
    \caption{The RGB colour composite image of 666~MHz (red), 160~$\mu$m (green), and 8.0~$\mu$m (blue). The circles correspond to the position of ATLASGAL clumps, with the bigger circles corresponding to high mass star-formation (see Figure~\ref{fig:13}). The black cross corresponds to OH maser, the black squares correspond to the methanol masers, and the '+'s correspond to the H$_2$O masers identified towards the region.}
    \label{fig:11}
\end{figure}

\begin{figure*}
    \centering
    \includegraphics[width=\textwidth]{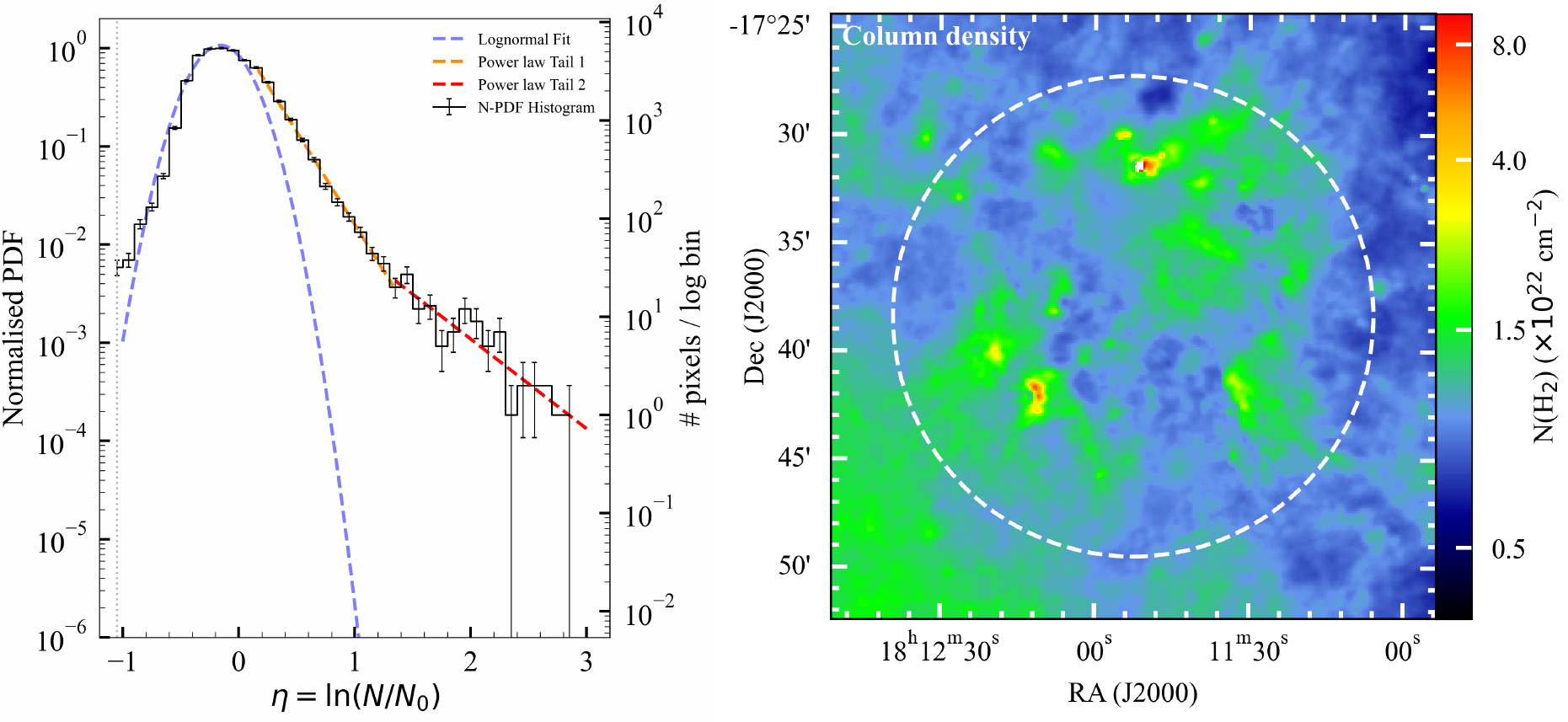}
    \caption{\textit{(Left)} Normalised probability distribution function of column density constructed for the cloud, for the region enclosed within the white dotted circle, as shown in the PPMAP column density image on the right. 
    }
    \label{fig:12}
\end{figure*}

The G12.79+0.43 region exhibits multiple observational signatures indicative of ongoing massive star formation, including extended ionised gas emission, dense molecular material traced by $^{12}$CO and $^{13}$CO, and far-infrared and submillimeter dust continuum emission. The presence of bright mid-infrared emission further points to warm dust heated by young massive stars. A population of YSOs is detected across the region, reinforcing the view that G12.79+0.43 is an active star-forming complex. In addition to diffuse radio emission, several compact radio sources are detected toward the region (Section 3.3).

The diffuse radio emission fills the central cavity ($\alpha_{J2000}=18^{h}11^{m}54.29^{s}, \delta_{J2000}=-17^{\circ}39'16.9''$) spanning $\sim10.5'$ (7.3~pc at a distance of 2.4 kpc). In contrast, the mid- and far-infrared emission traced by warm dust predominantly outlines the periphery of the cavity (Figure~\ref{fig:11}), producing a complementary morphology in which ionised gas fills the interior while dust emission traces the surrounding shell. This structure is characteristic of an evolved or partially open \hii~region with an associated hot dissociation region. Toward the southern side of the cavity, the infrared shell appears disrupted or open, and the radio emission shows a tail-like extension extending beyond the cavity boundary. This morphology is suggestive of ionised gas preferentially escaping along a lower-density path, possibly driven by density gradients in the surrounding medium. Localised radio peaks are also observed along the cavity boundary and are spatially coincident with individual star-forming regions embedded within the shell.

We constructed the column density (N) probability density function (N-PDF) using the PPMAP column density map toward the G12.79 cloud. A circular region encompassing the five sub-regions was considered for the N-PDF estimation after background subtraction (Figure~\ref{fig:12}) Previous studies have shown that N-PDFs of molecular clouds can provide insights into the effects of turbulence, gravity, feedback, and magnetic fields, with these processes often reflected in features such as double log-normal distributions and single or double power-law tails \citep{2014ApJ...781...91G, 2015A&A...576L...1L}. We fitted a log-normal function and two power-law tails to the N-PDF following the grid-search method described in \citet{2018ApJ...859..162C}. The deviation points from the log-normal component to the first power-law tail, and from the first to the second power-law tail, were determined by minimizing the sum of the squared residuals for different functional fits.

The best-fit log-normal function has $\mu=-0.15 \pm 0.01$ and $\sigma=0.22 \pm 0.01$. The power-law tails have slopes of $s_1 = -4.35 \pm 0.098$ and $s_2 = -2.10 \pm 0.31$. The deviation points are located at 0.15 and 1.35. These values are in broad agreement with those obtained toward other star-forming regions \citep{2015MNRAS.453L..41S, 2022A&A...666A.165S}. The presence of power-law tails suggests that star-formation activity has commenced in the cloud. This interpretation is supported by other star-formation tracers in the cloud, such as H{\sc ii} regions, YSOs, and masers. We find that the second power-law tail, which occurs at higher column densities, is shallower than the first. Several simulation studies have shown that a power-law tail can emerge at high column densities from an initially log-normal N-PDF due to the onset of star formation through free-fall collapse in molecular clouds \citep{2011ApJ...727L..20K, 2020ApJ...903L...2J}. The development of a shallower power-law tail at the highest column densities, as reported for a few clouds, has been attributed to feedback from young stars, such as ionization-driven compression, or to strong magnetic fields in initially subcritical clouds \citep{2024MNRAS.528..432V}.

Towards the central cavity filled with radio emission, we find 1 Class I YSO (MG012.7705+00.4182, see Table~\ref{tab:C_2}) and three other Class II YSOs - MG012.7580+00.4255, MG012.8031+00.4415, and MG012.7935+00.4028. The Class I YSO has a 70~$\mu$m counterpart but no associated compact radio emission. We also find two compact radio sources (S. No. 36 and 38 in Table~\ref{tab:C_1}) with non-thermal spectral indices ($-0.6$ and $-0.4$). In the absence of infrared counterparts, it is difficult to ascertain whether they belong to the G12.79+0.43 region or are extragalactic sources.

The G12.79 cloud complex is particularly interesting from the perspective of star formation due to the presence of multiple \hii~regions and several molecular velocity components along the line of sight. Methanol, hydroxyl, and water masers have been detected toward IRAS~18089$-$1732 and IRAS~18092$-$1742, indicating active massive star-forming sites \citep{2000ApJS..129..159A, 2005A&A...432..737P, 2010MNRAS.409..913G}. The ATLASGAL survey \citep{2022MNRAS.510.3389U} identified a total of 18 dense clumps in this region (listed in Table~\ref{tab:4}), with catalogue distances ranging between 1.8 and 3.0~kpc. Mass estimates are available for 17 of these clumps and lie in the range 27–405~M$_\odot$, while size estimates (radii) are reported for 13 clumps, spanning 0.11–0.44~pc.

\begin{figure}
    \centering
    \includegraphics[width=\columnwidth]{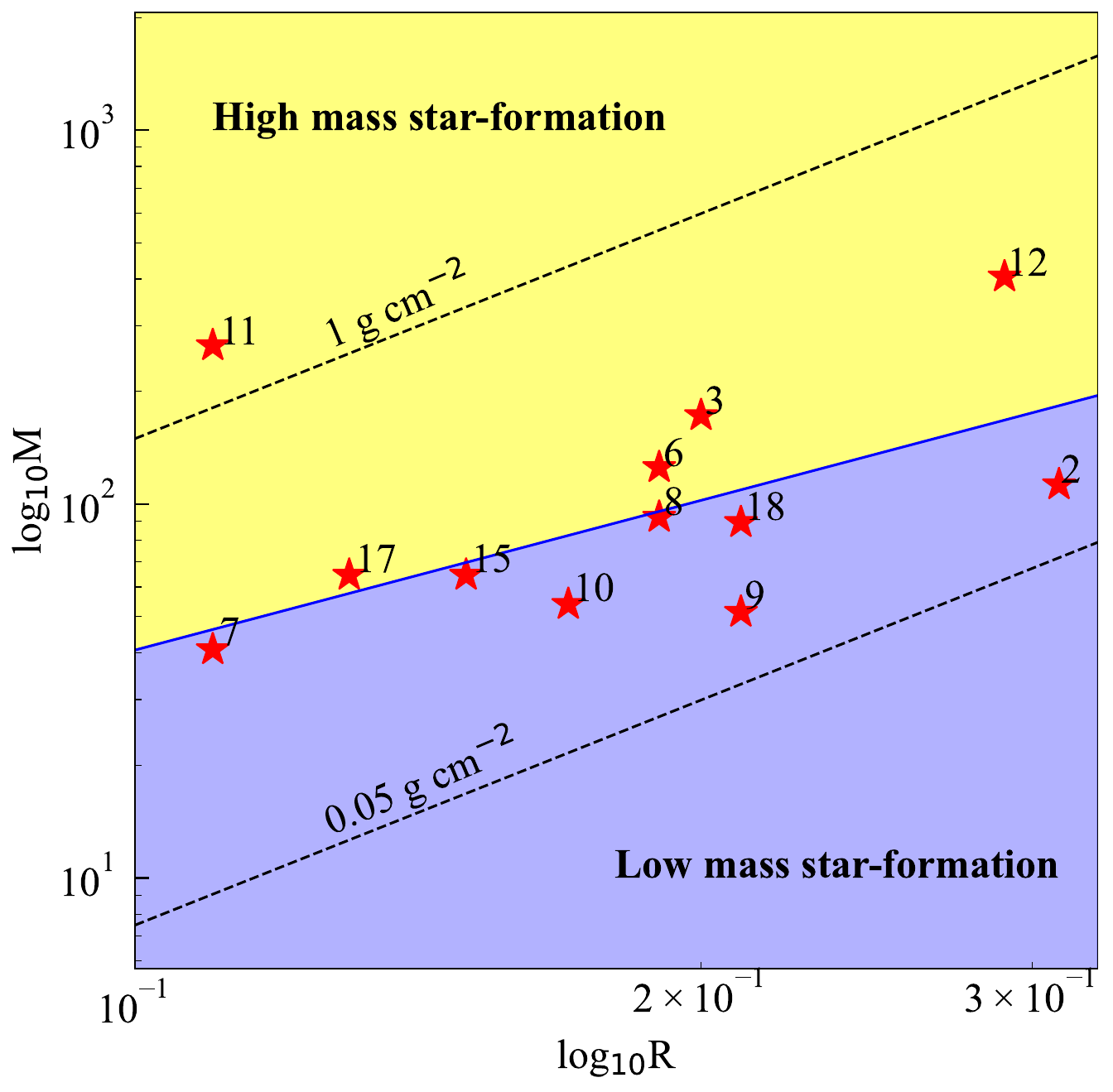}
    \caption{The clump masses are plotted against their radii. The blue shaded region is associated with low mass star-formation and satisfies the relation $m(r) < 870$ M$_{\odot}$ (R$_{eff}$/pc)$^{1.33}$ \citep{2010ApJ...716..433K}, $R_{eff}$ being the effective radius of the clump, defined as the geometric mean of the major and minor axes estimate of the clump.} The two dotted lines correspond to surface densities of 0.05 and 1~g~cm$^{-2}$. This figure is a subset of the one given in \citet{2018MNRAS.473.1059U}.
    \label{fig:13}
\end{figure}

We examine the mass–radius relation for these clumps and find that five lie within the parameter space typically associated with high-mass star formation (Figure~\ref{fig:13}). These massive clumps are highlighted as larger circles in Figure~\ref{fig:11}, while the locations of masers are indicated by crosses. The presence of masers, YSOs, and radio continuum emission toward the nebulous regions N, NW, SW, SE1, and SE2 supports the interpretation that these locations represent active sites of massive star formation. Additional details on the star formation activity in these regions are provided in Appendix~\ref{sec:notes}. In contrast, the origin of the diffuse radio emission toward the central cavity remains uncertain with the current observational constraints. While this emission could reflect the cumulative effect of past massive star formation activity, such as a dispersed or evolved stellar population, additional data would be required to robustly assess this scenario.

\subsection{The gas kinematics for the region}
\begin{figure}
    \centering
    \includegraphics[width=\columnwidth]{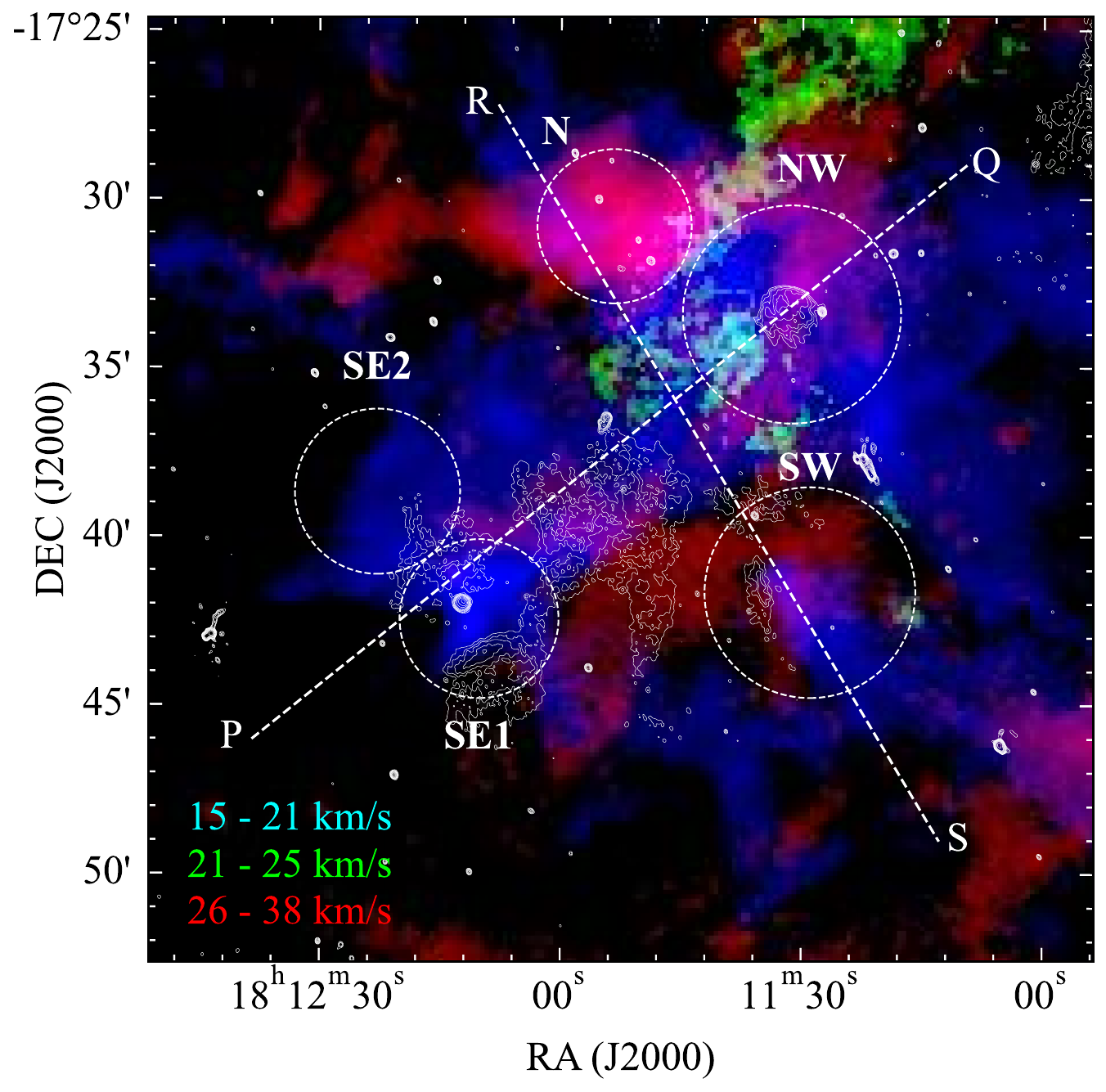}
    \caption{The different molecular cloud components observed towards the region, shown as a colour-composite image. The component $V_1$ (15 - 21~km/s) is shown in blue, component $V_2$ (21 - 25~km/s) in green and component $V_3$ (26 - 38~km/s) in red.} The PV slices are shown as the dashed white lines PQ and RS. The white contours correspond to the uGMRT band-4 (550-750~MHz) emission. 
    \label{fig:14}
\end{figure}

In this subsection, we examine the molecular gas kinematics in conjunction with the ionised gas emission to understand how the local gas dynamics are structured within the G12.79+0.43 region. As discussed earlier, the molecular gas toward the region exhibits three dominant velocity components: $V_1$ (15–21~km~s$^{-1}$), $V_2$ (21–25~km~s$^{-1}$), and $V_3$ (26–38~km~s$^{-1}$). An RGB composite image of these components is shown in Figure~\ref{fig:14}, with $V_1$ displayed in red, $V_2$ in green, and $V_3$ in blue.

The $V_1$ and $V_3$ components cover most of the field, while the $V_2$ component is more spatially confined and is predominantly detected toward the north-western (NW) region. The $V_1$ emission peaks toward the SE1 region, whereas $V_3$ is strongest toward the northern (N) part of the complex. A notable feature is the relative lack of $V_1$ emission toward the SW and parts of SE1, where the emission appears to be replaced or filled by the $V_3$ component. Ionised gas emission traced by the radio continuum extends into this region of reduced $V_1$ emission. In addition, $V_1$ emission is observed to the north and west of the radio peak toward SE1. Toward the NW region, the radio emission spatially overlaps with both $V_1$ and $V_3$, while the $V_2$ component appears to lie to the west of, and partially encircle, the radio emission.

\begin{figure*}
    \centering
    \includegraphics[width=\textwidth]{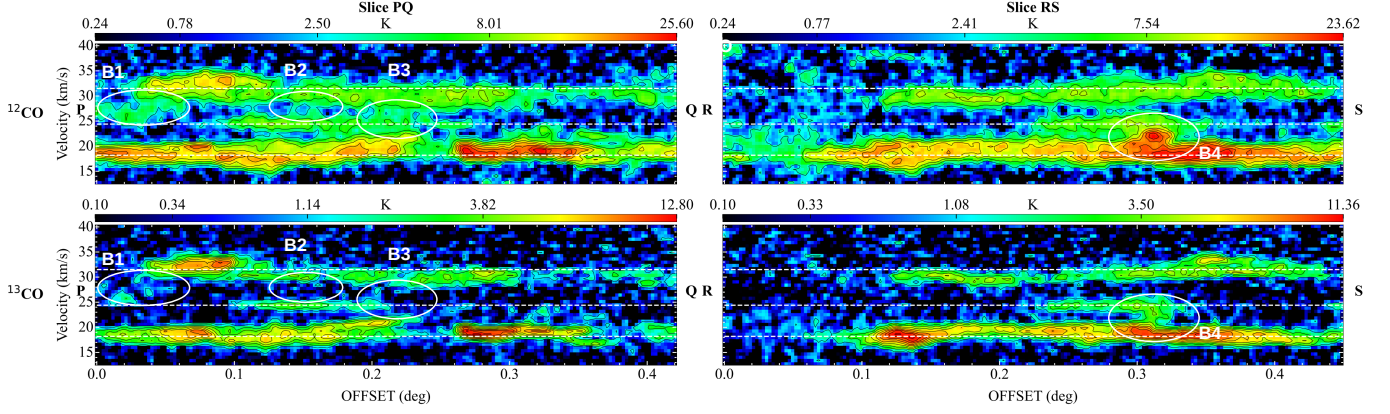}
    \caption{The PV diagrams along the slices PQ (\textit{left}) and RS (\textit{right}) showing bridging features, encircled by white ellipses and between the three components in the case of $^{12}$CO (\textit{top}) and $^{13}$CO (\textit{bottom}), which are designated as B1, B2, B3, and B4. The dotted horizontal white lines represent the LSR velocities of the three components.}
    \label{fig:15}
\end{figure*}

The coexistence and partial spatial overlap of multiple velocity components raise the question of whether dynamical interactions are occurring between these components. Such interactions may arise from a variety of processes, including cloud–cloud collisions, expansion driven by feedback, or other large-scale dynamical motions \citep[e.g.,][]{{2016ApJ...820...26F},{2025AJ....169..181K},{2025A&A...701A.244C}}. To further investigate the gas kinematics, we constructed position–velocity (PV) diagrams of the $^{12}$CO and $^{13}$CO emission along two cuts, PQ and RS, indicated in Figure~\ref{fig:14}. The resulting PV diagrams are shown in Figure~\ref{fig:15}. The $^{12}$CO PV diagrams reveal the presence of bridging features, named B1, B2, B3, and B4 connecting the velocity components, highlighted by white ellipses in Figure~\ref{fig:15}. The most prominent bridges are observed between the $V_1$ and $V_2$ components, while weaker and more diffuse connections are seen between $V_2$ and $V_3$. The lower intensity B2 and B3 features are not detected in $^{13}$CO. This is likely because $^{12}$CO traces low density gas, while $^{13}$CO traces relatively higher density gas. These features are indicative of kinematic continuity between the components and suggest interaction among them. The bridging structures are spatially located near the intersection of the two PV cuts, corresponding to regions where the molecular components overlap in projection.

Although the velocity separation between the $V_1$ and $V_3$ components would correspond to different kinematic distances under the assumption of purely circular Galactic rotation, such an interpretation is unlikely to be valid in this region. The presence of bridging features and the spatial overlap of the components instead suggest that the observed velocity spread arises from dynamical processes, such as expansion, shear or cloud-cloud interactions rather than true line-of-sight distance differences. Under this interpretation, the molecular components are likely physically associated. Although the local gas kinematics suggest interaction among the molecular components, their broader context becomes clearer when examined within the large-scale molecular environment of the region, which we investigate in the following subsection.

\begin{deluxetable}{ccccc}
    \label{tab:4}
    \tablecaption{Catalogue of clumps identified towards the region by \citet{2018MNRAS.473.1059U}}
    \tablehead{\colhead{No.} & \colhead{Name} & \colhead{R} & \colhead{M} & \colhead{Evolutionary Stage} \\ \colhead{} & \colhead{} & \colhead{(pc)} & \colhead{($M_{\odot}$)} & \colhead{}}
    \startdata
    1 & AGAL012.694+00.349 & - & 27.5 & Protostellar \\
    2 & AGAL012.709+00.474 & 0.31 & 112.72 & PDR \\
    3 & AGAL012.771+00.336 & 0.2 & 172.58 & PDR \\
    4 & AGAL012.781+00.351 & - & 27.35 & Quiescent \\
    5 & AGAL012.781+00.382 & - & 42.46 & Quiescent \\
    6 & AGAL012.818+00.322 & 0.19 & 125.31 & Protostellar \\
    7 & AGAL012.824+00.377 & 0.11 & 40.83 & Protostellar \\
    8 & AGAL012.836+00.396 & 0.19 & 92.26 & Quiescent \\
    9 & AGAL012.836+00.552 & 0.21 & 51.40 & PDR \\
    10 & AGAL012.856+00.522 & 0.17 & 53.95 & Ambiguous \\
    11 & AGAL012.888+00.489 & 0.11 & 265.46 & HII region (radio) \\
    12 & AGAL012.889+00.516 & 0.29 & 405.51 & YSO \\
    13 & AGAL012.889+00.537 & - & 176.20 & Ambiguous \\
    14 & AGAL012.901+00.484 & 0.44 & - & Ambiguous \\
    15 & AGAL012.916+00.492 & 0.15 & 64.57 & HII region \\
    16 & AGAL012.929+00.436 & - & 120.23 & Ambiguous \\
    17 & AGAL012.936+00.356 & 0.13 & 64.57 & Protostellar \\
    18 & AGAL012.988+00.354 & 0.21 & 89.33 & Protostellar \\
    \enddata
\end{deluxetable}

\subsection{The Large-Scale Environment of G12.79+0.43 Complex}
\begin{figure*}
    \centering
    \includegraphics[scale=0.5]{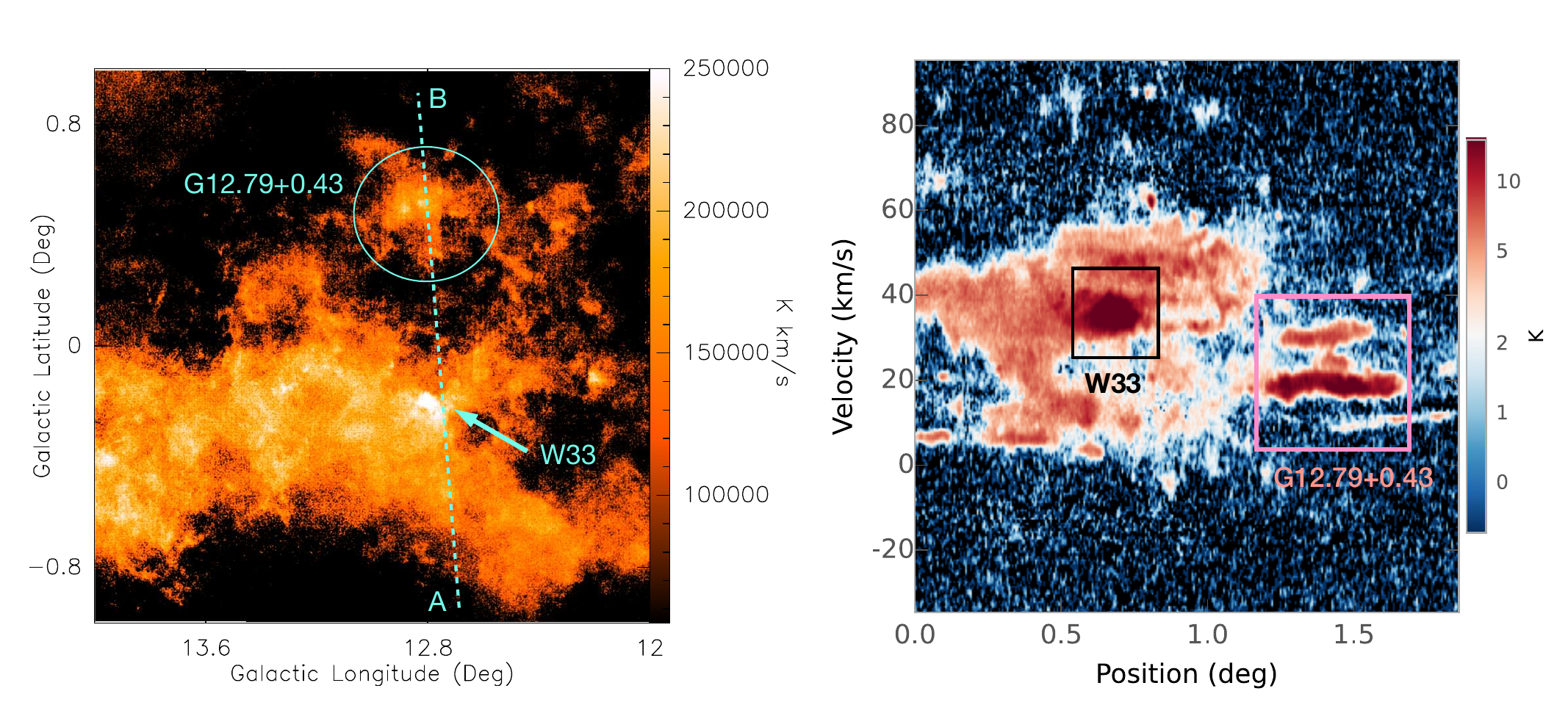}
    \caption{(Left): Integrated intensity map of the $^{12}$CO(1--0) emission in the velocity range 15 to 40~km/s. The region of interest, G12.79+0.43, is marked with a cyan circle. The dashed line AB indicates the PV cut used to extract the corresponding diagram. (Right): PV diagram along the AB cut, revealing kinematic structure of the region.}
    \label{fig:16}
\end{figure*}

To investigate the kinematic properties of the parent cloud and its large-scale structure, we use the FUGIN survey images to present an integrated intensity map of the $^{12}$CO emission spanning $2^\circ\times2^\circ$ in the full velocity range of the cloud (including velocity components $V_1$, $V_2$, and $V_3$), from 13 to 39~km/s, shown in Figure~\ref{fig:16}(left). The CO emission map reveals extended emission over more than $1^\circ$, with the most prominent feature being a bubble-like structure centered near $l\sim12.8^\circ$. This feature has an angular extent of $1.2^\circ\times1.0^\circ$, and the G12.79+0.43 region is located along its northern rim. To study the velocity structure and possible kinematic association with the bubble, we constructed a PV diagram along a cut AB, presented in Figure~\ref{fig:16}(right). The diagram shows an extended emission feature spanning $\sim2^\circ$ in position and from 0 to 60~km/s in velocity. The morphology of the PV features suggests a broken shell-like pattern consistent with expansion \citep[e.g.,][]{2011ApJ...742..105A}. From the velocity of the brightest clump (35~km/s), a kinematic distance of $\sim$3.7~kpc is inferred. However, the nearby massive star-forming complex W33 lies to the south of this region. W33 has been assigned a kinematic distance of 3.7~kpc, but more reliable VLBI parallax measurements yield a distance of 2.4~kpc \citep{2013A&A...553A.117I}. Given the spatial and morphological correlation between W33, the superbubble structure, and the G12.79+0.43 region, we adopt the parallax-based distance of 2.4~kpc for the superbubble. At this distance, the bubble has a physical diameter of $\sim50.2$~pc (centered at $l=12.83^\circ, b=0.17^\circ$) and an expansion velocity of $\sim$25~km/s, as estimated from the PV diagram.

The resulting size is significantly larger than typical H II regions or stellar wind bubbles ($\sim$a few to 10~pc), placing this feature in the category of superbubbles, large expanding cavities, tens to hundreds of parsecs across, produced by the collective feedback of massive stars and supernovae \citep[e.g.,][]{2006ApJ...636..685S,  2019A&A...631A..52J, 2021ApJ...918...36Y, 2024JPhCS2796a2004S}. To estimate the mass of the shell, we first estimated H$_2$ column densities from $^{13}$CO emission using the method outlined in Section 3.1. A spatial mask, defined based on the shell morphology, was applied to select only the pixels associated with the shell. We further converted the obtained column densities to the mass of the expanding shell using the expression.

\begin{equation}
M=N(\textrm{H}_2)\,m_\mathrm{H}\,\mu\,A_{pixel}
\end{equation}

\noindent where $m_\mathrm{H}$ is the mass of the hydrogen atom ($1.67\times10^{-24}$~g), $\mu=2.8$ is the mean molecular weight, and $A_{pixel}$ is the pixel area. Substituting these values,  we estimate the mass of the CO shell to be $2.9\times10^5$~M$_\odot$. We calculate the kinetic energy of the shell using the estimated mass and an expansion velocity of 25~km/s, that is $1.8\times10^{51}$~erg. This is consistent with the cumulative feedback of several massive stars and one or more supernova explosions.\citep[$\sim10^{51}$~erg, e.g.,][]{2005ApJ...631..947M}. Assuming a constant expansion velocity, the dynamical age of the superbubble can be estimated using the expression

\begin{equation}
 t=\eta\frac{R_s}{v_s}   
\end{equation}

\noindent where $R_s$ is the radius of the bubble (25.1~pc) and $v_s$ is the expansion velocity (25~km/s), and $\eta$ is the scaling constant assumed to be 0.25 for momentum-driven system \citep{2023A&A...676A..67W}. Using these values, the dynamical age of the G12.83+0.17 superbubble is estimated to be $\sim$0.3~Myr. Fourteen smaller infrared bubbles are cataloged within 20$'$ of its center \citep{2012MNRAS.424.2442S}, the largest being MWP1G012773+003549 (diameter 9.2 pc). While the dynamical age is shorter than the typical timescales required for the classical “collect-and-collapse” mechanism, the presence of young protostellar candidates and compact infrared bubbles suggests that the expansion is already influencing the surrounding molecular gas. In particular, the compression of pre-existing dense clumps can accelerate collapse on sub-Myr timescales \citep{2018PASJ...70S..53I}. The PV diagram shows a pair of coherent velocity components (position offsets $1.1^\circ-1.6^\circ$) separated by $\sim$10~km/s, consistent with the approaching and receding sides of an expanding shell. This spatial and kinematic association indicates that the G12.79+0.43 subregion is physically connected to the larger superbubble. These observations hint that the expanding shell could be contributing to the conditions for star formation at its periphery, although further evidence is needed to establish a causal link.

\section{Conclusions} \label{sec:conclusions}
We present a multiwavelength study of the molecular cloud complex G12.79+0.43, combining radio, infrared, and molecular line data to investigate its structure, gas dynamics, and star formation acitivty. Our main conclusions are summarised below.

\begin{enumerate}
       \item The G12.79+0.43 cloud complex exhibits five prominent infrared-bright regions (N, NW, SW, SE1, and SE2) distributed around a central cavity that is comparatively faint in the infrared but filled with diffuse radio and molecular gas emission.

    \item Molecular line observations reveal three dominant velocity components along the line of sight: $V_1$ (15--21~km~s$^{-1}$), $V_2$ (21--25~km~s$^{-1}$), and $V_3$ (26--38~km~s$^{-1}$), with peak column densities of order $\sim10^{23}$~cm$^{-2}$ toward SE1, NW, and N, respectively.

    \item Position--velocity analysis of the $^{12}$CO emission reveals bridging features between the $V_2$ component and the adjacent velocity components, particularly toward the NW region, indicating kinematic interaction among the molecular gas components. 
    
    \item Cold dust emission traced by \textit{Herschel} reveals filamentary structures toward the central region and clumpy emission toward the five infrared-bright regions. The hydrogen column densities derived from dust emission are $\sim10^{22}$~cm$^{-2}$ in the bright regions, while dust temperatures range from $\sim18$ to 25~K, peaking at $\sim27$~K toward SE1.

    \item Mid-infrared observations from \textit{Spitzer} GLIMPSE show bright emission toward the five regions and dark lanes toward the centre, spatially coincident with cold dust filaments. Diffuse emission is also detected in the 24~$\mu$m band toward the central cavity.
    
    \item Radio continuum observations reveal bright emission associated with the five regions and extended emission (7.3~pc) filling the central cavity. A total of 70 compact radio sources and six \hii~regions are identified. The derived Lyman continuum photon rates indicate that the \hii~regions, aged 0.19--4.85 Myr, are ionised by early B-type ZAMS stars.
    
    \item A total of 82 YSO candidates, including 28 Class~I sources, are identified using \textit{Spitzer} colour--colour criteria. Both Class~I and Class~II YSOs are present toward the central cavity, consistent with ongoing star formation within the cloud.
    
    \item On larger scales, G12.79+0.43 is located along the rim of a $\sim50$~pc molecular superbubble traced in $^{12}$CO emission. The shell-like morphology observed in the PV diagram, together with the estimated kinetic energy ($\sim10^{51}$~erg) and dynamical age ($\sim0.3$~Myr), is consistent with a young superbubble produced by the collective feedback of massive stars and supernovae.

\end{enumerate}

In summary, G12.79+0.43 is a dynamically complex molecular cloud complex characterised by multiple velocity components, embedded \hii~regions, and ongoing star formation, and is situated at the rim of a young molecular superbubble shaped by large-scale feedback in the Galactic plane.

\section{Acknowledgements}
We thank the anonymous referee for the comments, which has improved this article. We thank the staff of the GMRT that made these observations possible. GMRT is run by the National Centre for Radio Astrophysics of the Tata Institute of Fundamental Research. This work has made use of images from the Digitised Sky Survey (DSS). This work is based [in part] on observations made with the \textit{Spitzer} Space Telescope, which was operated by the Jet Propulsion Laboratory, California Institute of Technology under a contract with NASA. This publication makes use of data from \textit{Herschel} is an ESA space observatory with science instruments provided by European-led Principal Investigator consortia and with important participation from NASA. This publication makes use of data from FUGIN, FOREST Unbiased Galactic plane Imaging survey with the Nobeyama 45-m telescope, a legacy project in the Nobeyama 45-m radio telescope. This work has also made use of the images from the SARAO MeerKAT 1.3 GHz Galactic Plane Survey (SMGPS). AS and SV acknowledge financial support from the Department of Science and Technology–Science and Engineering Research Board (SERB) grant CRG/2019/002581. VVS acknowledges the support of the Department of Atomic Energy, Government of India, under Project Identification No. RTI 4012.

\bibliography{RCW155}{}

@ARTICLE{1960MNRAS.121..103R,
       author = {{Rodgers}, A.~W. and {Campbell}, C.~T. and {Whiteoak}, J.~B.},
        title = "{A catalogue of H{\ensuremath{\alpha}}-emission regions in the southern Milky Way}",
      journal = {\mnras},
         year = 1960,
        month = jan,
       volume = {121},
        pages = {103},
          doi = {10.1093/mnras/121.1.103},
       adsurl = {https://ui.adsabs.harvard.edu/abs/1960MNRAS.121..103R}
}

@ARTICLE{2022A&A...668A..44S,
       author = {{Suad}, L.~A. and {Molina Lera}, J.~A. and {Cichowolski}, S.},
        title = "{GS 121-05-037: A new Galactic chimney candidate with signs of triggered star formation}",
      journal = {\aap},
         year = 2022,
        month = dec,
       volume = {668},
          eid = {A44},
        pages = {A44},
          doi = {10.1051/0004-6361/202243942},
       adsurl = {https://ui.adsabs.harvard.edu/abs/2022A&A...668A..44S}
}

@ARTICLE{2008MNRAS.387...31D,
       author = {{Dawson}, J.~R. and {Mizuno}, N. and {Onishi}, T. and {McClure-Griffiths}, N.~M. and {Fukui}, Y.},
        title = "{The `Carina Flare' supershell: probing the atomic and molecular ISM in a Galactic chimney}",
      journal = {\mnras},
         year = 2008,
        month = jun,
       volume = {387},
       number = {1},
        pages = {31-44},
          doi = {10.1111/j.1365-2966.2008.13152.x},
archivePrefix = {arXiv},
       eprint = {0802.4463},
 primaryClass = {astro-ph},
       adsurl = {https://ui.adsabs.harvard.edu/abs/2008MNRAS.387...31D}
}

@ARTICLE{2021A&A...654A.109K,
       author = {{Koumpia}, E. and {de Wit}, W. -J. and {Oudmaijer}, R.~D. and {Frost}, A.~J. and {Lumsden}, S. and {Caratti o Garatti}, A. and {Goodwin}, S.~P. and {Stecklum}, B. and {Mendigut{\'\i}a}, I. and {Ilee}, J.~D. and {Vioque}, M.},
        title = "{The first interferometric survey of massive YSOs in the K-band. Hot dust, ionised gas, and binarity at au scales}",
      journal = {\aap},
         year = 2021,
        month = oct,
       volume = {654},
          eid = {A109},
        pages = {A109},
          doi = {10.1051/0004-6361/202141373},
archivePrefix = {arXiv},
       eprint = {2108.02868},
 primaryClass = {astro-ph.SR},
       adsurl = {https://ui.adsabs.harvard.edu/abs/2021A&A...654A.109K}
}

@ARTICLE{2008ApJ...681.1341W,
       author = {{Watson}, C. and {Povich}, M.~S. and {Churchwell}, E.~B. and {Babler}, B.~L. and {Chunev}, G. and {Hoare}, M. and {Indebetouw}, R. and {Meade}, M.~R. and {Robitaille}, T.~P. and {Whitney}, B.~A.},
        title = "{Infrared Dust Bubbles: Probing the Detailed Structure and Young Massive Stellar Populations of Galactic H II Regions}",
      journal = {\apj},
         year = 2008,
        month = jul,
       volume = {681},
       number = {2},
        pages = {1341-1355},
          doi = {10.1086/588005},
archivePrefix = {arXiv},
       eprint = {0806.0609},
 primaryClass = {astro-ph},
       adsurl = {https://ui.adsabs.harvard.edu/abs/2008ApJ...681.1341W}
}

@ARTICLE{2006ApJ...649..759C,
       author = {{Churchwell}, E. and {Povich}, M.~S. and {Allen}, D. and {Taylor}, M.~G. and {Meade}, M.~R. and {Babler}, B.~L. and {Indebetouw}, R. and {Watson}, C. and {Whitney}, B.~A. and {Wolfire}, M.~G. and {Bania}, T.~M. and {Benjamin}, R.~A. and {Clemens}, D.~P. and {Cohen}, M. and {Cyganowski}, C.~J. and {Jackson}, J.~M. and {Kobulnicky}, H.~A. and {Mathis}, J.~S. and {Mercer}, E.~P. and {Stolovy}, S.~R. and {Uzpen}, B. and {Watson}, D.~F. and {Wolff}, M.~J.},
        title = "{The Bubbling Galactic Disk}",
      journal = {\apj},
         year = 2006,
        month = oct,
       volume = {649},
       number = {2},
        pages = {759-778},
          doi = {10.1086/507015},
       adsurl = {https://ui.adsabs.harvard.edu/abs/2006ApJ...649..759C}
}

@ARTICLE{2013A&A...553A.117I,
       author = {{Immer}, K. and {Reid}, M.~J. and {Menten}, K.~M. and {Brunthaler}, A. and {Dame}, T.~M.},
        title = "{Trigonometric parallaxes of massive star forming regions: G012.88+0.48 and W33}",
      journal = {\aap},
         year = 2013,
        month = may,
       volume = {553},
          eid = {A117},
        pages = {A117},
          doi = {10.1051/0004-6361/201220793},
archivePrefix = {arXiv},
       eprint = {1304.2041},
 primaryClass = {astro-ph.GA},
       adsurl = {https://ui.adsabs.harvard.edu/abs/2013A&A...553A.117I}
}

@ARTICLE{2012MNRAS.424.2442S,
       author = {{Simpson}, R.~J. and {Povich}, M.~S. and {Kendrew}, S. and {Lintott}, C.~J. and {Bressert}, E. and {Arvidsson}, K. and {Cyganowski}, C. and {Maddison}, S. and {Schawinski}, K. and {Sherman}, R. and {Smith}, A.~M. and {Wolf-Chase}, G.},
        title = "{The Milky Way Project First Data Release: a bubblier Galactic disc}",
      journal = {\mnras},
         year = 2012,
        month = aug,
       volume = {424},
       number = {4},
        pages = {2442-2460},
          doi = {10.1111/j.1365-2966.2012.20770.x},
archivePrefix = {arXiv},
       eprint = {1201.6357},
 primaryClass = {astro-ph.GA},
       adsurl = {https://ui.adsabs.harvard.edu/abs/2012MNRAS.424.2442S}
}

@ARTICLE{2006ApJ...636..685S,
       author = {{Sakamoto}, Kazushi and {Ho}, Paul T.~P. and {Iono}, Daisuke and {Keto}, Eric R. and {Mao}, Rui-Qing and {Matsushita}, Satoki and {Peck}, Alison B. and {Wiedner}, Martina C. and {Wilner}, David J. and {Zhao}, Jun-Hui},
        title = "{Molecular Superbubbles in the Starburst Galaxy NGC 253}",
      journal = {\apj},
         year = 2006,
        month = jan,
       volume = {636},
       number = {2},
        pages = {685-697},
          doi = {10.1086/498075},
archivePrefix = {arXiv},
       eprint = {astro-ph/0509430},
 primaryClass = {astro-ph},
       adsurl = {https://ui.adsabs.harvard.edu/abs/2006ApJ...636..685S}
}

@ARTICLE{2019A&A...631A..52J,
       author = {{Joubaud}, T. and {Grenier}, I.~A. and {Ballet}, J. and {Soler}, J.~D.},
        title = "{Gas shells and magnetic fields in the Orion-Eridanus superbubble}",
      journal = {\aap},
         year = 2019,
        month = nov,
       volume = {631},
          eid = {A52},
        pages = {A52},
          doi = {10.1051/0004-6361/201936239},
archivePrefix = {arXiv},
       eprint = {1909.10083},
 primaryClass = {astro-ph.HE},
       adsurl = {https://ui.adsabs.harvard.edu/abs/2019A&A...631A..52J}
}

@ARTICLE{2016MNRAS.456.2425V,
       author = {{Veena}, V.~S. and {Vig}, S. and {Tej}, A. and {Varricatt}, W.~P. and {Ghosh}, S.~K. and {Chandrasekhar}, T. and {Ashok}, N.~M.},
        title = "{Star formation towards the southern cometary H II region IRAS 17256-3631}",
      journal = {\mnras},
         year = 2016,
        month = mar,
       volume = {456},
       number = {3},
        pages = {2425-2445},
          doi = {10.1093/mnras/stv2832},
archivePrefix = {arXiv},
       eprint = {1512.00987},
 primaryClass = {astro-ph.GA},
       adsurl = {https://ui.adsabs.harvard.edu/abs/2016MNRAS.456.2425V}
}

@ARTICLE{2025A&A...701A.244C,
       author = {{Cosentino}, G. and {Jim{\'e}nez-Serra}, I. and {Liu}, R. and {Law}, C.-Y. and {Tan}, J.~C. and {Henshaw}, J.~D. and {Barnes}, A.~T. and {Fontani}, F. and {Caselli}, P. and {Viti}, S.},
        title = "{Low-velocity large-scale shocks in the infrared dark cloud G035.39-00.33: Bubble-driven cloud-cloud collisions}",
      journal = {\aap},
         year = 2025,
        month = sep,
       volume = {701},
          eid = {A244},
        pages = {A244},
          doi = {10.1051/0004-6361/202556057},
archivePrefix = {arXiv},
       eprint = {2508.06081},
 primaryClass = {astro-ph.GA},
       adsurl = {https://ui.adsabs.harvard.edu/abs/2025A&A...701A.244C}
}

@ARTICLE{2025AJ....169..181K,
       author = {{Kohno}, Mikito and {Yamada}, Rin I. and {Tachihara}, Kengo and {Fujita}, Shinji and {Enokiya}, Rei and {Tokuda}, Kazuki and {Habe}, Asao and {Sano}, Hidetoshi and {Hayakawa}, Takahiro and {Demachi}, Fumika and {Ito}, Takuto and {Tsuge}, Kisetsu and {Nishimura}, Atsushi and {Kobayashi}, Masato I.~N. and {Yamamoto}, Hiroaki and {Fukui}, Yasuo},
        title = "{Giant Molecular Clouds in RCW 106 (G333): Galactic Mini-starbursts and Massive Star Formation Induced by Supersonic Cloud─Cloud Collisions}",
      journal = {\aj},
         year = 2025,
        month = mar,
       volume = {169},
       number = {3},
          eid = {181},
        pages = {181},
          doi = {10.3847/1538-3881/adae87},
archivePrefix = {arXiv},
       eprint = {2501.14575},
 primaryClass = {astro-ph.GA},
       adsurl = {https://ui.adsabs.harvard.edu/abs/2025AJ....169..181K}
}

@ARTICLE{2016ApJ...820...26F,
       author = {{Fukui}, Y. and {Torii}, K. and {Ohama}, A. and {Hasegawa}, K. and {Hattori}, Y. and {Sano}, H. and {Ohashi}, S. and {Fujii}, K. and {Kuwahara}, S. and {Mizuno}, N. and {Dawson}, J.~R. and {Yamamoto}, H. and {Tachihara}, K. and {Okuda}, T. and {Onishi}, T. and {Mizuno}, A.},
        title = "{The Two Molecular Clouds in RCW 38: Evidence for the Formation of the Youngest Super Star Cluster in the Milky Way Triggered by Cloud-Cloud Collision}",
      journal = {\apj},
         year = 2016,
        month = mar,
       volume = {820},
       number = {1},
          eid = {26},
        pages = {26},
          doi = {10.3847/0004-637X/820/1/26},
archivePrefix = {arXiv},
       eprint = {1504.05391},
 primaryClass = {astro-ph.SR},
       adsurl = {https://ui.adsabs.harvard.edu/abs/2016ApJ...820...26F}
}

@ARTICLE{2016AJ....152..146N,
       author = {{Nandakumar}, G. and {Veena}, V.~S. and {Vig}, S. and {Tej}, A. and {Ghosh}, S.~K. and {Ojha}, D.~K.},
        title = "{Star-forming Activity in the H II Regions Associated with the IRAS 17160-3707 Complex}",
      journal = {\aj},
         year = 2016,
        month = nov,
       volume = {152},
       number = {5},
          eid = {146},
        pages = {146},
          doi = {10.3847/0004-6256/152/5/146},
archivePrefix = {arXiv},
       eprint = {1607.01267},
 primaryClass = {astro-ph.GA},
       adsurl = {https://ui.adsabs.harvard.edu/abs/2016AJ....152..146N}
}

@ARTICLE{2021ApJ...918...36Y,
       author = {{Yamane}, Y. and {Sano}, H. and {Filipovi{\'c}}, M.~D. and {Tokuda}, K. and {Fujii}, K. and {Babazaki}, Y. and {Mitsuishi}, I. and {Inoue}, T. and {Aharonian}, F. and {Inaba}, T. and {Inutsuka}, S. and {Maxted}, N. and {Mizuno}, N. and {Onishi}, T. and {Rowell}, G. and {Tsuge}, K. and {Voisin}, F. and {Yoshiike}, S. and {Fukuda}, T. and {Kawamura}, A. and {Bamba}, A. and {Tachihara}, K. and {Fukui}, Y.},
        title = "{Associated Molecular and Atomic Clouds with X-Ray Shell of Superbubble 30 Doradus C in the LMC}",
      journal = {\apj},
         year = 2021,
        month = sep,
       volume = {918},
       number = {1},
          eid = {36},
        pages = {36},
          doi = {10.3847/1538-4357/ac0adb},
archivePrefix = {arXiv},
       eprint = {2106.09916},
 primaryClass = {astro-ph.HE},
       adsurl = {https://ui.adsabs.harvard.edu/abs/2021ApJ...918...36Y}
}

@INPROCEEDINGS{2024JPhCS2796a2004S,
       author = {{Su{\'a}rez-L{\'o}pez}, O. and {Villares}, A.~S. and {Banda-Barrag{\'a}n}, W.~E.},
        title = "{Galactic Superbubbles in 3D: Wind Formation and Cloud Shielding}",
    booktitle = {Journal of Physics Conference Series},
         year = 2024,
       series = {Journal of Physics Conference Series},
       volume = {2796},
        month = jul,
    publisher = {IOP},
          eid = {012004},
        pages = {012004},
          doi = {10.1088/1742-6596/2796/1/012004},
archivePrefix = {arXiv},
       eprint = {2404.13498},
 primaryClass = {astro-ph.GA},
       adsurl = {https://ui.adsabs.harvard.edu/abs/2024JPhCS2796a2004S}
}

@ARTICLE{2018PASJ...70S..53I,
       author = {{Inoue}, Tsuyoshi and {Hennebelle}, Patrick and {Fukui}, Yasuo and {Matsumoto}, Tomoaki and {Iwasaki}, Kazunari and {Inutsuka}, Shu-ichiro},
        title = "{The formation of massive molecular filaments and massive stars triggered by a magnetohydrodynamic shock wave}",
      journal = {\pasj},
         year = 2018,
        month = may,
       volume = {70},
          eid = {S53},
        pages = {S53},
          doi = {10.1093/pasj/psx089},
archivePrefix = {arXiv},
       eprint = {1707.02035},
 primaryClass = {astro-ph.GA},
       adsurl = {https://ui.adsabs.harvard.edu/abs/2018PASJ...70S..53I}
}

@ARTICLE{2005ApJ...631..947M,
       author = {{Moriguchi}, Y. and {Tamura}, K. and {Tawara}, Y. and {Sasago}, H. and {Yamaoka}, K. and {Onishi}, T. and {Fukui}, Y.},
        title = "{A Detailed Study of Molecular Clouds toward the TeV Gamma-Ray Supernova Remnant G347.3-0.5}",
      journal = {\apj},
         year = 2005,
        month = oct,
       volume = {631},
       number = {2},
        pages = {947-963},
          doi = {10.1086/432653},
archivePrefix = {arXiv},
       eprint = {astro-ph/0506489},
 primaryClass = {astro-ph},
       adsurl = {https://ui.adsabs.harvard.edu/abs/2005ApJ...631..947M}
}

@ARTICLE{2011ApJ...742..105A,
       author = {{Arce}, H{\'e}ctor G. and {Borkin}, Michelle A. and {Goodman}, Alyssa A. and {Pineda}, Jaime E. and {Beaumont}, Christopher N.},
        title = "{A Bubbling Nearby Molecular Cloud: COMPLETE Shells in Perseus}",
      journal = {\apj},
         year = 2011,
        month = dec,
       volume = {742},
       number = {2},
          eid = {105},
        pages = {105},
          doi = {10.1088/0004-637X/742/2/105},
archivePrefix = {arXiv},
       eprint = {1109.3368},
 primaryClass = {astro-ph.SR},
       adsurl = {https://ui.adsabs.harvard.edu/abs/2011ApJ...742..105A}
}

@ARTICLE{2023A&A...676A..67W,
       author = {{Watkins}, E.~J. and {Kreckel}, K. and {Groves}, B. and {Glover}, S.~C.~O. and {Whitmore}, B.~C. and {Leroy}, A.~K. and {Schinnerer}, E. and {Meidt}, S.~E. and {Egorov}, O.~V. and {Barnes}, A.~T. and {Lee}, J.~C. and {Bigiel}, F. and {Boquien}, M. and {Chandar}, R. and {Chevance}, M. and {Dale}, D.~A. and {Grasha}, K. and {Klessen}, R.~S. and {Kruijssen}, J.~M.~D. and {Larson}, K.~L. and {Li}, J. and {M{\'e}ndez-Delgado}, J.~E. and {Pessa}, I. and {Saito}, T. and {Sanchez-Blazquez}, P. and {Sarbadhicary}, S.~K. and {Scheuermann}, F. and {Thilker}, D.~A. and {Williams}, T.~G.},
        title = "{Quantifying the energetics of molecular superbubbles in PHANGS galaxies}",
      journal = {\aap},
         year = 2023,
        month = aug,
       volume = {676},
          eid = {A67},
        pages = {A67},
          doi = {10.1051/0004-6361/202346075},
archivePrefix = {arXiv},
       eprint = {2302.03699},
 primaryClass = {astro-ph.GA},
       adsurl = {https://ui.adsabs.harvard.edu/abs/2023A&A...676A..67W}
}

@ARTICLE{2023ApJ...944L..22B,
       author = {{Barnes}, Ashley. T. and {Watkins}, Elizabeth J. and {Meidt}, Sharon E. and {Kreckel}, Kathryn and {Sormani}, Mattia C. and {Tre{\ss}}, Robin G. and {Glover}, Simon C.~O. and {Bigiel}, Frank and {Chandar}, Rupali and {Emsellem}, Eric and {Lee}, Janice C. and {Leroy}, Adam K. and {Sandstrom}, Karin M. and {Schinnerer}, Eva and {Rosolowsky}, Erik and {Belfiore}, Francesco and {Blanc}, Guillermo A. and {Boquien}, M{\'e}d{\'e}ric and {Brok}, Jakob den and {Cao}, Yixian and {Chevance}, M{\'e}lanie and {Dale}, Daniel A. and {Egorov}, Oleg V. and {Eibensteiner}, Cosima and {Grasha}, Kathryn and {Groves}, Brent and {Hassani}, Hamid and {Henshaw}, Jonathan D. and {Jeffreson}, Sarah and {Jim{\'e}nez-Donaire}, Mar{\'\i}a J. and {Keller}, Benjamin W. and {Klessen}, Ralf S. and {Koch}, Eric W. and {Kruijssen}, J.~M. Diederik and {Larson}, Kirsten L. and {Li}, Jing and {Liu}, Daizhong and {Lopez}, Laura A. and {Murphy}, Eric J. and {Neumann}, Lukas and {Pety}, J{\'e}r{\^o}me and {Pinna}, Francesca and {Querejeta}, Miguel and {Renaud}, Florent and {Saito}, Toshiki and {Sarbadhicary}, Sumit K. and {Sardone}, Amy and {Smith}, Rowan J. and {Stuber}, Sophia K. and {Sun}, Jiayi and {Thilker}, David A. and {Usero}, Antonio and {Whitmore}, Bradley C. and {Williams}, Thomas G.},
        title = "{PHANGS-JWST First Results: Multiwavelength View of Feedback-driven Bubbles (the Phantom Voids) across NGC 628}",
      journal = {\apjl},
         year = 2023,
        month = feb,
       volume = {944},
       number = {2},
          eid = {L22},
        pages = {L22},
          doi = {10.3847/2041-8213/aca7b9},
archivePrefix = {arXiv},
       eprint = {2212.00812},
 primaryClass = {astro-ph.GA},
       adsurl = {https://ui.adsabs.harvard.edu/abs/2023ApJ...944L..22B}
}

@ARTICLE{2012MNRAS.421.3159M,
       author = {{Moss}, V.~A. and {McClure-Griffiths}, N.~M. and {Braun}, R. and {Hill}, A.~S. and {Madsen}, G.~J.},
        title = "{GSH 006-15+7: a local Galactic supershell featuring transition from H I emission to absorption}",
      journal = {\mnras},
         year = 2012,
        month = apr,
       volume = {421},
       number = {4},
        pages = {3159-3169},
          doi = {10.1111/j.1365-2966.2012.20538.x},
archivePrefix = {arXiv},
       eprint = {1201.2700},
 primaryClass = {astro-ph.GA},
       adsurl = {https://ui.adsabs.harvard.edu/abs/2012MNRAS.421.3159M}
}

@ARTICLE{2011BSRSL..80..297C,
       author = {{Chu}, You-Hua and {Gruendl}, Robert A.},
        title = "{Feedback from Massive YSOs and Massive Stars}",
      journal = {Bulletin de la Societe Royale des Sciences de Liege},
         year = 2011,
        month = jan,
       volume = {80},
        pages = {297-309},
          doi = {10.48550/arXiv.1012.1373},
archivePrefix = {arXiv},
       eprint = {1012.1373},
 primaryClass = {astro-ph.SR},
       adsurl = {https://ui.adsabs.harvard.edu/abs/2011BSRSL..80..297C}
}

@ARTICLE{2003PASP..115..953B,
       author = {{Benjamin}, Robert A. and {Churchwell}, E. and {Babler}, Brian L. and {Bania}, T.~M. and {Clemens}, Dan P. and {Cohen}, Martin and {Dickey}, John M. and {Indebetouw}, R{\'e}my and {Jackson}, James M. and {Kobulnicky}, Henry A. and {Lazarian}, Alex and {Marston}, A.~P. and {Mathis}, John S. and {Meade}, Marilyn R. and {Seager}, Sara and {Stolovy}, S.~R. and {Watson}, C. and {Whitney}, Barbara A. and {Wolff}, Michael J. and {Wolfire}, Mark G.},
        title = "{GLIMPSE. I. An SIRTF Legacy Project to Map the Inner Galaxy}",
      journal = {\pasp},
         year = 2003,
        month = aug,
       volume = {115},
       number = {810},
        pages = {953-964},
          doi = {10.1086/376696},
archivePrefix = {arXiv},
       eprint = {astro-ph/0306274},
 primaryClass = {astro-ph},
       adsurl = {https://ui.adsabs.harvard.edu/abs/2003PASP..115..953B}
}

@ARTICLE{2016AnA...591A.149M,
       author = {{Molinari}, S. and {Schisano}, E. and {Elia}, D. and {Pestalozzi}, M. and {Traficante}, A. and {Pezzuto}, S. and {Swinyard}, B.~M. and {Noriega-Crespo}, A. and {Bally}, J. and {Moore}, T.~J.~T. and {Plume}, R. and {Zavagno}, A. and {di Giorgio}, A.~M. and {Liu}, S.~J. and {Pilbratt}, G.~L. and {Mottram}, J.~C. and {Russeil}, D. and {Piazzo}, L. and {Veneziani}, M. and {Benedettini}, M. and {Calzoletti}, L. and {Faustini}, F. and {Natoli}, P. and {Piacentini}, F. and {Merello}, M. and {Palmese}, A. and {Del Grande}, R. and {Polychroni}, D. and {Rygl}, K.~L.~J. and {Polenta}, G. and {Barlow}, M.~J. and {Bernard}, J. -P. and {Martin}, P.~G. and {Testi}, L. and {Ali}, B. and {Andr{\'e}}, P. and {Beltr{\'a}n}, M.~T. and {Billot}, N. and {Carey}, S. and {Cesaroni}, R. and {Compi{\`e}gne}, M. and {Eden}, D. and {Fukui}, Y. and {Garcia-Lario}, P. and {Hoare}, M.~G. and {Huang}, M. and {Joncas}, G. and {Lim}, T.~L. and {Lord}, S.~D. and {Martinavarro-Armengol}, S. and {Motte}, F. and {Paladini}, R. and {Paradis}, D. and {Peretto}, N. and {Robitaille}, T. and {Schilke}, P. and {Schneider}, N. and {Schulz}, B. and {Sibthorpe}, B. and {Strafella}, F. and {Thompson}, M.~A. and {Umana}, G. and {Ward-Thompson}, D. and {Wyrowski}, F.},
        title = "{Hi-GAL, the Herschel infrared Galactic Plane Survey: photometric maps and compact source catalogues. First data release for the inner Milky Way: +68{\textdegree} {\ensuremath{\geq}} l {\ensuremath{\geq}} -70{\textdegree}}",
      journal = {\aap},
         year = 2016,
        month = jul,
       volume = {591},
          eid = {A149},
        pages = {A149},
          doi = {10.1051/0004-6361/201526380},
archivePrefix = {arXiv},
       eprint = {1604.05911},
 primaryClass = {astro-ph.GA},
       adsurl = {https://ui.adsabs.harvard.edu/abs/2016A&A...591A.149M}
}

@ARTICLE{2017PASJ...69...78U,
       author = {{Umemoto}, Tomofumi and {Minamidani}, Tetsuhiro and {Kuno}, Nario and {Fujita}, Shinji and {Matsuo}, Mitsuhiro and {Nishimura}, Atsushi and {Torii}, Kazufumi and {Tosaki}, Tomoka and {Kohno}, Mikito and {Kuriki}, Mika and {Tsuda}, Yuya and {Hirota}, Akihiko and {Ohashi}, Satoshi and {Yamagishi}, Mitsuyoshi and {Handa}, Toshihiro and {Nakanishi}, Hiroyuki and {Omodaka}, Toshihiro and {Koide}, Nagito and {Matsumoto}, Naoko and {Onishi}, Toshikazu and {Tokuda}, Kazuki and {Seta}, Masumichi and {Kobayashi}, Yukinori and {Tachihara}, Kengo and {Sano}, Hidetoshi and {Hattori}, Yusuke and {Onodera}, Sachiko and {Oasa}, Yumiko and {Kamegai}, Kazuhisa and {Tsuboi}, Masato and {Sofue}, Yoshiaki and {Higuchi}, Aya E. and {Chibueze}, James O. and {Mizuno}, Norikazu and {Honma}, Mareki and {Muller}, Erik and {Inoue}, Tsuyoshi and {Morokuma-Matsui}, Kana and {Shinnaga}, Hiroko and {Ozawa}, Takeaki and {Takahashi}, Ryo and {Yoshiike}, Satoshi and {Costes}, Jean and {Kuwahara}, Sho},
        title = "{FOREST unbiased Galactic plane imaging survey with the Nobeyama 45 m telescope (FUGIN). I. Project overview and initial results}",
      journal = {\pasj},
         year = 2017,
        month = oct,
       volume = {69},
       number = {5},
          eid = {78},
        pages = {78},
          doi = {10.1093/pasj/psx061},
archivePrefix = {arXiv},
       eprint = {1707.05981},
 primaryClass = {astro-ph.GA},
       adsurl = {https://ui.adsabs.harvard.edu/abs/2017PASJ...69...78U}
}

@ARTICLE{2017CSci..113..707G,
       author = {{Gupta}, Y. and {Ajithkumar}, B. and {Kale}, H.~S. and {Nayak}, S. and {Sabhapathy}, S. and {Sureshkumar}, S. and {Swami}, R.~V. and {Chengalur}, J.~N. and {Ghosh}, S.~K. and {Ishwara-Chandra}, C.~H. and {Joshi}, B.~C. and {Kanekar}, N. and {Lal}, D.~V. and {Roy}, S.},
        title = "{The upgraded GMRT: opening new windows on the radio Universe}",
      journal = {Current Science},
         year = 2017,
        month = aug,
       volume = {113},
       number = {4},
        pages = {707-714},
          doi = {10.18520/cs/v113/i04/707-714},
       adsurl = {https://ui.adsabs.harvard.edu/abs/2017CSci..113..707G}
}

@ARTICLE{2018ApJ...852...93V,
       author = {{Veena}, V.~S. and {Vig}, S. and {Mookerjea}, B. and {S{\'a}nchez-Monge}, {\'A}. and {Tej}, A. and {Ishwara-Chandra}, C.~H.},
        title = "{Probing the Massive Star-forming Environment: A Multiwavelength Investigation of the Filamentary IRDC G333.73+0.37}",
      journal = {\apj},
         year = 2018,
        month = jan,
       volume = {852},
       number = {2},
          eid = {93},
        pages = {93},
          doi = {10.3847/1538-4357/aa9aef},
archivePrefix = {arXiv},
       eprint = {1711.06398},
 primaryClass = {astro-ph.GA},
       adsurl = {https://ui.adsabs.harvard.edu/abs/2018ApJ...852...93V}
}

@ARTICLE{2019MNRAS.488.1141J,
       author = {{Jayasinghe}, Tharindu and {Dixon}, Don and {Povich}, Matthew S. and {Binder}, Breanna and {Velasco}, Jose and {Lepore}, Denise M. and {Xu}, Duo and {Offner}, Stella and {Kobulnicky}, Henry A. and {Anderson}, Loren D. and {Kendrew}, Sarah and {Simpson}, Robert J.},
        title = "{The Milky Way Project second data release: bubbles and bow shocks}",
      journal = {\mnras},
         year = 2019,
        month = sep,
       volume = {488},
       number = {1},
        pages = {1141-1165},
          doi = {10.1093/mnras/stz1738},
archivePrefix = {arXiv},
       eprint = {1905.12625},
 primaryClass = {astro-ph.GA},
       adsurl = {https://ui.adsabs.harvard.edu/abs/2019MNRAS.488.1141J}
}

@ARTICLE{2002ApJ...566..931S,
       author = {{Sridharan}, T.~K. and {Beuther}, H. and {Schilke}, P. and {Menten}, K.~M. and {Wyrowski}, F.},
        title = "{High-Mass Protostellar Candidates. I. The Sample and Initial Results}",
      journal = {\apj},
         year = 2002,
        month = feb,
       volume = {566},
       number = {2},
        pages = {931-944},
          doi = {10.1086/338332},
archivePrefix = {arXiv},
       eprint = {astro-ph/0110363},
 primaryClass = {astro-ph},
       adsurl = {https://ui.adsabs.harvard.edu/abs/2002ApJ...566..931S}
}

@ARTICLE{2016ApJ...822...59S,
       author = {{Svoboda}, Brian E. and {Shirley}, Yancy L. and {Battersby}, Cara and {Rosolowsky}, Erik W. and {Ginsburg}, Adam G. and {Ellsworth-Bowers}, Timothy P. and {Pestalozzi}, Michele R. and {Dunham}, Miranda K. and {Evans}, Neal J., II and {Bally}, John and {Glenn}, Jason},
        title = "{The Bolocam Galactic Plane Survey. XIV. Physical Properties of Massive Starless and Star-forming Clumps}",
      journal = {\apj},
         year = 2016,
        month = may,
       volume = {822},
       number = {2},
          eid = {59},
        pages = {59},
          doi = {10.3847/0004-637X/822/2/59},
archivePrefix = {arXiv},
       eprint = {1511.08810},
 primaryClass = {astro-ph.GA},
       adsurl = {https://ui.adsabs.harvard.edu/abs/2016ApJ...822...59S}
}

@ARTICLE{2006AJ....131.2525H,
       author = {{Helfand}, David J. and {Becker}, Robert H. and {White}, Richard L. and {Fallon}, Adam and {Tuttle}, Sarah},
        title = "{MAGPIS: A Multi-Array Galactic Plane Imaging Survey}",
      journal = {\aj},
         year = 2006,
        month = may,
       volume = {131},
       number = {5},
        pages = {2525-2537},
          doi = {10.1086/503253},
       adsurl = {https://ui.adsabs.harvard.edu/abs/2006AJ....131.2525H}
}

@ARTICLE{2011ApJS..194...32A,
       author = {{Anderson}, L.~D. and {Bania}, T.~M. and {Balser}, Dana S. and {Rood}, Robert T.},
        title = "{The Green Bank Telescope H II Region Discovery Survey. II. The Source Catalog}",
      journal = {\apjs},
         year = 2011,
        month = jun,
       volume = {194},
       number = {2},
          eid = {32},
        pages = {32},
          doi = {10.1088/0067-0049/194/2/32},
archivePrefix = {arXiv},
       eprint = {1103.5085},
 primaryClass = {astro-ph.GA},
       adsurl = {https://ui.adsabs.harvard.edu/abs/2011ApJS..194...32A}
}

@ARTICLE{2016A&A...588A.143S,
       author = {{Schmiedeke}, A. and {Schilke}, P. and {M{\"o}ller}, Th. and {S{\'a}nchez-Monge}, {\'A}. and {Bergin}, E. and {Comito}, C. and {Csengeri}, T. and {Lis}, D.~C. and {Molinari}, S. and {Qin}, S. -L. and {Rolffs}, R.},
        title = "{The physical and chemical structure of Sagittarius B2. I. Three-dimensional thermal dust and free-free continuum modeling on 100 au to 45 pc scales}",
      journal = {\aap},
         year = 2016,
        month = apr,
       volume = {588},
          eid = {A143},
        pages = {A143},
          doi = {10.1051/0004-6361/201527311},
archivePrefix = {arXiv},
       eprint = {1602.02274},
 primaryClass = {astro-ph.GA},
       adsurl = {https://ui.adsabs.harvard.edu/abs/2016A&A...588A.143S}
}

@ARTICLE{2006ApJ...653.1226Q,
       author = {{Quireza}, Cintia and {Rood}, Robert T. and {Bania}, T.~M. and {Balser}, Dana S. and {Maciel}, Walter J.},
        title = "{The Electron Temperature Gradient in the Galactic Disk}",
      journal = {\apj},
         year = 2006,
        month = dec,
       volume = {653},
       number = {2},
        pages = {1226-1240},
          doi = {10.1086/508803},
archivePrefix = {arXiv},
       eprint = {astro-ph/0609006},
 primaryClass = {astro-ph},
       adsurl = {https://ui.adsabs.harvard.edu/abs/2006ApJ...653.1226Q}
}

@ARTICLE{2018MNRAS.473.1059U,
       author = {{Urquhart}, J.~S. and {K{\"o}nig}, C. and {Giannetti}, A. and {Leurini}, S. and {Moore}, T.~J.~T. and {Eden}, D.~J. and {Pillai}, T. and {Thompson}, M.~A. and {Braiding}, C. and {Burton}, M.~G. and {Csengeri}, T. and {Dempsey}, J.~T. and {Figura}, C. and {Froebrich}, D. and {Menten}, K.~M. and {Schuller}, F. and {Smith}, M.~D. and {Wyrowski}, F.},
        title = "{ATLASGAL - properties of a complete sample of Galactic clumps}",
      journal = {\mnras},
         year = 2018,
        month = jan,
       volume = {473},
       number = {1},
        pages = {1059-1102},
          doi = {10.1093/mnras/stx2258},
archivePrefix = {arXiv},
       eprint = {1709.00392},
 primaryClass = {astro-ph.GA},
       adsurl = {https://ui.adsabs.harvard.edu/abs/2018MNRAS.473.1059U}
}

@MISC{2021MNRAS.500.3027D,
       author = {{Duarte-Cabral}, A. and {Colombo}, D. and {Urquhart}, J.~S. and {Ginsburg}, A. and {Russeil}, D. and {Schuller}, F. and {Anderson}, L.~D. and {Barnes}, P.~J. and {Beltr{\'a}n}, M.~T. and {Beuther}, H. and {Bontemps}, S. and {Bronfman}, L. and {Csengeri}, T. and {Dobbs}, C.~L. and {Eden}, D. and {Giannetti}, A. and {Kauffmann}, J. and {Mattern}, M. and {Medina}, S. -N.~X. and {Menten}, K.~M. and {Lee}, M. -Y. and {Pettitt}, A.~R. and {Riener}, M. and {Rigby}, A.~J. and {Traficante}, A. and {Veena}, V.~S. and {Wienen}, M. and {Wyrowski}, F. and {Agurto}, C. and {Azagra}, F. and {Cesaroni}, R. and {Finger}, R. and {Gonzalez}, E. and {Henning}, T. and {Hernandez}, A.~K. and {Kainulainen}, J. and {Leurini}, S. and {Lopez}, S. and {Mac-Auliffe}, F. and {Mazumdar}, P. and {Molinari}, S. and {Motte}, F. and {Muller}, E. and {Nguyen-Luong}, Q. and {Parra}, R. and {Perez-Beaupuits}, J. -P. and {Montenegro-Montes}, F.~M. and {Moore}, T.~J.~T. and {Ragan}, S.~E. and {S{\'a}nchez-Monge}, A. and {Sanna}, A. and {Schilke}, P. and {Schisano}, E. and {Schneider}, N. and {Suri}, S. and {Testi}, L. and {Torstensson}, K. and {Venegas}, P. and {Wang}, K. and {Zavagno}, A.},
        title = "{The SEDIGISM survey: molecular clouds in the inner Galaxy}",
         year = 2021,
        month = jan,
        pages = {3027-3049},
          doi = {10.1093/mnras/staa2480},
archivePrefix = {arXiv},
       eprint = {2012.01502},
 primaryClass = {astro-ph.GA},
    publisher = {OUP},
       adsurl = {https://ui.adsabs.harvard.edu/abs/2021MNRAS.500.3027D}
}

@ARTICLE{2014ApJS..212....1A,
       author = {{Anderson}, L.~D. and {Bania}, T.~M. and {Balser}, Dana S. and {Cunningham}, V. and {Wenger}, T.~V. and {Johnstone}, B.~M. and {Armentrout}, W.~P.},
        title = "{The WISE Catalog of Galactic H II Regions}",
      journal = {\apjs},
         year = 2014,
        month = may,
       volume = {212},
       number = {1},
          eid = {1},
        pages = {1},
          doi = {10.1088/0067-0049/212/1/1},
archivePrefix = {arXiv},
       eprint = {1312.6202},
 primaryClass = {astro-ph.GA},
       adsurl = {https://ui.adsabs.harvard.edu/abs/2014ApJS..212....1A}
}

@ARTICLE{2004ApJ...616L..19B,
       author = {{Beuther}, H. and {Zhang}, Q. and {Hunter}, T.~R. and {Sridharan}, T.~K. and {Zhao}, J. -H. and {Sollins}, P. and {Ho}, P.~T.~P. and {Liu}, S. -Y. and {Ohashi}, N. and {Su}, Y.~N. and {Lim}, J.},
        title = "{Submillimeter Array Multiline observations of the Massive Star-forming region IRAS 18089-1732}",
      journal = {\apjl},
         year = 2004,
        month = nov,
       volume = {616},
       number = {1},
        pages = {L19-L22},
          doi = {10.1086/422500},
archivePrefix = {arXiv},
       eprint = {astro-ph/0406020},
 primaryClass = {astro-ph},
       adsurl = {https://ui.adsabs.harvard.edu/abs/2004ApJ...616L..19B}
}

@ARTICLE{2004ApJ...616L..23B,
       author = {{Beuther}, H. and {Hunter}, T.~R. and {Zhang}, Q. and {Sridharan}, T.~K. and {Zhao}, J. -H. and {Sollins}, P. and {Ho}, P.~T.~P. and {Ohashi}, N. and {Su}, Y.~N. and {Lim}, J. and {Liu}, S. -Y.},
        title = "{Submillimeter Array Outflow/Disk Studies in the Massive Star-forming Region IRAS 18089-1732}",
      journal = {\apjl},
         year = 2004,
        month = nov,
       volume = {616},
       number = {1},
        pages = {L23-L26},
          doi = {10.1086/383570},
archivePrefix = {arXiv},
       eprint = {astro-ph/0402505},
 primaryClass = {astro-ph},
       adsurl = {https://ui.adsabs.harvard.edu/abs/2004ApJ...616L..23B}
}

@ARTICLE{2014MNRAS.443.2252G,
       author = {{Green}, C. -E. and {Green}, J.~A. and {Burton}, M.~G. and {Horiuchi}, S. and {Tothill}, N.~F.~H. and {Walsh}, A.~J. and {Purcell}, C.~R. and {Lovell}, J.~E.~J. and {Millar}, T.~J.},
        title = "{New detections of HC$_{5}$N towards hot cores associated with 6.7 GHz methanol masers}",
      journal = {\mnras},
         year = 2014,
        month = sep,
       volume = {443},
       number = {3},
        pages = {2252-2263},
          doi = {10.1093/mnras/stu1349},
archivePrefix = {arXiv},
       eprint = {1407.1699},
 primaryClass = {astro-ph.GA},
       adsurl = {https://ui.adsabs.harvard.edu/abs/2014MNRAS.443.2252G}
}

@ARTICLE{2015ApJS..219....2J,
       author = {{Jin}, Mihwa and {Lee}, Jeong-Eun and {Kim}, Kee-Tae},
        title = "{The HCN/HNC Abundance Ratio Toward Different Evolutionary Phases of Massive Star Formation}",
      journal = {\apjs},
         year = 2015,
        month = jul,
       volume = {219},
       number = {1},
          eid = {2},
        pages = {2},
          doi = {10.1088/0067-0049/219/1/2},
archivePrefix = {arXiv},
       eprint = {1505.00849},
 primaryClass = {astro-ph.SR},
       adsurl = {https://ui.adsabs.harvard.edu/abs/2015ApJS..219....2J}
}

@ARTICLE{2017ApJ...844...68T,
       author = {{Taniguchi}, Kotomi and {Saito}, Masao and {Hirota}, Tomoya and {Ozeki}, Hiroyuki and {Miyamoto}, Yusuke and {Kaneko}, Hiroyuki and {Minamidani}, Tetsuhiro and {Shimoikura}, Tomomi and {Nakamura}, Fumitaka and {Dobashi}, Kazuhito},
        title = "{Observations of Cyanopolyynes toward Four High-mass Star-forming Regions Containing Hot Cores}",
      journal = {\apj},
         year = 2017,
        month = jul,
       volume = {844},
       number = {1},
          eid = {68},
        pages = {68},
          doi = {10.3847/1538-4357/aa7899},
archivePrefix = {arXiv},
       eprint = {1706.02465},
 primaryClass = {astro-ph.GA},
       adsurl = {https://ui.adsabs.harvard.edu/abs/2017ApJ...844...68T}
}

@ARTICLE{2010A&A...517A..56F,
       author = {{Fontani}, F. and {Cesaroni}, R. and {Furuya}, R.~S.},
        title = "{Class I and Class II methanol masers in high-mass star-forming regions}",
      journal = {\aap},
         year = 2010,
        month = jul,
       volume = {517},
          eid = {A56},
        pages = {A56},
          doi = {10.1051/0004-6361/200913679},
archivePrefix = {arXiv},
       eprint = {1004.3689},
 primaryClass = {astro-ph.GA},
       adsurl = {https://ui.adsabs.harvard.edu/abs/2010A&A...517A..56F}
}

@ARTICLE{2002A&A...390..289B,
       author = {{Beuther}, H. and {Walsh}, A. and {Schilke}, P. and {Sridharan}, T.~K. and {Menten}, K.~M. and {Wyrowski}, F.},
        title = "{CH$_{3}$OH and H$_{2}$O masers in high-mass star-forming regions}",
      journal = {\aap},
         year = 2002,
        month = jul,
       volume = {390},
        pages = {289-298},
          doi = {10.1051/0004-6361:20020710},
archivePrefix = {arXiv},
       eprint = {astro-ph/0205348},
 primaryClass = {astro-ph},
       adsurl = {https://ui.adsabs.harvard.edu/abs/2002A&A...390..289B}
}

@ARTICLE{2007A&A...465..865E,
       author = {{Edris}, K.~A. and {Fuller}, G.~A. and {Cohen}, R.~J.},
        title = "{A survey of OH masers towards high mass protostellar objects}",
      journal = {\aap},
         year = 2007,
        month = apr,
       volume = {465},
       number = {3},
        pages = {865-877},
          doi = {10.1051/0004-6361:20066280},
archivePrefix = {arXiv},
       eprint = {astro-ph/0701652},
 primaryClass = {astro-ph},
       adsurl = {https://ui.adsabs.harvard.edu/abs/2007A&A...465..865E}
}

@ARTICLE{2022MNRAS.510.3389U,
       author = {{Urquhart}, J.~S. and {Wells}, M.~R.~A. and {Pillai}, T. and {Leurini}, S. and {Giannetti}, A. and {Moore}, T.~J.~T. and {Thompson}, M.~A. and {Figura}, C. and {Colombo}, D. and {Yang}, A.~Y. and {K{\"o}nig}, C. and {Wyrowski}, F. and {Menten}, K.~M. and {Rigby}, A.~J. and {Eden}, D.~J. and {Ragan}, S.~E.},
        title = "{ATLASGAL - evolutionary trends in high-mass star formation}",
      journal = {\mnras},
         year = 2022,
        month = mar,
       volume = {510},
       number = {3},
        pages = {3389-3407},
          doi = {10.1093/mnras/stab3511},
archivePrefix = {arXiv},
       eprint = {2111.12816},
 primaryClass = {astro-ph.GA},
       adsurl = {https://ui.adsabs.harvard.edu/abs/2022MNRAS.510.3389U}
}

@ARTICLE{2011ApJ...731...90D,
       author = {{Dunham}, Miranda K. and {Robitaille}, Thomas P. and {Evans}, Neal J., II and {Schlingman}, Wayne M. and {Cyganowski}, Claudia J. and {Urquhart}, James},
        title = "{A Mid-infrared Census of Star Formation Activity in Bolocam Galactic Plane Survey Sources}",
      journal = {\apj},
         year = 2011,
        month = apr,
       volume = {731},
       number = {2},
          eid = {90},
        pages = {90},
          doi = {10.1088/0004-637X/731/2/90},
archivePrefix = {arXiv},
       eprint = {1102.1032},
 primaryClass = {astro-ph.GA},
       adsurl = {https://ui.adsabs.harvard.edu/abs/2011ApJ...731...90D}
}

@ARTICLE{2022A&A...658A.160Y,
       author = {{Yang}, A.~Y. and {Urquhart}, J.~S. and {Wyrowski}, F. and {Thompson}, M.~A. and {K{\"o}nig}, C. and {Colombo}, D. and {Menten}, K.~M. and {Duarte-Cabral}, A. and {Schuller}, F. and {Csengeri}, T. and {Eden}, D. and {Barnes}, P. and {Traficante}, A. and {Bronfman}, L. and {Sanchez-Monge}, A. and {Ginsburg}, A. and {Cesaroni}, R. and {Lee}, M. -Y. and {Beuther}, H. and {Medina}, S. -N.~X. and {Mazumdar}, P. and {Henning}, T.},
        title = "{The SEDIGISM survey: A search for molecular outflows}",
      journal = {\aap},
         year = 2022,
        month = feb,
       volume = {658},
          eid = {A160},
        pages = {A160},
          doi = {10.1051/0004-6361/202142039},
archivePrefix = {arXiv},
       eprint = {2111.10850},
 primaryClass = {astro-ph.GA},
       adsurl = {https://ui.adsabs.harvard.edu/abs/2022A&A...658A.160Y}
}

@ARTICLE{2016MNRAS.460...82S,
       author = {{Sz{\H{u}}cs}, L{\'a}szl{\'o} and {Glover}, Simon C.~O. and {Klessen}, Ralf S.},
        title = "{How well does CO emission measure the H$_{2}$ mass of MCs?}",
      journal = {\mnras},
         year = 2016,
        month = jul,
       volume = {460},
       number = {1},
        pages = {82-102},
          doi = {10.1093/mnras/stw912},
archivePrefix = {arXiv},
       eprint = {1604.04545},
 primaryClass = {astro-ph.GA},
       adsurl = {https://ui.adsabs.harvard.edu/abs/2016MNRAS.460...82S}
}

@ARTICLE{1996A&AS..115...81B,
       author = {{Bronfman}, L. and {Nyman}, L. -A. and {May}, J.},
        title = "{A CS(2-1) survey of IRAS point sources with color characteristics of ultra-compact HII regions.}",
      journal = {\aaps},
         year = 1996,
        month = jan,
       volume = {115},
        pages = {81},
       adsurl = {https://ui.adsabs.harvard.edu/abs/1996A&AS..115...81B}
}

@ARTICLE{2017MNRAS.469.2163E,
       author = {{Eden}, D.~J. and {Moore}, T.~J.~T. and {Plume}, R. and {Urquhart}, J.~S. and {Thompson}, M.~A. and {Parsons}, H. and {Dempsey}, J.~T. and {Rigby}, A.~J. and {Morgan}, L.~K. and {Thomas}, H.~S. and {Berry}, D. and {Buckle}, J. and {Brunt}, C.~M. and {Butner}, H.~M. and {Carretero}, D. and {Chrysostomou}, A. and {Currie}, M.~J. and {deVilliers}, H.~M. and {Fich}, M. and {Gibb}, A.~G. and {Hoare}, M.~G. and {Jenness}, T. and {Manser}, G. and {Mottram}, J.~C. and {Natario}, C. and {Olguin}, F. and {Peretto}, N. and {Pestalozzi}, M. and {Polychroni}, D. and {Redman}, R.~O. and {Salji}, C. and {Summers}, L.~J. and {Tahani}, K. and {Traficante}, A. and {diFrancesco}, J. and {Evans}, A. and {Fuller}, G.~A. and {Johnstone}, D. and {Joncas}, G. and {Longmore}, S.~N. and {Martin}, P.~G. and {Richer}, J.~S. and {Weferling}, B. and {White}, G.~J. and {Zhu}, M.},
        title = "{The JCMT Plane Survey: first complete data release - emission maps and compact source catalogue}",
      journal = {\mnras},
         year = 2017,
        month = aug,
       volume = {469},
       number = {2},
        pages = {2163-2183},
          doi = {10.1093/mnras/stx874},
archivePrefix = {arXiv},
       eprint = {1704.02982},
 primaryClass = {astro-ph.GA},
       adsurl = {https://ui.adsabs.harvard.edu/abs/2017MNRAS.469.2163E}
}

@ARTICLE{2008AJ....136.2413R,
       author = {{Robitaille}, Thomas P. and {Meade}, Marilyn R. and {Babler}, Brian L. and {Whitney}, Barbara A. and {Johnston}, Katharine G. and {Indebetouw}, R{\'e}my and {Cohen}, Martin and {Povich}, Matthew S. and {Sewilo}, Marta and {Benjamin}, Robert A. and {Churchwell}, Edward},
        title = "{Intrinsically Red Sources Observed by Spitzer in the Galactic Midplane}",
      journal = {\aj},
         year = 2008,
        month = dec,
       volume = {136},
       number = {6},
        pages = {2413-2440},
          doi = {10.1088/0004-6256/136/6/2413},
archivePrefix = {arXiv},
       eprint = {0809.1654},
 primaryClass = {astro-ph},
       adsurl = {https://ui.adsabs.harvard.edu/abs/2008AJ....136.2413R}
}

@ARTICLE{2021ApJS..254...33K,
       author = {{Kuhn}, Michael A. and {de Souza}, Rafael S. and {Krone-Martins}, Alberto and {Castro-Ginard}, Alfred and {Ishida}, Emille E.~O. and {Povich}, Matthew S. and {Hillenbrand}, Lynne A. and {COIN Collaboration}},
        title = "{SPICY: The Spitzer/IRAC Candidate YSO Catalog for the Inner Galactic Midplane}",
      journal = {\apjs},
         year = 2021,
        month = jun,
       volume = {254},
       number = {2},
          eid = {33},
        pages = {33},
          doi = {10.3847/1538-4365/abe465},
archivePrefix = {arXiv},
       eprint = {2011.12961},
 primaryClass = {astro-ph.GA},
       adsurl = {https://ui.adsabs.harvard.edu/abs/2021ApJS..254...33K}
}

@ARTICLE{2013MNRAS.435..663B,
       author = {{Bhavya}, B. and {Subramaniam}, Annapurni and {Kuriakose}, V.~C.},
        title = "{Study of young stellar objects and associated filamentary structures in the inner Galaxy}",
      journal = {\mnras},
         year = 2013,
        month = oct,
       volume = {435},
       number = {1},
        pages = {663-678},
          doi = {10.1093/mnras/stt1324},
archivePrefix = {arXiv},
       eprint = {1307.5225},
 primaryClass = {astro-ph.SR},
       adsurl = {https://ui.adsabs.harvard.edu/abs/2013MNRAS.435..663B}
}

@ARTICLE{2005A&A...432..737P,
       author = {{Pestalozzi}, M.~R. and {Minier}, V. and {Booth}, R.~S.},
        title = "{A general catalogue of 6.7-GHz methanol masers. I. Data.}",
      journal = {\aap},
         year = 2005,
        month = mar,
       volume = {432},
       number = {2},
        pages = {737-742},
          doi = {10.1051/0004-6361:20035855},
archivePrefix = {arXiv},
       eprint = {astro-ph/0411564},
 primaryClass = {astro-ph},
       adsurl = {https://ui.adsabs.harvard.edu/abs/2005A&A...432..737P}
}

@ARTICLE{2010MNRAS.409..913G,
       author = {{Green}, J.~A. and {Caswell}, J.~L. and {Fuller}, G.~A. and {Avison}, A. and {Breen}, S.~L. and {Ellingsen}, S.~P. and {Gray}, M.~D. and {Pestalozzi}, M. and {Quinn}, L. and {Thompson}, M.~A. and {Voronkov}, M.~A.},
        title = "{The 6-GHz methanol multibeam maser catalogue - II. Galactic longitudes 6{\textdegree} to 20{\textdegree}}",
      journal = {\mnras},
         year = 2010,
        month = dec,
       volume = {409},
       number = {3},
        pages = {913-935},
          doi = {10.1111/j.1365-2966.2010.17376.x},
archivePrefix = {arXiv},
       eprint = {1007.3050},
 primaryClass = {astro-ph.GA},
       adsurl = {https://ui.adsabs.harvard.edu/abs/2010MNRAS.409..913G}
}

@ARTICLE{2000ApJS..129..159A,
       author = {{Argon}, A.~L. and {Reid}, M.~J. and {Menten}, Karl M.},
        title = "{Interstellar Hydroxyl Masers in the Galaxy. I. The VLA Survey}",
      journal = {\apjs},
         year = 2000,
        month = jul,
       volume = {129},
       number = {1},
        pages = {159-227},
          doi = {10.1086/313406},
       adsurl = {https://ui.adsabs.harvard.edu/abs/2000ApJS..129..159A}
}

@ARTICLE{2010ApJ...716..433K,
       author = {{Kauffmann}, J. and {Pillai}, T. and {Shetty}, R. and {Myers}, P.~C. and {Goodman}, A.~A.},
        title = "{The Mass-size Relation from Clouds to Cores. II. Solar Neighborhood Clouds}",
      journal = {\apj},
         year = 2010,
        month = jun,
       volume = {716},
       number = {1},
        pages = {433-445},
          doi = {10.1088/0004-637X/716/1/433},
archivePrefix = {arXiv},
       eprint = {1004.1170},
 primaryClass = {astro-ph.GA},
       adsurl = {https://ui.adsabs.harvard.edu/abs/2010ApJ...716..433K}
}

@ARTICLE{2017MNRAS.471..100E,
       author = {{Elia}, Davide and {Molinari}, S. and {Schisano}, E. and {Pestalozzi}, M. and {Pezzuto}, S. and {Merello}, M. and {Noriega-Crespo}, A. and {Moore}, T.~J.~T. and {Russeil}, D. and {Mottram}, J.~C. and {Paladini}, R. and {Strafella}, F. and {Benedettini}, M. and {Bernard}, J.~P. and {Di Giorgio}, A. and {Eden}, D.~J. and {Fukui}, Y. and {Plume}, R. and {Bally}, J. and {Martin}, P.~G. and {Ragan}, S.~E. and {Jaffa}, S.~E. and {Motte}, F. and {Olmi}, L. and {Schneider}, N. and {Testi}, L. and {Wyrowski}, F. and {Zavagno}, A. and {Calzoletti}, L. and {Faustini}, F. and {Natoli}, P. and {Palmeirim}, P. and {Piacentini}, F. and {Piazzo}, L. and {Pilbratt}, G.~L. and {Polychroni}, D. and {Baldeschi}, A. and {Beltr{\'a}n}, M.~T. and {Billot}, N. and {Cambr{\'e}sy}, L. and {Cesaroni}, R. and {Garc{\'\i}a-Lario}, P. and {Hoare}, M.~G. and {Huang}, M. and {Joncas}, G. and {Liu}, S.~J. and {Maiolo}, B.~M.~T. and {Marsh}, K.~A. and {Maruccia}, Y. and {M{\`e}ge}, P. and {Peretto}, N. and {Rygl}, K.~L.~J. and {Schilke}, P. and {Thompson}, M.~A. and {Traficante}, A. and {Umana}, G. and {Veneziani}, M. and {Ward-Thompson}, D. and {Whitworth}, A.~P. and {Arab}, H. and {Bandieramonte}, M. and {Becciani}, U. and {Brescia}, M. and {Buemi}, C. and {Bufano}, F. and {Butora}, R. and {Cavuoti}, S. and {Costa}, A. and {Fiorellino}, E. and {Hajnal}, A. and {Hayakawa}, T. and {Kacsuk}, P. and {Leto}, P. and {Li Causi}, G. and {Marchili}, N. and {Martinavarro-Armengol}, S. and {Mercurio}, A. and {Molinaro}, M. and {Riccio}, G. and {Sano}, H. and {Sciacca}, E. and {Tachihara}, K. and {Torii}, K. and {Trigilio}, C. and {Vitello}, F. and {Yamamoto}, H.},
        title = "{The Hi-GAL compact source catalogue - I. The physical properties of the clumps in the inner Galaxy (-71.0{\textdegree} < {\ensuremath{\ell}} < 67.0{\textdegree})}",
      journal = {\mnras},
         year = 2017,
        month = oct,
       volume = {471},
       number = {1},
        pages = {100-143},
          doi = {10.1093/mnras/stx1357},
archivePrefix = {arXiv},
       eprint = {1706.01046},
 primaryClass = {astro-ph.GA},
       adsurl = {https://ui.adsabs.harvard.edu/abs/2017MNRAS.471..100E}
}

@ARTICLE{1994ARA&A..32..191W,
       author = {{Wilson}, T.~L. and {Rood}, R.},
        title = "{Abundances in the Interstellar Medium}",
      journal = {\araa},
         year = 1994,
        month = jan,
       volume = {32},
        pages = {191-226},
          doi = {10.1146/annurev.aa.32.090194.001203},
       adsurl = {https://ui.adsabs.harvard.edu/abs/1994ARA&A..32..191W}
}

@ARTICLE{1982ApJ...262..590F,
       author = {{Frerking}, M.~A. and {Langer}, W.~D. and {Wilson}, R.~W.},
        title = "{The relationship between carbon monoxide abundance and visual extinction in interstellar clouds.}",
      journal = {\apj},
         year = 1982,
        month = nov,
       volume = {262},
        pages = {590-605},
          doi = {10.1086/160451},
       adsurl = {https://ui.adsabs.harvard.edu/abs/1982ApJ...262..590F}
}

@ARTICLE{1973AJ.....78..929P,
       author = {{Panagia}, Nino},
        title = "{Some Physical parameters of early-type stars}",
      journal = {\aj},
         year = 1973,
        month = nov,
       volume = {78},
        pages = {929-934},
          doi = {10.1086/111498},
       adsurl = {https://ui.adsabs.harvard.edu/abs/1973AJ.....78..929P}
}

@ARTICLE{2018ARA&A..56...41M,
       author = {{Motte}, Fr{\'e}d{\'e}rique and {Bontemps}, Sylvain and {Louvet}, Fabien},
        title = "{High-Mass Star and Massive Cluster Formation in the Milky Way}",
      journal = {\araa},
         year = 2018,
        month = sep,
       volume = {56},
        pages = {41-82},
          doi = {10.1146/annurev-astro-091916-055235},
archivePrefix = {arXiv},
       eprint = {1706.00118},
 primaryClass = {astro-ph.GA},
       adsurl = {https://ui.adsabs.harvard.edu/abs/2018ARA&A..56...41M}
}

@ARTICLE{2025arXiv250116866B,
       author = {{Beuther}, H. and {Kuiper}, R. and {Tafalla}, M.},
        title = "{Star formation from low to high mass: A comparative view}",
      journal = {arXiv e-prints},
         year = 2025,
        month = jan,
          eid = {arXiv:2501.16866},
        pages = {arXiv:2501.16866},
          doi = {10.48550/arXiv.2501.16866},
archivePrefix = {arXiv},
       eprint = {2501.16866},
 primaryClass = {astro-ph.GA},
       adsurl = {https://ui.adsabs.harvard.edu/abs/2025arXiv250116866B}
}

@INPROCEEDINGS{2023ASPC..534..153H,
       author = {{Hacar}, A. and {Clark}, S.~E. and {Heitsch}, F. and {Kainulainen}, J. and {Panopoulou}, G.~V. and {Seifried}, D. and {Smith}, R.},
        title = "{Initial Conditions for Star Formation: a Physical Description of the Filamentary ISM}",
    booktitle = {Protostars and Planets VII},
         year = 2023,
       editor = {{Inutsuka}, S. and {Aikawa}, Y. and {Muto}, T. and {Tomida}, K. and {Tamura}, M.},
       series = {Astronomical Society of the Pacific Conference Series},
       volume = {534},
        month = jul,
        pages = {153},
          doi = {10.48550/arXiv.2203.09562},
archivePrefix = {arXiv},
       eprint = {2203.09562},
 primaryClass = {astro-ph.GA},
       adsurl = {https://ui.adsabs.harvard.edu/abs/2023ASPC..534..153H}
}

@ARTICLE{2021A&A...652A..71S,
       author = {{Sabatini}, G. and {Bovino}, S. and {Giannetti}, A. and {Grassi}, T. and {Brand}, J. and {Schisano}, E. and {Wyrowski}, F. and {Leurini}, S. and {Menten}, K.~M.},
        title = "{Establishing the evolutionary timescales of the massive star formation process through chemistry}",
      journal = {\aap},
         year = 2021,
        month = aug,
       volume = {652},
          eid = {A71},
        pages = {A71},
          doi = {10.1051/0004-6361/202140469},
archivePrefix = {arXiv},
       eprint = {2106.00692},
 primaryClass = {astro-ph.GA},
       adsurl = {https://ui.adsabs.harvard.edu/abs/2021A&A...652A..71S}
}

@ARTICLE{2019A&A...629A..81T,
       author = {{Trevi{\~n}o-Morales}, S.~P. and {Fuente}, A. and {S{\'a}nchez-Monge}, {\'A}. and {Kainulainen}, J. and {Didelon}, P. and {Suri}, S. and {Schneider}, N. and {Ballesteros-Paredes}, J. and {Lee}, Y. -N. and {Hennebelle}, P. and {Pilleri}, P. and {Gonz{\'a}lez-Garc{\'\i}a}, M. and {Kramer}, C. and {Garc{\'\i}a-Burillo}, S. and {Luna}, A. and {Goicoechea}, J.~R. and {Tremblin}, P. and {Geen}, S.},
        title = "{Dynamics of cluster-forming hub-filament systems. The case of the high-mass star-forming complex Monoceros R2}",
      journal = {\aap},
         year = 2019,
        month = sep,
       volume = {629},
          eid = {A81},
        pages = {A81},
          doi = {10.1051/0004-6361/201935260},
archivePrefix = {arXiv},
       eprint = {1907.03524},
 primaryClass = {astro-ph.GA},
       adsurl = {https://ui.adsabs.harvard.edu/abs/2019A&A...629A..81T}
}

@ARTICLE{2024MNRAS.531..649G,
       author = {{Goedhart}, S. and {Cotton}, W.~D. and {Camilo}, F. and {Thompson}, M.~A. and {Umana}, G. and {Bietenholz}, M. and {Woudt}, P.~A. and {Anderson}, L.~D. and {Bordiu}, C. and {Buckley}, D.~A.~H. and {Buemi}, C.~S. and {Bufano}, F. and {Cavallaro}, F. and {Chen}, H. and {Chibueze}, J.~O. and {Egbo}, D. and {Frank}, B.~S. and {Hoare}, M.~G. and {Ingallinera}, A. and {Irabor}, T. and {Kraan-Korteweg}, R.~C. and {Kurapati}, S. and {Leto}, P. and {Loru}, S. and {Mutale}, M. and {Obonyo}, W.~O. and {Plavin}, A. and {Rajohnson}, S.~H.~A. and {Rigby}, A. and {Riggi}, S. and {Seidu}, M. and {Serra}, P. and {Smart}, B.~M. and {Stappers}, B.~W. and {Steyn}, N. and {Surnis}, M. and {Trigilio}, C. and {Williams}, G.~M. and {Abbott}, T.~D. and {Adam}, R.~M. and {Asad}, K.~M.~B. and {Baloyi}, T. and {Bauermeister}, E.~F. and {Bennet}, T.~G.~H. and {Bester}, H. and {Botha}, A.~G. and {Brederode}, L.~R.~S. and {Buchner}, S. and {Burger}, J.~P. and {Cheetham}, T. and {Cloete}, K. and {de Villiers}, M.~S. and {de Villiers}, D.~I.~L. and {du Toit}, L.~J. and {Esterhuyse}, S.~W.~P. and {Fanaroff}, B.~L. and {Fourie}, D.~J. and {Gamatham}, R.~R.~G. and {Gatsi}, T.~G. and {Geyer}, M. and {Gouws}, M. and {Gumede}, S.~C. and {Heywood}, I. and {Hokwana}, A. and {Hoosen}, S.~W. and {Horn}, D.~M. and {Horrell}, L.~M.~G. and {Hugo}, B.~V. and {Isaacson}, A.~I. and {J{\'o}zsa}, G.~I.~G. and {Jonas}, J.~L. and {Jordaan}, J.~D.~B.~L. and {Joubert}, A.~F. and {Julie}, R.~P.~M. and {Kapp}, F.~B. and {Kriek}, N. and {Kriel}, H. and {Krishnan}, V.~K. and {Kusel}, T.~W. and {Legodi}, L.~S. and {Lehmensiek}, R. and {Lord}, R.~T. and {Macfarlane}, P.~S. and {Magnus}, L.~G. and {Magozore}, C. and {Main}, J.~P.~L. and {Malan}, J.~A. and {Manley}, J.~R. and {Marais}, S.~J. and {Maree}, M.~D.~J. and {Martens}, A. and {Maruping}, P. and {McAlpine}, K. and {Merry}, B.~C. and {Mgodeli}, M. and {Millenaar}, R.~P. and {Mokone}, O.~J. and {Monama}, T.~E. and {New}, W.~S. and {Ngcebetsha}, B. and {Ngoasheng}, K.~J. and {Nicolson}, G.~D. and {Ockards}, M.~T. and {Oozeer}, N. and {Passmoor}, S.~S. and {Patel}, A.~A. and {Peens-Hough}, A. and {Perkins}, S.~J. and {Ramaila}, A.~J.~T. and {Ratcliffe}, S.~M. and {Renil}, R. and {Richter}, L.~L. and {Salie}, S. and {Sambu}, N. and {Schollar}, C.~T.~G. and {Schwardt}, L.~C. and {Schwartz}, R.~L. and {Serylak}, M. and {Siebrits}, R. and {Sirothia}, S.~K. and {Slabber}, M.~J. and {Smirnov}, O.~M. and {Tiplady}, A.~J. and {van Balla}, T.~J. and {van der Byl}, A. and {Van Tonder}, V. and {Venter}, A.~J. and {Venter}, M. and {Welz}, M.~G. and {Williams}, L.~P.},
        title = "{The SARAO MeerKAT 1.3 GHz Galactic Plane Survey}",
      journal = {\mnras},
         year = 2024,
        month = jun,
       volume = {531},
       number = {1},
        pages = {649-681},
          doi = {10.1093/mnras/stae1166},
archivePrefix = {arXiv},
       eprint = {2312.07275},
 primaryClass = {astro-ph.GA},
       adsurl = {https://ui.adsabs.harvard.edu/abs/2024MNRAS.531..649G}
}

@ARTICLE{2016PASA...33...30R,
       author = {{Rathborne}, J.~M. and {Whitaker}, J.~S. and {Jackson}, J.~M. and {Foster}, J.~B. and {Contreras}, Y. and {Stephens}, I.~W. and {Guzm{\'a}n}, A.~E. and {Longmore}, S.~N. and {Sanhueza}, P. and {Schuller}, F. and {Wyrowski}, F. and {Urquhart}, J.~S.},
        title = "{Molecular Line Emission Towards High-Mass Clumps: The MALT90 Catalogue}",
      journal = {\pasa},
         year = 2016,
        month = jul,
       volume = {33},
          eid = {e030},
        pages = {e030},
          doi = {10.1017/pasa.2016.23},
       adsurl = {https://ui.adsabs.harvard.edu/abs/2016PASA...33...30R}
}

@ARTICLE{2013ApJS..209....2S,
       author = {{Shirley}, Yancy L. and {Ellsworth-Bowers}, Timothy P. and {Svoboda}, Brian and {Schlingman}, Wayne M. and {Ginsburg}, Adam and {Rosolowsky}, Erik and {Gerner}, Thomas and {Mairs}, Steven and {Battersby}, Cara and {Stringfellow}, Guy and {Dunham}, Miranda K. and {Glenn}, Jason and {Bally}, John},
        title = "{The Bolocam Galactic Plane Survey. X. A Complete Spectroscopic Catalog of Dense Molecular Gas Observed toward 1.1 mm Dust Continuum Sources with 7.{\textdegree}5 <= l <= 194{\textdegree}}",
      journal = {\apjs},
         year = 2013,
        month = nov,
       volume = {209},
       number = {1},
          eid = {2},
        pages = {2},
          doi = {10.1088/0067-0049/209/1/2},
archivePrefix = {arXiv},
       eprint = {1308.4149},
 primaryClass = {astro-ph.GA},
       adsurl = {https://ui.adsabs.harvard.edu/abs/2013ApJS..209....2S}
}

@ARTICLE{2011ApJ...733...25X,
       author = {{Xu}, Y. and {Moscadelli}, L. and {Reid}, M.~J. and {Menten}, K.~M. and {Zhang}, B. and {Zheng}, X.~W. and {Brunthaler}, A.},
        title = "{Trigonometric Parallaxes of Massive Star-forming Regions. VIII. G12.89+0.49, G15.03-0.68 (M17), and G27.36-0.16}",
      journal = {\apj},
         year = 2011,
        month = may,
       volume = {733},
       number = {1},
          eid = {25},
        pages = {25},
          doi = {10.1088/0004-637X/733/1/25},
archivePrefix = {arXiv},
       eprint = {1103.3139},
 primaryClass = {astro-ph.GA},
       adsurl = {https://ui.adsabs.harvard.edu/abs/2011ApJ...733...25X}
}

@ARTICLE{2024MNRAS.528.2199R,
       author = {{Rawat}, Vineet and {Samal}, M.~R. and {Walker}, D.~L. and {Ojha}, D.~K. and {Tej}, A. and {Zavagno}, A. and {Zhang}, C.~P. and {Elia}, Davide and {Dutta}, S. and {Jose}, J. and {Eswaraiah}, C. and {Sharma}, E.},
        title = "{The Giant Molecular Cloud G148.24+00.41: gas properties, kinematics, and cluster formation at the nexus of filamentary flows}",
      journal = {\mnras},
         year = 2024,
        month = feb,
       volume = {528},
       number = {2},
        pages = {2199-2219},
          doi = {10.1093/mnras/stae060},
archivePrefix = {arXiv},
       eprint = {2401.03202},
 primaryClass = {astro-ph.GA},
       adsurl = {https://ui.adsabs.harvard.edu/abs/2024MNRAS.528.2199R}
}

@ARTICLE{2025A&A...699A.137B,
       author = {{Berdikhan}, Dilda and {Esimbek}, Jarken and {Henkel}, Christian and {Xu}, Ye and {Zhou}, Jianjun and {Liu}, De-Jian and {Abdikamalov}, Ernazar and {Ma}, Yingxiu and {Komesh}, Toktarkhan and {He}, Yuxin and {Zhang}, Wenjun and {Tang}, Xindi and {Wu}, Gang and {Li}, Dalei and {Zhou}, Dongdong and {Tursun}, Kadirya and {Shen}, Hailiang and {Imanaly}, Ernar and {Jandaolet}, Qaynar and {Manapbayeva}, Arailym and {Tuiakbayeva}, Duriya},
        title = "{Cloud-cloud collision and star formation in G013.313+0.193}",
      journal = {\aap},
         year = 2025,
        month = jul,
       volume = {699},
          eid = {A137},
        pages = {A137},
          doi = {10.1051/0004-6361/202453285},
archivePrefix = {arXiv},
       eprint = {2504.14943},
 primaryClass = {astro-ph.GA},
       adsurl = {https://ui.adsabs.harvard.edu/abs/2025A&A...699A.137B}
}

@ARTICLE{2025MNRAS.tmp.1315P,
       author = {{Porel}, Puja and {Soam}, Archana and {Karoly}, Janik and {Chung}, Eun Jung and {Lee}, Chang Won},
        title = "{Investigating the Kinematics of Molecular Gas in Cometary Globule L1616}",
      journal = {\mnras},
         year = 2025,
        month = aug,
          doi = {10.1093/mnras/staf1361},
archivePrefix = {arXiv},
       eprint = {2508.12246},
 primaryClass = {astro-ph.GA},
       adsurl = {https://ui.adsabs.harvard.edu/abs/2025MNRAS.tmp.1315P}
}

@ARTICLE{2017MNRAS.471.2730M,
       author = {{Marsh}, K.~A. and {Whitworth}, A.~P. and {Lomax}, O. and {Ragan}, S.~E. and {Becciani}, U. and {Cambr{\'e}sy}, L. and {Di Giorgio}, A. and {Eden}, D. and {Elia}, D. and {Kacsuk}, P. and {Molinari}, S. and {Palmeirim}, P. and {Pezzuto}, S. and {Schneider}, N. and {Sciacca}, E. and {Vitello}, F.},
        title = "{Multitemperature mapping of dust structures throughout the Galactic Plane using the PPMAP tool with Herschel Hi-GAL data}",
      journal = {\mnras},
         year = 2017,
        month = nov,
       volume = {471},
       number = {3},
        pages = {2730-2742},
          doi = {10.1093/mnras/stx1723},
archivePrefix = {arXiv},
       eprint = {1707.03808},
 primaryClass = {astro-ph.GA},
       adsurl = {https://ui.adsabs.harvard.edu/abs/2017MNRAS.471.2730M}
}

@ARTICLE{2024MNRAS.527.4244S,
       author = {{Seshadri}, Arun and {Vig}, S. and {Ghosh}, S.~K. and {Ojha}, D.~K.},
        title = "{Massive star formation in the hub-filament system of RCW 117}",
      journal = {\mnras},
         year = 2024,
        month = jan,
       volume = {527},
       number = {2},
        pages = {4244-4259},
          doi = {10.1093/mnras/stad3385},
archivePrefix = {arXiv},
       eprint = {2311.00477},
 primaryClass = {astro-ph.GA},
       adsurl = {https://ui.adsabs.harvard.edu/abs/2024MNRAS.527.4244S}
}

@ARTICLE{2011A&A...529L...6A,
       author = {{Arzoumanian}, D. and {Andr{\'e}}, Ph. and {Didelon}, P. and {K{\"o}nyves}, V. and {Schneider}, N. and {Men'shchikov}, A. and {Sousbie}, T. and {Zavagno}, A. and {Bontemps}, S. and {di Francesco}, J. and {Griffin}, M. and {Hennemann}, M. and {Hill}, T. and {Kirk}, J. and {Martin}, P. and {Minier}, V. and {Molinari}, S. and {Motte}, F. and {Peretto}, N. and {Pezzuto}, S. and {Spinoglio}, L. and {Ward-Thompson}, D. and {White}, G. and {Wilson}, C.~D.},
        title = "{Characterizing interstellar filaments with Herschel in IC 5146}",
      journal = {\aap},
         year = 2011,
        month = may,
       volume = {529},
          eid = {L6},
        pages = {L6},
          doi = {10.1051/0004-6361/201116596},
archivePrefix = {arXiv},
       eprint = {1103.0201},
 primaryClass = {astro-ph.GA},
       adsurl = {https://ui.adsabs.harvard.edu/abs/2011A&A...529L...6A}
}

@ARTICLE{2018A&A...615A.125B,
       author = {{Bresnahan}, D. and {Ward-Thompson}, D. and {Kirk}, J.~M. and {Pattle}, K. and {Eyres}, S. and {White}, G.~J. and {K{\"o}nyves}, V. and {Men'shchikov}, A. and {Andr{\'e}}, Ph. and {Schneider}, N. and {Di Francesco}, J. and {Arzoumanian}, D. and {Benedettini}, M. and {Ladjelate}, B. and {Palmeirim}, P. and {Bracco}, A. and {Molinari}, S. and {Pezzuto}, S. and {Spinoglio}, L.},
        title = "{The dense cores and filamentary structure of the molecular cloud in Corona Australis: Herschel SPIRE and PACS observations from the Herschel Gould Belt Survey}",
      journal = {\aap},
         year = 2018,
        month = jul,
       volume = {615},
          eid = {A125},
        pages = {A125},
          doi = {10.1051/0004-6361/201730515},
archivePrefix = {arXiv},
       eprint = {1801.07805},
 primaryClass = {astro-ph.GA},
       adsurl = {https://ui.adsabs.harvard.edu/abs/2018A&A...615A.125B}
}

@ARTICLE{2013ApJ...766...85R,
       author = {{Rivera-Ingraham}, A. and {Martin}, P.~G. and {Polychroni}, D. and {Motte}, F. and {Schneider}, N. and {Bontemps}, S. and {Hennemann}, M. and {Men'shchikov}, A. and {Nguyen Luong}, Q. and {Andr{\'e}}, Ph. and {Arzoumanian}, D. and {Bernard}, J. -Ph. and {Di Francesco}, J. and {Elia}, D. and {Fallscheer}, C. and {Hill}, T. and {Li}, J.~Z. and {Minier}, V. and {Pezzuto}, S. and {Roy}, A. and {Rygl}, K.~L.~J. and {Sadavoy}, S.~I. and {Spinoglio}, L. and {White}, G.~J. and {Wilson}, C.~D.},
        title = "{Herschel Observations of the W3 GMC: Clues to the Formation of Clusters of High-mass Stars}",
      journal = {\apj},
         year = 2013,
        month = apr,
       volume = {766},
       number = {2},
          eid = {85},
        pages = {85},
          doi = {10.1088/0004-637X/766/2/85},
archivePrefix = {arXiv},
       eprint = {1301.3805},
 primaryClass = {astro-ph.GA},
       adsurl = {https://ui.adsabs.harvard.edu/abs/2013ApJ...766...85R}
}

@ARTICLE{2015A&A...584A..91K,
       author = {{K{\"o}nyves}, V. and {Andr{\'e}}, Ph. and {Men'shchikov}, A. and {Palmeirim}, P. and {Arzoumanian}, D. and {Schneider}, N. and {Roy}, A. and {Didelon}, P. and {Maury}, A. and {Shimajiri}, Y. and {Di Francesco}, J. and {Bontemps}, S. and {Peretto}, N. and {Benedettini}, M. and {Bernard}, J. -Ph. and {Elia}, D. and {Griffin}, M.~J. and {Hill}, T. and {Kirk}, J. and {Ladjelate}, B. and {Marsh}, K. and {Martin}, P.~G. and {Motte}, F. and {Nguy{\^e}n Luong}, Q. and {Pezzuto}, S. and {Roussel}, H. and {Rygl}, K.~L.~J. and {Sadavoy}, S.~I. and {Schisano}, E. and {Spinoglio}, L. and {Ward-Thompson}, D. and {White}, G.~J.},
        title = "{A census of dense cores in the Aquila cloud complex: SPIRE/PACS observations from the Herschel Gould Belt survey}",
      journal = {\aap},
         year = 2015,
        month = dec,
       volume = {584},
          eid = {A91},
        pages = {A91},
          doi = {10.1051/0004-6361/201525861},
archivePrefix = {arXiv},
       eprint = {1507.05926},
 primaryClass = {astro-ph.GA},
       adsurl = {https://ui.adsabs.harvard.edu/abs/2015A&A...584A..91K}
}

@ARTICLE{2013ApJ...773..102F,
       author = {{Fallscheer}, C. and {Reid}, M.~A. and {Di Francesco}, J. and {Martin}, P.~G. and {Hill}, T. and {Hennemann}, M. and {Nguyen-Luong}, Q. and {Motte}, F. and {Men'shchikov}, A. and {Andr{\'e}}, Ph. and {Ward-Thompson}, D. and {Griffin}, M. and {Kirk}, J. and {Konyves}, V. and {Rygl}, K.~L.~J. and {Sadavoy}, S. and {Sauvage}, M. and {Schneider}, N. and {Anderson}, L.~D. and {Benedettini}, M. and {Bernard}, J. -P. and {Bontemps}, S. and {Ginsburg}, A. and {Molinari}, S. and {Polychroni}, D. and {Rivera-Ingraham}, A. and {Roussel}, H. and {Testi}, L. and {White}, G. and {Williams}, J.~P. and {Wilson}, C.~D. and {Wong}, M. and {Zavagno}, A.},
        title = "{Herschel Reveals Massive Cold Clumps in NGC 7538}",
      journal = {\apj},
         year = 2013,
        month = aug,
       volume = {773},
       number = {2},
          eid = {102},
        pages = {102},
          doi = {10.1088/0004-637X/773/2/102},
archivePrefix = {arXiv},
       eprint = {1307.0022},
 primaryClass = {astro-ph.SR},
       adsurl = {https://ui.adsabs.harvard.edu/abs/2013ApJ...773..102F}
}

@ARTICLE{2015AJ....149...64G,
       author = {{Gutermuth}, Robert A. and {Heyer}, Mark},
        title = "{A 24{\,}{\ensuremath{\mu}}m Point Source Catalog of the Galactic Plane from Spitzer/MIPSGAL}",
      journal = {\aj},
         year = 2015,
        month = feb,
       volume = {149},
       number = {2},
          eid = {64},
        pages = {64},
          doi = {10.1088/0004-6256/149/2/64},
archivePrefix = {arXiv},
       eprint = {1412.4751},
 primaryClass = {astro-ph.SR},
       adsurl = {https://ui.adsabs.harvard.edu/abs/2015AJ....149...64G}
}

@ARTICLE{2024A&A...688A.203M,
       author = {{Marton}, G. and {Gezer}, I. and {Madar{\'a}sz}, M. and {Dionatos}, O. and {Audard}, M. and {Roquette}, J. and {Hernandez}, D. and {Paladini}, R. and {Altieri}, B.},
        title = "{The new Herschel/PACS Point Source Catalogue}",
      journal = {\aap},
         year = 2024,
        month = aug,
       volume = {688},
          eid = {A203},
        pages = {A203},
          doi = {10.1051/0004-6361/202450032},
archivePrefix = {arXiv},
       eprint = {2406.03116},
 primaryClass = {astro-ph.IM},
       adsurl = {https://ui.adsabs.harvard.edu/abs/2024A&A...688A.203M}
}

@BOOK{1997pism.book.....D,
       author = {{Dyson}, J.~E. and {Williams}, D.~A.},
        title = "{The physics of the interstellar medium}",
         year = 1997,
          doi = {10.1201/9780585368115},
       adsurl = {https://ui.adsabs.harvard.edu/abs/1997pism.book.....D}
}

@ARTICLE{2006AJ....131..939Z,
       author = {{Zapata}, Luis A. and {Rodr{\'\i}guez}, Luis F. and {Ho}, Paul T.~P. and {Beuther}, Henrik and {Zhang}, Qizhou},
        title = "{In Search of Circumstellar Disks around Young Massive Stars}",
      journal = {\aj},
         year = 2006,
        month = feb,
       volume = {131},
       number = {2},
        pages = {939-950},
          doi = {10.1086/499156},
archivePrefix = {arXiv},
       eprint = {astro-ph/0510761},
 primaryClass = {astro-ph},
       adsurl = {https://ui.adsabs.harvard.edu/abs/2006AJ....131..939Z}
}

@ARTICLE{2004ApJS..154..379M,
       author = {{Muzerolle}, J. and {Megeath}, S.~T. and {Gutermuth}, R.~A. and {Allen}, L.~E. and {Pipher}, J.~L. and {Hartmann}, L. and {Gordon}, K.~D. and {Padgett}, D.~L. and {Noriega-Crespo}, A. and {Myers}, P.~C. and et al.},
        title = "{The 24 Micron View of Embedded Star Formation in NGC 7129}",
      journal = {\apjs},
         year = 2004,
        month = sep,
       volume = {154},
       number = {1},
        pages = {379-384},
          doi = {10.1086/422451},
archivePrefix = {arXiv},
       eprint = {astro-ph/0406065},
 primaryClass = {astro-ph},
       adsurl = {https://ui.adsabs.harvard.edu/abs/2004ApJS..154..379M}
}

@INPROCEEDINGS{2005IAUS..227..111K,
       author = {{Kurtz}, Stan},
        title = "{Hypercompact HII regions}",
    booktitle = {Massive Star Birth: A Crossroads of Astrophysics},
         year = 2005,
       editor = {{Cesaroni}, R. and {Felli}, M. and {Churchwell}, E. and {Walmsley}, M.},
       series = {IAU Symposium},
       volume = {227},
        month = jan,
        pages = {111-119},
          doi = {10.1017/S1743921305004424},
       adsurl = {https://ui.adsabs.harvard.edu/abs/2005IAUS..227..111K}
}

@ARTICLE{2015MNRAS.453L..41S,
       author = {{Schneider}, N. and {Bontemps}, S. and {Girichidis}, P. and {Rayner}, T. and {Motte}, F. and {Andr{\'e}}, P. and {Russeil}, D. and {Abergel}, A. and {Anderson}, L. and {Arzoumanian}, D. and et al.},
        title = "{Detection of two power-law tails in the probability distribution functions of massive GMCs.}",
      journal = {\mnras},
         year = 2015,
        month = nov,
       volume = {453},
        pages = {L41-L45},
          doi = {10.1093/mnrasl/slv101},
archivePrefix = {arXiv},
       eprint = {1507.08869},
 primaryClass = {astro-ph.GA},
       adsurl = {https://ui.adsabs.harvard.edu/abs/2015MNRAS.453L..41S}
}

@ARTICLE{2018ApJ...859..162C,
       author = {{Chen}, Hope How-Huan and {Burkhart}, Blakesley and {Goodman}, Alyssa and {Collins}, David C.},
        title = "{The Anatomy of the Column Density Probability Distribution Function (N-PDF)}",
      journal = {\apj},
         year = 2018,
        month = jun,
       volume = {859},
       number = {2},
          eid = {162},
        pages = {162},
          doi = {10.3847/1538-4357/aabaf6},
archivePrefix = {arXiv},
       eprint = {1707.09356},
 primaryClass = {astro-ph.GA},
       adsurl = {https://ui.adsabs.harvard.edu/abs/2018ApJ...859..162C}
}

@ARTICLE{2022A&A...666A.165S,
       author = {{Schneider}, N. and {Ossenkopf-Okada}, V. and {Clarke}, S. and {Klessen}, R.~S. and {Kabanovic}, S. and {Veltchev}, T. and {Bontemps}, S. and {Dib}, S. and {Csengeri}, T. and {Federrath}, C. and et al.},
        title = "{Understanding star formation in molecular clouds. IV. Column density PDFs from quiescent to massive molecular clouds}",
      journal = {\aap},
         year = 2022,
        month = oct,
       volume = {666},
          eid = {A165},
        pages = {A165},
          doi = {10.1051/0004-6361/202039610},
archivePrefix = {arXiv},
       eprint = {2207.14604},
 primaryClass = {astro-ph.GA},
       adsurl = {https://ui.adsabs.harvard.edu/abs/2022A&A...666A.165S}
}

@ARTICLE{2020ApJ...903L...2J,
       author = {{Jaupart}, Etienne and {Chabrier}, Gilles},
        title = "{Evolution of the Density PDF in Star-forming Clouds: The Role of Gravity}",
      journal = {\apjl},
         year = 2020,
        month = nov,
       volume = {903},
       number = {1},
          eid = {L2},
        pages = {L2},
          doi = {10.3847/2041-8213/abbda8},
archivePrefix = {arXiv},
       eprint = {2010.00603},
 primaryClass = {astro-ph.GA},
       adsurl = {https://ui.adsabs.harvard.edu/abs/2020ApJ...903L...2J}
}

@ARTICLE{2006PASP..118..590R,
       author = {{Rosolowsky}, Erik and {Leroy}, Adam},
        title = "{Bias-free Measurement of Giant Molecular Cloud Properties}",
      journal = {\pasp},
         year = 2006,
        month = apr,
       volume = {118},
       number = {842},
        pages = {590-610},
          doi = {10.1086/502982},
archivePrefix = {arXiv},
       eprint = {astro-ph/0601706},
 primaryClass = {astro-ph},
       adsurl = {https://ui.adsabs.harvard.edu/abs/2006PASP..118..590R}
}

@ARTICLE{2009ApJ...692...91G,
       author = {{Goodman}, Alyssa A. and {Pineda}, Jaime E. and {Schnee}, Scott L.},
        title = "{The ``True'' Column Density Distribution in Star-Forming Molecular Clouds}",
      journal = {\apj},
         year = 2009,
        month = feb,
       volume = {692},
       number = {1},
        pages = {91-103},
          doi = {10.1088/0004-637X/692/1/91},
archivePrefix = {arXiv},
       eprint = {0806.3441},
 primaryClass = {astro-ph},
       adsurl = {https://ui.adsabs.harvard.edu/abs/2009ApJ...692...91G}
}

@ARTICLE{2022ApJ...931....9L,
       author = {{Lewis}, John Arban and {Lada}, Charles J. and {Dame}, T.~M.},
        title = "{Systematic Investigation of Dust and Gaseous CO in 12 Nearby Molecular Clouds}",
      journal = {\apj},
         year = 2022,
        month = may,
       volume = {931},
       number = {1},
          eid = {9},
        pages = {9},
          doi = {10.3847/1538-4357/ac5d58},
archivePrefix = {arXiv},
       eprint = {2203.07480},
 primaryClass = {astro-ph.GA},
       adsurl = {https://ui.adsabs.harvard.edu/abs/2022ApJ...931....9L}
}

@ARTICLE{2014ApJ...781...91G,
       author = {{Girichidis}, Philipp and {Konstandin}, Lukas and {Whitworth}, Anthony P. and {Klessen}, Ralf S.},
        title = "{On the Evolution of the Density Probability Density Function in Strongly Self-gravitating Systems}",
      journal = {\apj},
         year = 2014,
        month = feb,
       volume = {781},
       number = {2},
          eid = {91},
        pages = {91},
          doi = {10.1088/0004-637X/781/2/91},
archivePrefix = {arXiv},
       eprint = {1310.4346},
 primaryClass = {astro-ph.GA},
       adsurl = {https://ui.adsabs.harvard.edu/abs/2014ApJ...781...91G}
}

@ARTICLE{2015A&A...576L...1L,
       author = {{Lombardi}, Marco and {Alves}, Jo{\~a}o and {Lada}, Charles J.},
        title = "{Molecular clouds have power-law probability distribution functions}",
      journal = {\aap},
         year = 2015,
        month = apr,
       volume = {576},
          eid = {L1},
        pages = {L1},
          doi = {10.1051/0004-6361/201525650},
archivePrefix = {arXiv},
       eprint = {1502.03859},
 primaryClass = {astro-ph.SR},
       adsurl = {https://ui.adsabs.harvard.edu/abs/2015A&A...576L...1L}
}

@ARTICLE{2011ApJ...727L..20K,
       author = {{Kritsuk}, Alexei G. and {Norman}, Michael L. and {Wagner}, Rick},
        title = "{On the Density Distribution in Star-forming Interstellar Clouds}",
      journal = {\apjl},
         year = 2011,
        month = jan,
       volume = {727},
       number = {1},
          eid = {L20},
        pages = {L20},
          doi = {10.1088/2041-8205/727/1/L20},
archivePrefix = {arXiv},
       eprint = {1007.2950},
 primaryClass = {astro-ph.GA},
       adsurl = {https://ui.adsabs.harvard.edu/abs/2011ApJ...727L..20K}
}

@ARTICLE{2024MNRAS.528..432V,
       author = {{Veltchev}, Todor V. and {Girichidis}, Philipp and {Marinkova}, Lyubov and {Donkov}, Sava and {Stanchev}, Orlin and {Klessen}, Ralf S.},
        title = "{Multiple power-law tails in the density and column-density distribution in contracting star-forming clumps}",
      journal = {\mnras},
         year = 2024,
        month = feb,
       volume = {528},
       number = {1},
        pages = {432-443},
          doi = {10.1093/mnras/stae031},
archivePrefix = {arXiv},
       eprint = {2401.02148},
 primaryClass = {astro-ph.GA},
       adsurl = {https://ui.adsabs.harvard.edu/abs/2024MNRAS.528..432V}
}

@dataset{10.26131/irsa441,
doi = {10.26131/IRSA441},
url = {https://catcopy.ipac.caltech.edu/dois/doi.php?id=10.26131/IRSA441},
author = {{STScI}},
title = {Digitized Sky Survey (DSS)},
publisher = {IPAC},
year = {2020}
}

@dataset{10.26131/irsa405,
doi = {10.26131/IRSA405},
url = {https://catcopy.ipac.caltech.edu/dois/doi.php?id=10.26131/IRSA405},
author = {{GLIMPSE team}},
title = {Galactic Legacy Infrared Midplane Survey Extraordinaire (GLIMPSE)},
publisher = {IPAC},
year = {2020}
}

@dataset{10.26131/irsa435,
doi = {10.26131/IRSA435},
url = {https://catcopy.ipac.caltech.edu/dois/doi.php?id=10.26131/IRSA435},
author = {{MIPSGAL team}},
title = {A 24 and 70 Micron Survey of the Inner Galactic Disk with MIPS (MIPSGAL)},
publisher = {IPAC},
year = {2020}
}
\bibliographystyle{aasjournal}

\clearpage
\appendix
\restartappendixnumbering
\section{Molecular line parameters}

\begin{table*}[htb]
    \centering
    \begin{tabular}{cccccc}
        \hline \hline
         & & & Regions & & \\
        Parameters & N & NW & SW & SE1 & SE2 \\
        \hline
        $\sigma_1$ & $1.21 \pm 0.01$ & $1.23 \pm 0.02$ & $1.46 \pm 0.50$ & $1.33 \pm 0.02$ & $1.34 \pm 0.02$ \\
        $a_1$ & $2.98 \pm 0.03$ & $3.15 \pm 0.04$ & $0.28 \pm 0.08$ & $3.38 \pm 0.03$ & $3.38 \pm 0.05$ \\
        $v_1$ & $18.40 \pm 0.01$ & $18.33 \pm 0.02$ & $9.80 \pm 0.50$ & $18.05 \pm 0.02$ & $19.23 \pm 0.02$ \\
        $\sigma_2$ & $0.94 \pm 0.08$ & $1.20 \pm 0.09$ & $1.19 \pm 0.03$ & $1.25 \pm 0.06$ & $1.10 \pm 0.13$ \\
        $a_2$ & $0.38 \pm 0.03$ & $0.55 \pm 0.04$ & $2.93 \pm 0.07$ & $0.78 \pm 0.03$ & $0.45 \pm 0.05$ \\
        $v_2$ & $24.75 \pm 0.08$ & $23.99 \pm 0.09$ & $19.08 \pm 0.03$ & $30.19 \pm 0.06$ & $30.16 \pm 0.13$ \\
        $\sigma_3$ & $1.88 \pm 0.02$ & $1.57 \pm 0.05$ & $1.25 \pm 0.04$ &  --  & $3.00 \pm 0.29$ \\
        $a_3$ & $4.72 \pm 0.04$ & $1.48 \pm 0.04$ & $2.66 \pm 0.08$ &  --  & $0.43 \pm 0.08$ \\
        $v_3$ & $32.70 \pm 0.02$ & $31.41 \pm 0.05$ & $30.34 \pm 0.04$ &  --  & $46.63 \pm 0.63$ \\
        $\sigma_4$ &  --  &  --  &  --  &  --  & $1.09 \pm 0.18$ \\
        $a_4$ &  --  &  --  &  --  &  --  & $0.33 \pm 0.05$ \\
        $v_4$ &  --  &  --  &  --  &  --  & $58.29 \pm 0.18$ \\
        Reduced $\chi^2$ & $0.0002$ & $0.0003$ & $0.0013$ & $0.0002$ & $0.0006$ \\
    \hline \hline
    \end{tabular}
    \caption{Parameters obtained after fitting multiple gaussians to the $^{13}$CO spectrum, shown in Figure~\ref{fig:3}. Here, $\sigma$ corresponds to the standard deviation of the gaussian, $a$ corresponds to the amplitude of the gaussian, and $v$ corresponds to the mean velocity corresponding to the gaussian peak.}
    \label{tab:AppA}
\end{table*}

\section{Radio point source catalogue}
\begin{longrotatetable}
\movetabledown=15mm
\begin{deluxetable}{ccccccccccc}
    \tablecaption{The characteristics of the point sources identified from the uGMRT band-4 image. The last five columns show if there are any counterparts correponding to the wavelength bands \\ in the header.}
    \tablehead{\colhead{No.} & \colhead{RA(J2000)} & \colhead{Dec(J2000)} & \colhead{$\theta_{666}$} & \colhead{$\theta_{1300}$} & \colhead{$F_{666}$} & \colhead{$F_{1300}$}  & 
    \colhead{$\alpha_{666-1300}$} & \colhead{\textit{GLIMPSE}} & \colhead{\textit{MIPSGAL}} & \colhead{\textit{Hi-GAL}} \\ 
    \colhead{ } & \colhead{(h:m:s)} & \colhead{(d:m:s)} & \colhead{(arcsec)} & \colhead{(arcsec)} & \colhead{(mJy)} & \colhead{(mJy)} & \colhead{ } & \colhead{$3.6, 4.5, 5.8, 8.0~\mu$m} & \colhead{$24~\mu$m} & \colhead{$70~\mu$m}} 
    \startdata
        1 & 18:11:00.02 & -17:38:25.47 & 6.77 & 0 &  $0.298 \pm 0.030$  &  $-$  &  --  &  N  &  N  &  N \\
        2 & 18:11:00.37 & -17:49:28.53 & 8.88 & 9.87 &  $0.768 \pm 0.077$  &  $0.443 \pm 0.045$  & -0.77 &  N  &  N  &  N \\
        3 & 18:11:01.13 & -17:44:35.40 & 10.13 & 15.23 &  $1.305 \pm 0.131$  &  $4.449 \pm 0.445$  & 1.72 &  YYYY  &  Y  &  Y \\
        4 & 18:11:05.21 & -17:46:12.45 & 23.49 & 25.33 &  $96.619 \pm 9.662$  &  $49.193 \pm 4.919$  & -0.95 &  N  &  N  &  N \\
        5 & 18:11:08.36 & -17:40:05.11 & 7.78 & 0 &  $0.495 \pm 0.050$  &  $-$  &  --  &  N  &  N  &  N \\
        6 & 18:11:09.23 & -17:32:47.87 & 7.29 & 9.72 &  $0.310 \pm 0.032$  &  $0.309 \pm 0.032$  & 0 &  N  &  N  &  N \\
        7 & 18:11:11.76 & -17:40:57.51 & 10.83 & 10.7 &  $1.593 \pm 0.159$  &  $0.907 \pm 0.091$  & -0.79 &  N  &  N  &  N \\
        8 & 18:11:13.21 & -17:25:22.56 & 9.95 & 10.3 &  $0.727 \pm 0.073$  &  $0.546 \pm 0.055$  & -0.4 &  YNNN  &  N  &  N \\
        9 & 18:11:14.44 & -17:33:24.72 & 7.54 & 0 &  $0.204 \pm 0.021$  &  $-$  &  --  &  N  &  N  &  N \\
        10 & 18:11:15.20 & -17:27:52.63 & 14.71 & 14.46 &  $5.665 \pm 0.567$  &  $3.106 \pm 0.311$  & -0.84 &  YYNN  &  N  &  N \\
        11 & 18:11:15.27 & -17:31:35.83 & 10.92 & 9.72 &  $1.561 \pm 0.156$  &  $0.681 \pm 0.069$  & -1.16 &  N  &  N  &  N \\
        12 & 18:11:16.71 & -17:34:18.89 & 7.04 & 0 &  $0.164 \pm 0.017$  &  $-$  &  --  &  N  &  N  &  N \\
        13 & 18:11:17.84 & -17:25:04.18 & 11.96 & 12.32 &  $1.583 \pm 0.158$  &  $1.436 \pm 0.144$  & -0.14 &  N  &  N  &  N \\
        14 & 18:11:18.71 & -17:31:36.65 & 15.96 & 17.26 &  $8.595 \pm 0.860$  &  $6.107 \pm 0.611$  & -0.48 &  N  &  N  &  N \\
        15 & 18:11:20.94 & -17:31:39.87 & 8.56 & 0 &  $0.546 \pm 0.055$  &  $-$  &  --  &  N  &  N  &  N \\
        16 & 18:11:21.55 & -17:38:06.11 & 18.42 & 23.51 &  $57.810 \pm 5.781$  &  $31.797 \pm 3.180$  & -0.84 &  N  &  N  &  N \\
        17 & 18:11:22.35 & -17:37:43.84 & 16.53 & 20.73 &  $74.265 \pm 7.426$  &  $47.971 \pm 4.797$  & -0.61 &  N  &  N  &  N \\
        18 & 18:11:23.15 & -17:37:42.61 & 11.96 & 17.91 &  $24.739 \pm 2.474$  &  $25.807 \pm 2.581$  & 0.06 &  N  &  N  &  N \\
        19 & 18:11:25.09 & -17:30:30.51 & 10.04 & 10.3 &  $0.868 \pm 0.087$  &  $0.548 \pm 0.055$  & -0.64 &  N  &  N  &  N \\
        20 & 18:11:26.18 & -17:42:41.98 & 6.77 & 0 &  $0.303 \pm 0.031$  &  $-$  &  --  &  N  &  N  &  N \\
        21 & 18:11:27.60 & -17:33:20.55 & 16.02 & 19.59 &  $14.093 \pm 1.409$  &  $13.959 \pm 1.396$  & -0.01 &  N  &  N  &  N \\
        22 & 18:11:30.59 & -17:44:50.37 & 7.04 & 0 &  $0.174 \pm 0.018$  &  $-$  &  --  &  N  &  N  &  N \\
        23 & 18:11:31.13 & -17:35:22.33 & 7.42 & 0 &  $0.345 \pm 0.035$  &  $-$  &  --  &  N  &  N  &  N \\
        24 & 18:11:35.86 & -17:39:23.06 & 14.33 & 15.7 &  $3.582 \pm 0.358$  &  $2.002 \pm 0.200$  & -0.82 &  N  &  N  &  N \\
        \enddata
    \label{tab:C_1}
\end{deluxetable}
\end{longrotatetable}

\begin{longrotatetable}
\movetabledown=15mm
\begin{deluxetable}{ccccccccccc}
    \tablecaption{The characteristics of the point sources identified from the uGMRT band-4 image. The last five columns show if there are any counterparts correponding to the wavelength bands \\ in the header.}
    \tablehead{\colhead{No.} & \colhead{RA(J2000)} & \colhead{Dec(J2000)} & \colhead{$\theta_{666}$} & \colhead{$\theta_{1300}$} & \colhead{$F_{666}$} & \colhead{$F_{1300}$}  & 
    \colhead{$\alpha_{666-1300}$} & \colhead{\textit{GLIMPSE}} & \colhead{\textit{MIPSGAL}} & \colhead{\textit{Hi-GAL}} \\ 
    \colhead{ } & \colhead{(h:m:s)} & \colhead{(d:m:s)} & \colhead{(arcsec)} & \colhead{(arcsec)} & \colhead{(mJy)} & \colhead{(mJy)} & \colhead{ } & \colhead{$3.6, 4.5, 5.8, 8.0~\mu$m} & \colhead{$24~\mu$m} & \colhead{$70~\mu$m}} 
        \startdata
        25 & 18:11:36.53 & -17:42:12.63 & 9.38 & 10.16 &  $0.587 \pm 0.059$  &  $0.223 \pm 0.024$  & -1.36 &  YYNN  &  N  &  N \\
        26 & 18:11:37.94 & -17:46:03.54 & 6.77 & 0 &  $0.165 \pm 0.017$  &  $-$  &  --  &  N  &  N  &  N \\
        27 & 18:11:38.91 & -17:44:16.93 & 7.04 & 0 &  $0.231 \pm 0.024$  &  $-$  &  --  &  N  &  N  &  N \\
        28 & 18:11:39.01 & -17:43:05.48 & 7.66 & 10.97 &  $0.364 \pm 0.037$  &  $0.486 \pm 0.049$  & 0.4 &  N  &  N  &  N \\
        29 & 18:11:39.39 & -17:45:47.06 & 7.16 & 10.84 &  $0.354 \pm 0.036$  &  $0.415 \pm 0.042$  & 0.22 &  N  &  N  &  N \\
        30 & 18:11:41.95 & -17:36:45.36 & 9.86 & 12.32 &  $0.623 \pm 0.063$  &  $0.656 \pm 0.066$  & 0.07 &  N  &  N  &  N \\
        31 & 18:11:42.99 & -17:41:42.69 & 8.88 & 10.57 &  $0.487 \pm 0.049$  &  $0.548 \pm 0.056$  & 0.17 &  N  &  N  &  N \\
        32 & 18:11:48.88 & -17:31:50.34 & 15.44 & 17.51 &  $2.787 \pm 0.279$  &  $3.507 \pm 0.351$  & 0.32 &  N  &  N  &  N \\
        33 & 18:11:50.42 & -17:31:13.56 & 10.31 & 11.48 &  $1.145 \pm 0.115$  &  $1.154 \pm 0.116$  & 0.01 &  N  &  N  &  N \\
        34 & 18:11:51.45 & -17:31:29.86 & 6.9 & 0 &  $0.131 \pm 0.014$  &  $-$  &  --  &  YYYN  &  N  &  N \\
        35 & 18:11:51.53 & -17:31:40.87 & 7.04 & 0 &  $0.205 \pm 0.021$  &  $-$  &  --  &  N  &  N  &  N \\
        36 & 18:11:52.15 & -17:38:36.65 & 10.4 & 13.22 &  $0.742 \pm 0.075$  &  $0.478 \pm 0.049$  & -0.62 &  N  &  N  &  N \\
        37 & 18:11:52.54 & -17:32:04.42 & 11.88 & 17.26 &  $0.788 \pm 0.079$  &  $2.080 \pm 0.208$  & 1.36 &  YYYN  &  N  &  N \\
        38 & 18:11:53.64 & -17:39:31.20 & 8.01 & 9.87 &  $0.192 \pm 0.020$  &  $0.141 \pm 0.016$  & -0.44 &  N  &  N  &  N \\
        39 & 18:11:53.77 & -17:28:52.80 & 8.35 & 0 &  $0.590 \pm 0.059$  &  $-$  &  --  &  N  &  N  &  N \\
        40 & 18:11:54.36 & -17:36:29.31 & 18.76 & 17.26 &  $9.187 \pm 0.919$  &  $6.049 \pm 0.605$  & -0.59 &  N  &  N  &  N \\
        41 & 18:11:55.32 & -17:30:00.96 & 14.27 & 16.41 &  $2.473 \pm 0.247$  &  $3.180 \pm 0.318$  & 0.35 &  N  &  N  &  Y \\
        42 & 18:11:56.51 & -17:43:54.73 & 14.39 & 24.64 &  $3.353 \pm 0.335$  &  $7.065 \pm 0.707$  & 1.04 &  YYNN  &  N  &  Y \\
        43 & 18:11:58.31 & -17:28:39.36 & 13.13 & 14.56 &  $6.057 \pm 0.606$  &  $3.639 \pm 0.364$  & -0.71 &  N  &  N  &  N \\
        44 & 18:11:58.66 & -17:49:24.82 & 6.9 & 11.73 &  $0.291 \pm 0.030$  &  $0.434 \pm 0.044$  & 0.56 &  N  &  N  &  N \\
        45 & 18:11:59.94 & -17:51:33.16 & 6.9 & 0 &  $0.190 \pm 0.020$  &  $-$  &  --  &  N  &  N  &  N \\
        46 & 18:12:00.38 & -17:34:26.10 & 6.77 & 0 &  $0.316 \pm 0.032$  &  $-$  &  --  &  N  &  N  &  N \\
        47 & 18:12:03.59 & -17:48:09.33 & 10.58 & 12.21 &  $0.912 \pm 0.091$  &  $0.980 \pm 0.098$  & 0.1 &  YYNN  &  N  &  N \\
        48 & 18:12:04.27 & -17:41:47.51 & 9.86 & 12.21 &  $0.242 \pm 0.025$  &  $0.134 \pm 0.016$  & -0.83 &  N  &  N  &  N \\
        \enddata
    \label{tab:C_2}
\end{deluxetable}
\end{longrotatetable}

\begin{longrotatetable}
\movetabledown=15mm
\begin{deluxetable}{ccccccccccc}
    \tablecaption{The characteristics of the point sources identified from the uGMRT band-4 image. The last five columns show if there are any counterparts correponding to the wavelength bands \\ in the header.}
    \tablehead{\colhead{No.} & \colhead{RA(J2000)} & \colhead{Dec(J2000)} & \colhead{$\theta_{666}$} & \colhead{$\theta_{1300}$} & \colhead{$F_{666}$} & \colhead{$F_{1300}$}  & 
    \colhead{$\alpha_{666-1300}$} & \colhead{\textit{GLIMPSE}} & \colhead{\textit{MIPSGAL}} & \colhead{\textit{Hi-GAL}} \\ 
    \colhead{ } & \colhead{(h:m:s)} & \colhead{(d:m:s)} & \colhead{(arcsec)} & \colhead{(arcsec)} & \colhead{(mJy)} & \colhead{(mJy)} & \colhead{ } & \colhead{$3.6, 4.5, 5.8, 8.0~\mu$m} & \colhead{$24~\mu$m} & \colhead{$70~\mu$m}} 
        \startdata
         49 & 18:12:05.58 & -17:25:33.79 & 7.29 & 0 &  $0.327 \pm 0.033$  &  $-$  &  --  &  N  &  N  &  N \\
         50 & 18:12:11.33 & -17:49:57.53 & 10.58 & 10.57 &  $1.223 \pm 0.122$  &  $0.819 \pm 0.082$  & -0.56 &  YYNN  &  N  &  N \\
         51 & 18:12:12.20 & -17:30:57.14 & 6.9 & 0 &  $0.218 \pm 0.022$  &  $-$  &  --  &  YYNN  &  N  &  N \\
         52 & 18:12:15.39 & -17:32:25.95 & 12.41 & 14.95 &  $6.221 \pm 0.622$  &  $4.875 \pm 0.488$  & -0.34 &  N  &  N  &  N \\
         53 & 18:12:15.87 & -17:33:39.98 & 13.81 & 14.95 &  $13.960 \pm 1.396$  &  $4.459 \pm 0.446$  & -1.6 &  N  &  N  &  N \\
         54 & 18:12:17.00 & -17:24:17.46 & 7.9 & 10.01 &  $0.426 \pm 0.043$  &  $0.426 \pm 0.043$  & 0 &  N  &  N  &  N \\
         55 & 18:12:20.21 & -17:29:28.30 & 7.9 & 0 &  $0.507 \pm 0.051$  &  $-$  &  --  &  YYYN  &  N  &  N \\
         56 & 18:12:20.65 & -17:47:05.44 & 15.44 & 16.23 &  $7.997 \pm 0.800$  &  $5.114 \pm 0.512$  & -0.63 &  N  &  N  &  N \\
         57 & 18:12:21.20 & -17:34:07.83 & 15.02 & 16.15 &  $6.375 \pm 0.638$  &  $4.045 \pm 0.405$  & -0.64 &  N  &  N  &  N \\
         58 & 18:12:21.77 & -17:49:39.84 & 8.67 & 0 &  $0.713 \pm 0.072$  &  $-$  &  --  &  YYNN  &  N  &  N \\
         59 & 18:12:22.17 & -17:43:11.66 & 10.04 & 17.01 &  $1.091 \pm 0.109$  &  $2.448 \pm 0.245$  & 1.13 &  N  &  N  &  N \\
         60 & 18:12:27.38 & -17:52:09.69 & 11 & 0 &  $0.862 \pm 0.086$  &  $-$  &  --  &  N  &  N  &  N \\
         61 & 18:12:29.33 & -17:36:10.85 & 8.35 & 10.01 &  $0.537 \pm 0.054$  &  $0.359 \pm 0.037$  & -0.56 &  N  &  N  &  N \\
         62 & 18:12:30.15 & -17:52:01.18 & 10.13 & 0 &  $1.202 \pm 0.120$  &  $-$  &  --  &  YYNN  &  N  &  N \\
         63 & 18:12:30.60 & -17:35:11.20 & 13.4 & 15.14 &  $8.316 \pm 0.832$  &  $4.677 \pm 0.468$  & -0.81 &  N  &  N  &  N \\
         64 & 18:12:31.21 & -17:40:03.34 & 8.98 & 0 &  $0.312 \pm 0.032$  &  $-$  &  --  &  YYNN  &  N  &  N \\
         65 & 18:12:37.41 & -17:29:51.52 & 9.48 & 9.87 &  $1.146 \pm 0.115$  &  $0.738 \pm 0.074$  & -0.62 &  N  &  N  &  N \\
         66 & 18:12:38.34 & -17:33:52.79 & 7.04 & 11.6 &  $0.316 \pm 0.032$  &  $0.352 \pm 0.036$  & 0.15 &  N  &  N  &  N \\
         67 & 18:12:43.46 & -17:42:53.88 & 15.08 & 18.85 &  $70.411 \pm 7.041$  &  $40.394 \pm 4.039$  & -0.78 &  YYNN  &  N  &  N \\
         68 & 18:12:43.99 & -17:40:04.97 & 7.9 & 0 &  $0.306 \pm 0.031$  &  $-$  &  --  &  N  &  N  &  N \\
         69 & 18:12:44.13 & -17:42:59.85 & 12.63 & 14.16 &  $25.552 \pm 2.555$  &  $18.325 \pm 1.833$  & -0.47 &  N  &  N  &  N \\
         70 & 18:12:48.21 & -17:38:02.72 & 8.12 & 9.72 &  $0.433 \pm 0.044$  &  $0.472 \pm 0.048$  & 0.12 &  YYNN  &  N  &  N \\
        \enddata
    \label{tab:C_3}
\end{deluxetable}
\end{longrotatetable}

\section{Catalogue of YSOs identified}
\startlongtable
\begin{deluxetable*}{ccccc}
    \tablecaption{Catalogue of YSOs identified}
    \tablehead{\colhead{No.} & \colhead{Name} & \colhead{RA(J2000)} & \colhead{Dec(J2000)} & \colhead{Class} \\ 
    \colhead{} & \colhead{} & \colhead{(h:m:s)} & \colhead{(d:m:s)} & \colhead{}}
    \startdata
    1 & MG012.7586+00.6768 & 18:10:54.29 & -17:32:54.82 & Class I \\
    2 & MG012.7695+00.6737 & 18:10:56.29 & -17:32:26.07 & Class I \\
    3 & MG012.5049+00.5276 & 18:10:56.41 & -17:50:34.04 & Class II \\
    4 & MG012.6495+00.5880 & 18:11:00.63 & -17:41:13.27 & Class II \\
    5 & MG012.6012+00.5592 & 18:11:01.13 & -17:44:35.52 & Class I \\
    6 & MG012.6922+00.5810 & 18:11:07.37 & -17:39:10.81 & Class II \\
    7 & MG012.5370+00.4944 & 18:11:07.63 & -17:49:50.49 & Class I/II \\
    8 & MG012.5750+00.5121 & 18:11:08.34 & -17:47:19.74 & Class I \\
    9 & MG012.7029+00.5806 & 18:11:08.74 & -17:38:37.67 & Class I \\
    10 & MG012.6159+00.5282 & 18:11:09.75 & -17:44:42.81 & Hot excess \\
    11 & MG012.6704+00.5495 & 18:11:11.68 & -17:41:13.95 & Class I \\
    12 & MG012.5880+00.5001 & 18:11:12.58 & -17:46:59.62 & Class II \\
    13 & MG012.8352+00.6344 & 18:11:12.93 & -17:30:06.83 & Class II \\
    14 & MG012.6199+00.5142 & 18:11:13.34 & -17:44:54.46 & Class I \\
    15 & MG012.6563+00.5317 & 18:11:13.89 & -17:42:29.52 & Class II \\
    16 & MG012.8415+00.6211 & 18:11:16.61 & -17:30:10.18 & Class II \\
    17 & MG012.8125+00.6003 & 18:11:17.69 & -17:32:17.68 & Class I/II \\
    18 & MG012.8334+00.6090 & 18:11:18.31 & -17:30:56.51 & Class I \\
    19 & MG012.6796+00.5240 & 18:11:18.41 & -17:41:29.26 & Hot excess \\
    20 & MG012.8115+00.5906 & 18:11:19.71 & -17:32:37.59 & Class I/II \\
    21 & MG012.6922+00.5234 & 18:11:20.07 & -17:40:50.49 & Class II \\
    22 & MG012.7029+00.5209 & 18:11:21.92 & -17:40:21.00 & Class II \\
    23 & MG012.8681+00.5998 & 18:11:24.54 & -17:29:23.10 & Class I \\
    24 & MG012.9210+00.6262 & 18:11:25.12 & -17:25:50.28 & Hot excess \\
    25 & MG012.8716+00.5937 & 18:11:26.31 & -17:29:22.60 & Class II \\
    26 & MG012.6660+00.4781 & 18:11:26.92 & -17:43:31.64 & Hot excess \\
    27 & MG012.7688+00.5284 & 18:11:28.28 & -17:36:40.10 & Class I \\
    28 & MG012.7990+00.5364 & 18:11:30.15 & -17:34:50.92 & Class I/II \\
    29 & MG012.6961+00.4747 & 18:11:31.32 & -17:42:02.51 & Class I \\
    30 & MG012.7033+00.4761 & 18:11:31.85 & -17:41:37.50 & Class I \\
    31 & MG012.7866+00.5196 & 18:11:32.36 & -17:35:59.36 & Hot excess \\
    32 & MG012.6988+00.4685 & 18:11:33.00 & -17:42:05.00 & Class II \\
    33 & MG012.7140+00.4620 & 18:11:36.28 & -17:41:28.10 & Class II \\
    34 & MG012.8929+00.5557 & 18:11:37.27 & -17:29:21.44 & Class I \\
    35 & MG012.6411+00.4147 & 18:11:37.89 & -17:46:39.93 & Class I \\
    36 & MG012.8886+00.5481 & 18:11:38.42 & -17:29:48.06 & Class I \\
    37 & MG012.9214+00.5591 & 18:11:39.96 & -17:27:45.42 & Class II \\
    38 & MG012.7820+00.4782 & 18:11:40.93 & -17:37:25.56 & Class II \\
    39 & MG012.8317+00.4986 & 18:11:42.46 & -17:34:13.20 & Class I \\
    40 & MG012.7580+00.4255 & 18:11:49.67 & -17:40:12.63 & Class II \\
    41 & MG012.9529+00.5255 & 18:11:51.20 & -17:27:04.06 & Class II \\
    42 & MG012.8031+00.4415 & 18:11:51.61 & -17:37:22.36 & Class II \\
    43 & MG012.8168+00.4483 & 18:11:51.76 & -17:36:27.41 & Class II \\
    44 & MG012.7705+00.4182 & 18:11:52.80 & -17:39:45.82 & Class I \\
    45 & MG012.6807+00.3573 & 18:11:55.38 & -17:46:14.65 & Class II \\
    46 & MG012.6311+00.3162 & 18:11:58.45 & -17:50:02.33 & Class II \\
    47 & MG012.7935+00.4028 & 18:11:58.97 & -17:38:59.78 & Class II \\
    48 & MG012.6241+00.3094 & 18:11:59.10 & -17:50:35.99 & Class II \\
    49 & MG012.8271+00.4201 & 18:11:59.24 & -17:36:43.66 & Class I \\
    50 & MG012.6946+00.3465 & 18:11:59.44 & -17:45:49.41 & Class I \\
    51 & MG012.8752+00.4424 & 18:12:00.14 & -17:33:33.45 & Class I \\
    52 & MG012.9598+00.4849 & 18:12:00.98 & -17:27:52.61 & Class II \\
    53 & MG012.9164+00.4444 & 18:12:04.66 & -17:31:19.80 & Class I \\
    54 & MG012.9257+00.4455 & 18:12:05.55 & -17:30:48.43 & Class II \\
    55 & MG012.7777+00.3618 & 18:12:06.14 & -17:41:00.42 & Hot excess \\
    56 & MG012.9888+00.4772 & 18:12:06.19 & -17:26:34.31 & Class II \\
    57 & MG012.8580+00.4049 & 18:12:06.34 & -17:35:32.40 & Class II \\
    58 & MG012.8097+00.3764 & 18:12:06.78 & -17:38:54.20 & Class I/II \\
    59 & MG012.8990+00.4250 & 18:12:06.86 & -17:32:48.16 & Class II \\
    60 & MG012.8201+00.3806 & 18:12:07.11 & -17:38:14.11 & Class I/II \\
    61 & MG012.8224+00.3802 & 18:12:07.48 & -17:38:07.53 & Class II \\
    62 & MG012.8079+00.3641 & 18:12:09.29 & -17:39:21.12 & Class II \\
    63 & MG012.7349+00.3223 & 18:12:09.68 & -17:44:23.97 & Class II \\
    64 & MG012.9747+00.4486 & 18:12:10.80 & -17:28:08.29 & Hot excess \\
    65 & MG012.6452+00.2468 & 18:12:15.49 & -17:51:17.64 & Class II \\
    66 & MG012.8188+00.3389 & 18:12:16.15 & -17:39:30.18 & Class II \\
    67 & MG012.6514+00.2286 & 18:12:20.27 & -17:51:29.59 & Class II \\
    68 & MG012.8387+00.3210 & 18:12:22.53 & -17:38:58.43 & Class I \\
    69 & MG012.9995+00.3953 & 18:12:25.56 & -17:28:22.25 & Class II \\
    70 & MG012.9351+00.3557 & 18:12:26.53 & -17:32:54.11 & Class I \\
    71 & MG012.7089+00.2126 & 18:12:30.78 & -17:48:55.71 & Class II \\
    72 & MG012.9866+00.3552 & 18:12:32.85 & -17:30:12.26 & Class I \\
    73 & MG012.7251+00.1995 & 18:12:35.64 & -17:48:27.14 & Class II \\
    74 & MG012.7901+00.2283 & 18:12:37.13 & -17:44:12.18 & Class II \\
    75 & MG012.8427+00.2518 & 18:12:38.29 & -17:40:45.33 & Class I \\
    76 & MG012.7092+00.1759 & 18:12:38.93 & -17:49:58.39 & Class I \\
    77 & MG012.9178+00.2779 & 18:12:41.60 & -17:36:03.03 & Class II \\
    78 & MG012.7335+00.1730 & 18:12:42.51 & -17:48:46.59 & Class I \\
    79 & MG012.8390+00.2238 & 18:12:44.04 & -17:41:45.62 & Class II \\
    80 & MG012.8185+00.2092 & 18:12:44.79 & -17:43:15.55 & Class I \\
    81 & MG012.8115+00.1859 & 18:12:49.08 & -17:44:17.90 & Hot excess \\
    82 & MG012.9327+00.2489 & 18:12:49.82 & -17:36:06.20 & Class II \\
    \enddata
    \label{tab:B}
\end{deluxetable*}

\section{Notes on individual star-forming regions} \label{sec:notes}
\subsection{G12.79 - N} 
The G12.79 - N region is associated with the well studied object IRAS~18089-1732. \citet{2002ApJ...566..931S} identified a high mass protostellar object, and subsequent submillimetre investigations carried out by \citet{2004ApJ...616L..19B, 2004ApJ...616L..23B} revealed several molecules indicative of hot cores, as well as protostellar outflows. \citet{2006AJ....131..939Z} had identified four \hii regions around the IRAS object. Multiple studies have explored the astrochemistry of the region \citep{2014MNRAS.443.2252G, 2015ApJS..219....2J, 2017ApJ...844...68T}. Methanol, hydroxyl, as well as water masers have been detected \citep{2002A&A...390..289B, 2007A&A...465..865E, 2010A&A...517A..56F}. Two outflow candidates have been identified by \citet{2022A&A...658A.160Y}. No YSOs have been identified towards this region in the current work.

\subsection{G12.79 - NW}
\citet{2011ApJS..194...32A} discovered and classified this region as an irregular \hii region. Using radio recombination lines, they estimated the LSR velocity as $\sim 21.4$~km/s. Based on the MIR morphology, \citet{2019MNRAS.488.1141J} classified the region as an infrared bubble. \citet{2011ApJ...731...90D} identified 2 clumps in the region.

\subsection{G12.79 - SW}
The region in question appears an an arc in the GLIMPSE $5.8~\mu$m data and associated filamentary extinction features are observed in the optical as well as near-infrared (NIR) bands toward the centre. There are no radio sources identified in the region till date. \citet{2011ApJ...731...90D} detected three 1.1~mm cores towards the region, two of which are located on the arc. \citet{2022MNRAS.510.3389U} identified a single clump. Additionally, \citet{2022A&A...658A.160Y} identified an outflow in the region. 

\subsection{G12.79 - SE1}
The G12.79 - SE1 region hosts the IRAS object 18092-1742, which is classified as an UC \hii region by \citet{1996A&AS..115...81B}. This region is interesting owing to the presence of several structures in MIR maps, such as, arcs, bubbles, etc., as well as infrared extinction features (see Figure~\ref{fig:1}). \citet{2017MNRAS.469.2163E} detected 5 compact sources in the region, that are potential star-forming clumps. \citet{2008AJ....136.2413R} and \citet{2021ApJS..254...33K}  identified several intrinsically red sources and young stellar objects (YSOs), respectively, suggesting active star formation in the region.

\subsection{G12.79 - SE2}
This region is contiguous with G12.79 - SE1, located toward the southwest of the former. \citet{2022MNRAS.510.3389U} identified an ATLASGAL clump. Optical and near-infrared images reveal extinction features (see Figure~\ref{fig:1}), suggesting the presence of dust lanes. Using $^{13}$CO and C$^{18}$O data from the SEDIGISM survey, \citet{2022A&A...658A.160Y} identified an outflow in the region. \citet{2016ApJ...822...59S} detected two clumps, one possibly star-forming, and another starless, within the region. YSOs have also been detected towards the region in several studies, making the region a suitable candidate for further investigation.




\end{document}